\pdfoutput=1

\documentclass[10pt,journal,compsoc]{IEEEtran}
\usepackage{setspace}
\usepackage{amsmath}
\usepackage{amssymb}
\usepackage{mathtools}
\usepackage{amsthm}
\usepackage{multirow}

\usepackage{lineno}
\usepackage{graphicx}	
\usepackage{amsmath}
\usepackage{amssymb}
\usepackage{subfig}
\usepackage{wrapfig}
\usepackage{enumitem}
\usepackage[ruled,vlined]{algorithm2e}
\usepackage{booktabs}
\usepackage{xcolor}

\ifCLASSOPTIONcompsoc
  \usepackage[nocompress]{cite}
\else
  \usepackage{cite}
\fi
\ifCLASSINFOpdf
\else
\fi
\begin{document}
%
\title{Mesh Total Generalized Variation for Denoising}
%
%
%
%

\author{Zheng~Liu,
        Yanlei~Li,
        Weina~Wang,
        Ligang~Liu,
        and~Renjie~Chen$^{\dagger}$
\IEEEcompsocitemizethanks{
\IEEEcompsocthanksitem Z. Liu and Y. Li are with National Engineering Research Center of Geographic Information System, School of Geography and Information Engineering, China University of Geosciences (Wuhan).
\IEEEcompsocthanksitem W. Wang is with Department of Mathematics, Hangzhou Dianzi University.
\IEEEcompsocthanksitem R. Chen and L. Liu are with School of Mathematical Sciences, University of Science and Technology of China.}
\thanks{$^{\dagger}$Corresponding author. E-mail:renjiec@ustc.edu.cn.}}
\IEEEtitleabstractindextext{
\begin{abstract}
Recent studies have shown that the Total Generalized Variation (TGV) is highly effective in preserving sharp features as well as smooth transition variations for image processing tasks.
However, currently there is no existing work that is suitable for applying TGV to 3D data, in particular, triangular meshes.
In this paper, we develop a novel framework for discretizing second-order TGV on triangular meshes.
Further, we propose a TGV-based variational method for the denoising of face normal fields on triangular meshes.
The TGV regularizer in our method is composed of a first-order term and a second-order term, which are automatically balanced.
The first-order term allows our TGV regularizer to locate and preserve sharp features, while the second-order term allows our regularizer to recognize and recover smoothly curved regions.
To solve the optimization problem, we introduce an efficient iterative algorithm based on variable-splitting and augmented Lagrangian method.
Extensive results and comparisons on synthetic and real scanning data validate that the proposed method outperforms the state-of-the-art visually and numerically.
\end{abstract}
}
\maketitle

\IEEEdisplaynontitleabstractindextext

%
\IEEEpeerreviewmaketitle

\IEEEraisesectionheading{\section{Introduction}\label{sec:introduction}}
\IEEEPARstart{M}{esh} denoising is one of the most fundamental research topics in geometry processing.
With the rapid development of 3D scanning devices and depth cameras, it has become increasingly popular and common to acquire and reconstruct meshes from the real world automatically ~\cite{Liu2020AFeature}.
However, the acquired meshes are inevitably contaminated by noise because of local measurement errors in the scanning process and computational errors in the reconstruction algorithm used.
Noise not only degrades the quality of meshes, but also causes problems in downstream geometry processing applications \cite{Wang2012A}.
Thus, mesh denoising has been a widely studied topic in recent years,
whose main purpose is to remove noise while recovering geometric features as accurately as possible \cite{Wang2019Data}.
However, noise and geometric features are both of high frequency information, which makes it challenging to distinguish them from noisy input, especially in the presence of large noise.

To suppress noise while preserving geometric features, various mesh denoising methods have been investigated, including filtering-based methods \cite{Sun07Fast,Zheng11Bilateral,wei2014bi}, variational methods \cite{Lai13,Zhang15Variational,wu2015mesh,lu2015robust,Lu2017Robust,Liu2020Mesh}, nonlocal-based methods \cite{Li2018NonLocal,Wei2018Mesh,Chen2019Structure}, data-driven methods \cite{Wang2016Mesh,Wang2019Data,Wei2019Mesh,Li2020DNF}, etc.
Among them, variational methods have attracted much attention, as they can well preserve sharp features while suppressing noise significantly.

The variational method usually consists of a regularization term and a fidelity term.
The total variation (TV) regularizer is known for its excellent edge-preserving capability in image processing, and it has been extended by Zhang et al. \cite{Zhang15Variational} to restore the face normal field for triangular meshes.
However, as the TV regularizer uses the first-order operator, it tends to transform smooth transition variations into piecewise constant ones.
Hence, the TV regularizer often suffers from staircase artifacts in smooth regions.
These artifacts degrade the visual quality of the denoised result, which may induce false features that do not exist in smooth regions.
To reduce the undesired artifacts from TV, several higher-order regularizer \cite{Liu2019ANovel,Liu2019Triangulated} have been proposed. They can preserve geometric features and simultaneously prevent staircase artifacts.
Unfortunately, when the noise level is high, these higher-order method may blur geometric features in varying degrees.
To address these issues, it is natural to combine the first- and higher-order terms.
For example, Zhong et al. \cite{Zhong2019Robust} proposed a variational method, which combines a first- and a higher-order terms directly.
Although their method performs better than TV, it still has some artifacts near sharp features and flattens fine details more or less.
Thus, although some variational methods have been introduced, it is still challenging to find one regularization technique, which can effectively preserve sharp features in some parts of the surface while simultaneously recovering smooth regions in other parts.

Recently, the total generalized variation (TGV), proposed by Bredies et al. \cite{Bredies2010Total}, has become one of the most popular regularization technique in image processing.
TGV is composed of polynomials of arbitrary order, which can reconstruct piecewise polynomial functions with automatically balanced first- and higher-order variations rather than using fixed combination \cite{Ferstl2014Image,Jung2015Simultaneous}.
TGV can be interpreted as combining smoothness from the first-order up to arbitrary order variations.
It preserves sharp features via the first-order variations while effectively approximates smooth transition regions via the higher-order variations.
As a result, it does not produce staircase artifacts.
For most signal processing tasks, we believe that the second-order variant of TGV is sufficient.
The reason is that most signals can be approximated with piecewise linear functions, and relatively it is more difficult to discretize higher-order TGV.
Therefore, in this work, we focus on the second-order TGV, and we refer TGV in particular to its second-order version throughout the paper.

It is non-trivial to extend typical methods for 2D image processing to 3D mesh data because of the inherent data irregularities in meshes.
To the best of our knowledge, despite that TGV has achieved great success in image processing (e.g. image restoration \cite{Bredies2010Total}, depth upsampling \cite{Ferstl2014Image}, speckle reduction \cite{Feng2014Speckle}, texture decomposition \cite{Jung2015Simultaneous}, image reconstruction \cite{Knoll2011Second,Niu2014Sparse}), currently, there is no existing work that applies TGV to triangular meshes.
In this paper, we develop a numerical framework to discretize TGV over triangular meshes.
Based on this discretization, a vectorial TGV regularizer is proposed for face normal field.
Then, we introduce an efficient and effective algorithm to solve the problem.
The main contributions of this work include:
\begin{itemize}
  \item We define necessary discrete operators and further use these operators to apply TGV to triangular meshes. To the best of our knowledge, this is the first numerical framework for discretizing TGV on triangular meshes.
  \item We present a normal filter using TGV-based regularization. Our method is able to preserve sharp features and recover smooth regions, while preventing unnatural artifacts.
     We solve the optimization problem using variable-splitting and augmented Lagrangian method.
  \item  Qualitative and quantitative experiments on synthetic and scanned data show that our denoising method performs favorably against the state-of-the-art methods.

\end{itemize}

The rest of this paper is organized as follows.
A brief survey of mesh denoising methods is presented in section \ref{sec:2}.
Section \ref{sec:3} recalls TGV in image processing.
Section \ref{sec:4} presents the numerical framework for discretizing TGV and its vectorial version for meshes.
In section \ref{sec:5}, we propose a vectorial TGV based face normal filter, and we introduce an augmented Lagrangian method for solving the optimization problem.
Section \ref{sec:6} shows the results of our TGV-based method and further compares it to the state-of-the-art methods visually and quantitatively.
Finally, we conclude with remarks and discuss directions for future work in section \ref{sec:7}.

\section{Related work} \label{sec:2}
Due to the abundance of mesh denoising methods in the literature, it is beyond our scope to review all existing work.
In this section, we review four categories of research that are most relevant to this work.

\textbf{Filter-based methods}. The spatial filtering methods first compute filtering weights based on signal similarities, and then average the neighboring signals in each local region with the computed weights.
Early spatial filtering methods directly adopt isotropic smoothing \cite{Taubin1995A,desbrun1999implicit} or anisotropic smoothing \cite{Bajaj2003Anisotropic,fleishman2003bilateral,jones2003non,Wang2006Bilateral,pan2020hlo} to mesh vertices in order to remove noise.
Although anisotropic methods are more robust against noise compared to the isotropic smoothing methods,
they are still unable to preserve geometric features in the case of heavy noise.

Recently, it has become so widespread with normal filtering followed by vertex updating that it could arguably replace direct vertex position smoothing \cite{Sun07Fast,Wei2017Tensor}.
Zheng et al. \cite{Zheng11Bilateral} applied the bilateral filter to face normal field.
Although their method preserves geometric features, it may blur sharp features when the noise level increases.
To address this problem, Zhang et al.\cite{Zhang2015Guided} introduced a bilateral normal filter based on a well-designed guidance normal field.
Later on, Zhang et al. \cite{Zhang19Static} proposed a scale-aware normal filter using both static and dynamic guidance.
Yadav et al. \cite{Yadav18Mesh} proposed a normal filter based on tensor voting and binary optimization.
Furthermore, Yadav et al. \cite{Yadav2018Robust} developed a normal filter in the robust statistics framework that can preserve sharp features, but may smooth weak features and fine details.
Arvanitis et al. \cite{arvanitis2018feature} introduced a coarse-to-fine framework for the restoration of face normal field based on graph spectral processing.
Zhao et al. \cite{zhao2019graph} presented a feature-preserving normal filter, by first computing the guidance normal field using the graph-cut scheme, and then performing normal filtering using the guidance field.

\textbf{Variational methods}.
For mesh denoising, variational methods aim to find and apply appropriate priors in order to formulate the denoising process as an optimization problem.
Based on the prior that geometric features are sparse over the underlying surface, sparse regularizers are typically applied in the variational methods for the recovering of geometric features.

He and Schaefer \cite{He13Mesh} and Zhao et al. \cite{ZHAO2018Robust} applied the $\ell_0$ minimization to triangular meshes based on the piecewise constant prior.
These $\ell_0$ minimization methods achieve impressive results in preserving sharp features, but inevitably flatten weak features because of their high sparsity requirement.
Moreover, since the $\ell_0$ minimization is a NP-problem, the computation is time consuming.
Another popular sparse regularizer is the total variation (TV) minimization, which essentially imposes the first-order $\ell_1$ quasi-norm.
Zhang et al. \cite{Zhang15Variational} extended the TV regularizer for restoring the face normal field.
A commonly known drawback of the TV regularizer is that it tends to produce staircase artifacts in smoothly curved regions.
To address this problem, higher-order methods \cite{Liu2019Triangulated,Liu2019ANovel} have been introduced, which can prevent producing staircase artifacts.
Unfortunately, when noise level increases, these high-order methods tend to blur fine details and curve sharp features.
Another technique \cite{Zhong2019Robust} for reducing staircase artifacts is to combine a high-order term with the TV term.
This straightforward combined technique reduces staircase artifacts to some extent, but unnatural artifacts may still appear around sharp features.
Thus, it is still challenging to preserve sharp features while simultaneously recover smooth transition variations.

\textbf{Nonlocal-based methods}. Most of the above-mentioned variational methods are local (using local operators to formulate the problem). Based on the observation that pattern similarity may exist on the underlying surface, several researchers introduced nonlocal methods \cite{Li2018NonLocal,Wei2018Mesh,Chen2019Structure,lu2020low}.
These nonlocal methods first group similar patches together, and then perform a low-rank minimization on the patch group to recover pattern similarity of the underlying surface.
These methods can effectively recover surfaces using the pattern similarity prior.
However, due to the multi-patch collaborative mechanism, these nonlocal methods are computationally intensive, and tend to blur sharp features.

\textbf{Data-driven methods}. More recently, data-driven methods are receiving increased attention.
Wang et al. \cite{Wang2016Mesh} presented their pioneer work with cascaded normal regression (CNR) for face normal smoothing.
Their method first learns non-linear regression functions that map filtered normal descriptors to those of the ground-truth counterparts, and then applies the learned functions to filter normals.
In order to better recover details, Wang et al. \cite{Wang2019Data} and Wei et al. \cite{Wei2019Mesh} proposed a two-step denoising framework (denoising followed by refinement).
They first learn the mapping from noisy meshes to their ground-truth counterparts for smoothing face normals.
Then, they recover details by learning the mapping from the filtered normals to the ground-truth.
Later on, Li et al. \cite{Li2020Normal} proposed a normal filtering neural network, called NormalF-Net, which consists of a denoising and refinement subnetwork.
Li et al. \cite{Li2020DNF} presented an end-to-end convolutional neural network, named DNF-Net, to predict filtered normals from the noisy mesh.
The above mentioned data-driven methods can produce satisfactory denoising results using convolutional network.
However, the performance of these methods depends on the completeness of the training data set. Moreover, the computation cost of the training process for them is usually high.

\section{Background of Total Generalized Variation} \label{sec:3}
Rudin, Osher, and Fatemi proposed the total variation (TV) in their seminal work \cite{Rudin1992Nonlinear}, and it has since started the trend of applying variational methods for image processing.
TV has been widely used as a regularizer for edge recovering, which are considered as the key features of images.
For an image $u: \mathrm{\Omega}\rightarrow\mathbb{R}$, TV of $u$ is defined as follows:
\begin{equation} \label{eq:TV}
\mathrm{TV}(u) =\int_{\mathrm{\Omega}} |\nabla u|.
\end{equation}
A commonly known drawback of TV is it tends to produce staircase artifacts in smooth transition regions of the images, as TV favors solutions that are piecewise constant.
To address this problem, a more general variational method, called total generalized variation (TGV), was introduced by Bredies et al. \cite{Bredies2010Total}.
In theory, TGV can be used to measure image characteristics up to a certain order of differentiation.
As proved in \cite{Bredies2010Total}, the first-order TGV is equivalent to TV.
Thanks to its higher-order nature, TGV can eliminate staircase artifacts effectively.
However, for TGV in an order that is too high, it becomes difficult to discretize and computationally expensive.
Considering the trade-off between computational complexity and numerical accuracy, we focus on the second-order TGV in this work.

Given an image $u$, the second-order TGV of $u$ is formulated as
\begin{equation} \label{eq:TGV}
\mathrm{TGV}(u) = \min \limits_{v} \Big \{ \alpha_1 \int_{\mathrm{\Omega}} |\nabla u-v| + \alpha_0 \int_{\mathrm{\Omega}} |\xi(v)| \Big \},
\end{equation}
where $\alpha_1, \alpha_0 \in \mathbb{R}^{+}$ are weights, and $\xi(v) = \frac{1}{2} (\nabla v + \nabla v^{T}) $ denotes the distributional symmetrized derivative.
The 2-tensor $v$ is converted into a vector by concatenating its columns for computational convenience. 
In the following, we give a more intuitive explanation for TGV.
On one hand, in smooth transition regions of $u$, the second-order derivative $\nabla^{2} u$ is small locally, and the optimum of \eqref{eq:TGV} is obtained by choosing $\nabla u \approx v$ therein locally.
On the other hand, in regions near edges, $\nabla^{2} u$ is evidently larger than $\nabla u$, hence the minimum of \eqref{eq:TGV} tends to have $v \approx 0$ in these regions.
However, this is only an intuitive assumption, the actual values of minimum $v$ are located anywhere in the range of $[0, \nabla u]$. 
Under the help of edge-aware variable $v$, TGV automatically balances between first- and second-order variations, instead of having fixed combination of them.
We refer the interested reader to \cite{Bredies2010Total,Knoll2011Second,Niu2014Sparse} for further discussions about TGV.

For a $\mathfrak{N}$-channel image $\mathbf{u}: \mathrm{\Omega} \rightarrow \mathbb{R}^{\mathfrak{N}}$, where $\mathbf{u}=(u_{1}, u_{2}, \ldots, u_{\mathfrak{N}})$, the vectorial TGV of $\mathbf{u}$ is formulated as
\begin{equation} \label{eq:vTGV}
\mathbf{TGV}(\mathbf{u}) = \min \limits_{\mathbf{v}} \Big \{ \alpha_1 \int_{\mathrm{\Omega}} \|\nabla \mathbf{u}-\mathbf{v}\| + \alpha_0 \int_{\mathrm{\Omega}} \|\xi(\mathbf{v})\| \Big \},
\end{equation}
where $\| \nabla\mathbf{u} - \mathbf{v} \| = \big(\sum_{i=1}^{\mathfrak{N}}|\nabla u_{i}- v_{i}|^{2}\big)^{\frac{1}{2}} $, and $\| \xi(\mathbf{v}) \|= \sum_{j=1}^{4} \| \xi_{j}(\mathbf{v}) \|= \sum_{j=1}^{4} \big(\sum_{i=1}^{\mathfrak{N}}|\xi_{j}(v_{i})|^{2}\big)^{\frac{1}{2}} $.
As we can see, \eqref{eq:vTGV} can be applied to process multi-spectral images with the special case $\mathfrak{N}=3$ for RGB images.

\section{Discretization of Total Generalized Variation on triangular meshes} \label{sec:4}
In this section, we first introduce some basic notation.
Then, we elaborate on how to discretize TGV and its vectorial version over triangular meshes.
Finally, we discuss the related work \cite{Zhang15Variational,Liu2019ANovel} with our discretized TGV.

\subsection{Notation}
Let $\mathcal{M}$ be a compact triangulated surface of arbitrary topology with no degenerate triangles in $\mathbb{R}^3$.
The set of vertices, edges, and triangles of $\mathcal{M}$ are denoted as
$\{p_i:i=1,\cdots,\mathrm{P}\}$,
$\{e_i:i=1,\cdots,\mathrm{E}\}$, and
$\{\tau_i:i=1,\cdots,\mathrm{T}\}$, respectively.
Here $\mathrm{P}$, $\mathrm{E}$, and $\mathrm{T}$ are the numbers of
vertices, edges, and triangles of $\mathcal{M}$.
If $p$ is an endpoint of an edge $e$, then we write $p \prec e$.
Similarly, $e\prec \tau$ denotes that $e$ is an edge of
$\tau$, and $p \prec \tau$ denotes that $p$ is a vertex of $\tau$.

We further define the relative orientation of an edge $e$ w.r.t. a triangle $\tau$, denoted by $\mathrm{sgn}(e,\tau)$, as follows.
Assume all triangles are with counterclockwise orientation, while all edges are with randomly chosen orientations.
For an edge $e\prec\tau$, if its orientation is consistent with the orientation of $\tau$, then $\mathrm{sgn}(e,\tau)=1$; otherwise $\mathrm{sgn}(e,\tau)=-1$.

\subsection{Discretizing Total Generalized Variation}
In order to describe piecewise constant data (e.g., face normal field) on a triangular mesh $\mathcal{M}$, we introduce the piecewise constant function space.
Given mesh $\mathcal{M}$,  we define the space $U = \mathbb{R}^{\mathrm{T}}$, which is
isomorphic to the piecewise constant function space on $\mathcal{M}$.
For example, $u = (u_{1}, \cdots, u_{\mathrm{T}})\in U$, where the value of $u$ restricted to triangle $\tau$ is $u_{\tau}$, sometimes written as $u|_{\tau}$ for convenience.
In order to further describe function $v$ (see the definition of TGV \eqref{eq:TGV}), we propose the edge function space ${V} = \mathbb{R}^{\mathrm{E}}$, whose elements are functions defined at the edges of $\mathcal{M}$.
We sometimes also write $v_e$ as $v|_{e}$, to denote the component of $v \in V$, restricted to edge $e$.
Sometimes, we also refer to space $V$ as the edge function space $\mathcal{E}$.

We equip space $U$ and $V$ with the standard Euclidean inner product and norm as follows.
$\forall \ u^{1},u^{2}, u \!\in\! U$, we have:
\begin{equation}\label{FaceInner}
(u^{1},u^{2})_{U}=\sum\limits_{\tau}u^{1}|_{\tau}u^{2}|_{\tau}\mathrm{area}(\tau), \ \
\|u\|_{U} = \sqrt{(u,u)_{U}},
\end{equation}
where $\mathrm{area}(\tau)$ is the area of $\tau$. $\forall \ v^{1},v^{2}, v \in V$, we have:
\begin{equation} \label{EdgeInner}
(v^{1},v^{2})_{V}=\sum\limits_{e}v^{1}|_ev^{2}|_e \mathrm{len}(e), \ \ \|v\|_{V} = \sqrt{(v,v)_{V}},
\end{equation}
where $\mathrm{len}(e)$ is the length of $e$.

As discussed in \cite{Liu2020Mesh}, it is natural to define the first-order difference operator $\mathcal{D}_{\mathcal{M}}:U \rightarrow V$ on $\mathcal{M}$ as
\begin{equation} \label{firstOrderOperator}
(\mathcal{D}_{\mathcal{M}} u)|_e=\left\{\begin{array}{ll}
\sum\limits_{\mathclap{\tau,e\prec\tau}} u_{\tau} \mathrm{sgn}(e,\tau), \!\! & \!\! e \not\subset \partial \mathcal{M}\\
0, \!\! & \!\!  e \subset \partial \mathcal{M}
\end{array}\right.\!\!\!, \ \ \forall e.
\end{equation}
The adjoint operator of $\mathcal{D}_{\mathcal{M}}$, i.e., $\mathcal{D}^{\star}_{\mathcal{M}}: V \rightarrow U$, is given by
\begin{equation}\label{adjoint-firstOrderOperator}
(\mathcal{D}^{\star}_{\mathcal{M}}
v)|_{\tau}= -\frac{1}{\mathrm{area}(\tau)}\sum\limits_{\begin{subarray}{c}  e\prec\tau,  \\  e \not\subset \partial \mathcal{M}  \end{subarray}}
v_e \mathrm{sgn}(e,\tau)  \mathrm{len}(e), \ \ \forall \tau.
\end{equation}

In the discrete case, for each triangle $\tau$, there are three first-order differences over the edges along three different directions.
Thus, we can approximate the gradient operator in triangle $\tau$ as,
\begin{equation*}
  \nabla u|_{\tau} = (\mathcal{D}_{\mathcal{M}} u |_{e_{1,\tau}}, \mathcal{D}_{\mathcal{M}} u |_{e_{2,\tau}}, \mathcal{D}_{\mathcal{M}} u |_{e_{3,\tau}}),
\end{equation*}
where $e_{i,\tau} \prec \tau, i = 1,2,3$.
For convenience, we write the discrete gradient as $\nabla u = (\partial_1 u, \partial_2 u, \partial_3 u)$. It is natural to denote the second-order gradient in $\tau$ as,
\begin{equation} \label{eq:2orderGradientOfU}
\nabla^2 u |_{\tau}=
\left (
  \begin{array}{ccc}   
    \partial_1 \partial_1 u & \partial_1 \partial_2 u & \partial_1 \partial_3 u  \\  
    \partial_2 \partial_1 u & \partial_2 \partial_2 u & \partial_2 \partial_3 u \\  
    \partial_3 \partial_1 u & \partial_3 \partial_2 u & \partial_3 \partial_3 u \\ 
  \end{array}
\right),
\end{equation}
where the diagonal entries $\partial_i \partial_i u, i=\{1,2,3\}$ are the second-order directional derivatives in the same direction, while the off-diagonal entries $\partial_i \partial_j u, i\!\!\neq\!\!j$ are the second-order directional derivatives in two different directions.

To further discretize TGV, we need to define operators in the edge function space $\mathcal{E}$.
For each triangle $\tau$, $v|_{\tau} = (v_{e_{1,\tau}},v_{e_{2,\tau}},v_{e_{3,\tau}})$ denotes the values of $v$ restricted to the three edges of $\tau$.
For brevity, the values of $v$ in triangle $\tau$ can be written as $v|_{\tau}=(v_1, v_2, v_3)$. Then, the gradient (in three different directions) is given by
\begin{equation} \label{eq:gradientOfv}
\nabla v |_{\tau}=
\left(
  \begin{array}{ccc}   
    \partial_1 v_{1} & \partial_2 v_{1} & \partial_3 v_{1}  \\ 
    \partial_1 v_{2} & \partial_2 v_{2} & \partial_3 v_{2} \\  
    \partial_1 v_{3} & \partial_2 v_{3} & \partial_3 v_{3} \\  
  \end{array}
\right),
\end{equation}
where $\partial_i v_{j}$ is the first-order derivative.
Note that $\xi(v) = \frac{1}{2} (\nabla v + \nabla v^{T})$, the symmetrized tensor gradient $\xi(v)$ in $\tau$ can be directly derived as,
\begin{equation} \label{eq:TGVOperator}
 \xi(v)|_{\tau}=
\left(
  \begin{array}{ccc}  
    \partial_1 v_{1} & \frac{\partial_2 v_{1}+\partial_1 v_{2}}{2} & \frac{\partial_3 v_{1}+\partial_1 v_{3}}{2}  \\  
   \frac{\partial_2 v_{1}+\partial_1 v_{2}}{2} & \partial_2 v_{2} & \frac{\partial_3 v_{2}+\partial_2 v_{3}}{2}  \\  
    \frac{\partial_3 v_{1}+\partial_1 v_{3}}{2}  & \frac{\partial_3 v_{2}+\partial_2 v_{3}}{2} & \partial_3 v_{3} \\  
  \end{array}
\right).
\end{equation}

In the following, we discretize the symmetrized tensor gradient operator $\xi(\cdot)$, which is the core contribution of this paper.
As mentioned in Section \ref{sec:2}, in smooth transition regions,  $v$ has minimums whose values are close to $\nabla u$. Intuitively, this means,
\begin{equation} \label{eq:relationBetweenUAndV}
  v \approx \nabla u \rightarrow \nabla v \approx \nabla^2 u \rightarrow \partial_i v_i \approx \partial_i \partial_i u, \partial_i v_j \approx \partial_i \partial_j u,
\end{equation}
where $i,j=1,2,3$, ignoring the order of $i$ and $j$.
As we can see from \eqref{eq:gradientOfv} and \eqref{eq:TGVOperator}, we need two discretization forms of first-order derivatives w.r.t. $v$ (one for $\partial_i v_i$ and the other for $\partial_i v_j$).
From \eqref{eq:relationBetweenUAndV}, we can easily see that these two discretizations also intuitively determine the second-order derivatives w.r.t. $u$.

\begin{figure}[thb]
  \centering
  \subfloat[]{\label{fig:1-formOperator-a}\includegraphics[width=0.18\textwidth]{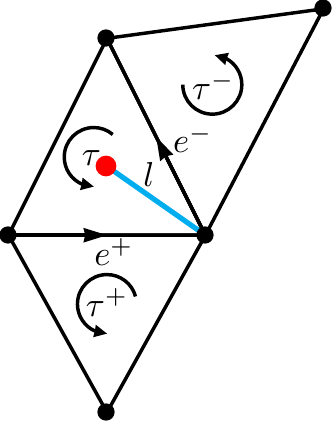}}
  \subfloat[]{\label{fig:1-formOperatpr-b}\includegraphics[width=0.16\textwidth]{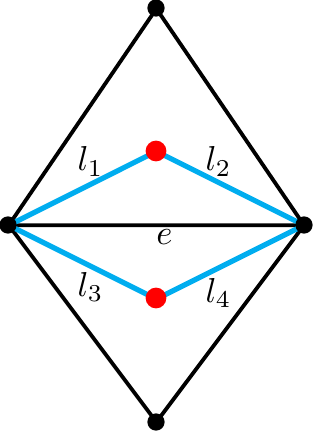}}
  \caption{
  (a) Illustration for the definition of the 1-form operator $\overline{\mathcal{D}}_{\mathcal{E}}$v.
  $[v]$ is the jump of $v$ over line $l$ plotted in cyan in triangle $\tau$ with its barycenter plotted in red.
  (b) Illustration for the definition of the adjoint operator $\overline{\mathcal{D}}^{\star}_{\mathcal{E}}
  \overline{w}$. $B_1(e)$ is a set containing four lines associated with edge $e$.
  \label{fig:1-formOoperator}}
\end{figure}

\begin{figure*}[thb]
  \centering
  \includegraphics[width=1.0\linewidth]{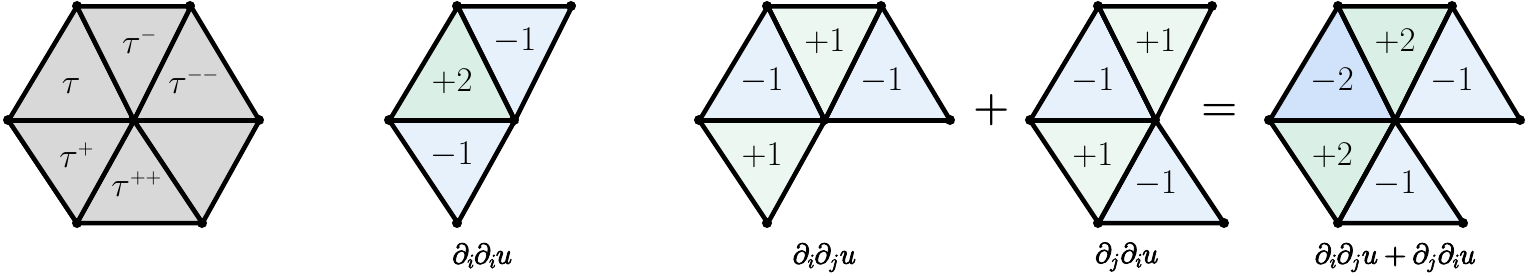}
  \caption{Illustration of the second-order directional derivatives for a triangle ($\tau$).
  \label{fig:illustration}}
\end{figure*}

Let $l$ be the line segment connecting the barycenter and a vertex of $\tau$.
Given $v \in V$, with the Neumann boundary condition, we define the 1-form jump of $v$ over $l$ as
\begin{equation}\label{eq:1-formJump}
[v]_{l} \!=\!
\left\{\begin{array}{ll}
\!\!\! v_{e^{+}} \mathrm{sgn}(e^{+}, \tau) + v_{e^{-}} \mathrm{sgn}(e^{-}, \tau), & e^{+} \text{ and } e^{-} \not\subset \partial \mathcal{M},\\
\!\!\! 0, & e^{+} \text{ or } e^{-} \subset \partial \mathcal{M},
\end{array}\right.
\end{equation}
where $e^+$ and $e^-$ are two edges sharing a common vertex of $l$.
$e^+$ enters the common vertex in counterclockwise direction, whereas $e^-$ leaves the common vertex in counterclockwise direction.
The two triangles sharing edges $e^+$ and $e^-$ are denoted as $\tau^+$ and $\tau^-$ respectively.
All the aforementioned descriptions are illustrated in Fig. \ref{fig:1-formOperator-a}.
Then, the discrete 1-form operator $\overline{\mathcal{D}}_{\mathcal{E}}$ is defined as
\begin{equation} \label{eq:1-formOperator}
    \overline{\mathcal{D}}_{\mathcal{E}}:V \rightarrow \overline{W}, \ \ (\overline{\mathcal{D}}_{\mathcal{E}}v)|_{l} =  {[v]_{l}},  \  \forall l, \ \mathrm{for} \ v \in V,
\end{equation}
where $\overline{W}= \mathbb{R}^{3\times \mathrm{T}}$.
The $\overline{W}$ space is equipped with the following inner product and norm:
\begin{equation}\label{eq:1-formInner}
(\overline{w}^{1},\overline{w}^{2})_{\overline{W}} \!=\! \sum\limits_{l} \overline{w}^{1}|_l \overline{w}^{2}|_l \mathrm{len}(l), \ \ \|\overline{w}\|_{\overline{W}} = \sqrt{(\overline{w},\overline{w})_{\overline{W}}},
\end{equation}
where $\overline{w}^{1},\overline{w}^{2}, \overline{w} \in \overline{W}$, and $\mathrm{len}(l)$ is the length of line $l$.
The adjoint operator of $\overline{\mathcal{D}}_{\mathcal{E}}$, that is $\overline{\mathcal{D}}^{\star}_{\mathcal{E}}: \overline{W} \rightarrow V$, can be derived using the inner products in $V$ and $\overline{W}$.
For $\overline{w} \in \overline{W}$, $\overline{\mathcal{D}}^{\star}_{\mathcal{E}}$ has the following form
\begin{equation}\label{eq:adjoint-1-formOperator}
(\overline{\mathcal{D}}^{\star}_{\mathcal{E}}
\overline{w})|_{e}= -\frac{1}{\mathrm{len}(e)}\sum\limits_{l \in B_{1}(e)}
\overline{w}_l \mathrm{sgn}(e,\tau_l)  \mathrm{len}(l), \ \ \forall e,
\end{equation}
where $B_{1}(e)$ is the set of lines associated with the edge $e$ (see Fig. \ref{fig:1-formOperatpr-b}) and $\tau_l$ is the triangle containing the line $l$.
Details for the derivation of $\overline{\mathcal{D}}^{\star}_{\mathcal{E}}$ can be found in lemma 1 in Part 1 of the supplementary material.

\textbf{Remark 1}. We give an intuitive interpretation of the discrete 1-form operator $\overline{\mathcal{D}}_{\mathcal{E}}$.
From \eqref{eq:1-formJump} and \eqref{eq:1-formOperator}, we can see that the 1-form operator w.r.t. $v$ depicts the variation of $v$ over the two adjacent edges.
Moreover, this operator can also be seen as an analogue of the second-order operator w.r.t. $u$ (in the same direction), which depicts
\begin{equation} \label{eq:1-formOperator-uv}
\begin{aligned}
(\overline{\mathcal{D}}_{\mathcal{E}} v)|_{l} = & v_{e^{+}} \mathrm{sgn}(e^{+}, \tau) + v_{e^{-}} \mathrm{sgn}(e^{-}, \tau) \\
                           \approx & (u_{\tau}\mathrm{sgn}(e^+, \tau) + u_{\tau^+}\mathrm{sgn}(e^+, \tau^+))\mathrm{sgn}(e^+, \tau) +  \\
            & (u_{\tau}\mathrm{sgn}(e^-, \tau) + u_{\tau^-}\mathrm{sgn}(e^-, \tau^-))\mathrm{sgn}(e^-, \tau)  \\
           =& \big( u_{\tau} - u_{\tau^+} \big) + \big(u_{\tau} - u_{\tau^-} \big) \\
           =& 2u_{\tau} - u_{\tau^+} - u_{\tau^-}.
\end{aligned}
\end{equation}
In summary, the 1-form operator $\overline{\mathcal{D}}_{\mathcal{E}}v$ can depict the first-order variations of $v$, which can be seen as an approximation of the first-order derivatives $\partial_i v_i$.
Besides, this operator also can describe the second-order variations of $u$, which can be seen as an approximation of the second-order directional derivatives $\partial_i \partial_i u$, whose discretization on meshes is demonstrated in Fig. \ref{fig:illustration}.

\begin{figure}[thb]
  \centering
  \includegraphics[width=0.25\textwidth]{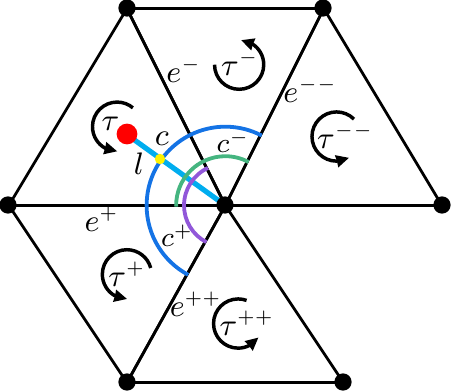}
  \caption{Illustration of the definition of 2-form operator $\widetilde{\mathcal{D}}_{\mathcal{E}}v$.
   $[[v]]$ is the 2-form jump over curve $c$ plotted in blue, which passes through four edges $(e^{--}, e^{-}, e^{+}, e^{++})$ and attaches itself to line $l$.
  Auxiliary curve $c^-$ passing through $(e^{--}, e^{+})$ is plotted in green, and curve $c^{+}$ passing through $(e^{-}, e^{++})$ is plotted in purple.
  \label{fig:2-formOperator}}
\end{figure}

Let $c$ be a curve passing through four edges $(e^{--}, e^{-}, e^{+}, e^{++})$ and associates itself to the line $l$ of $\tau$.
For $v \in V$, we define the 2-form jump of $v$ over $c$ as
\begin{equation} \label{eq:2-formJump}
\begin{aligned}
[[v]]_{c} = & [[v]]_{c^-} + [[v]]_{c^+} \\
          = & \Big( v_{e^{--}}\mathrm{sgn}(e^{--},\tau^-) + v_{e^+}\mathrm{sgn}(e^+,\tau^+) \Big) +  \\
          & \Big( v_{e^-}\mathrm{sgn}(e^-,\tau^-) + v_{e^{++}}\mathrm{sgn}(e^{++},\tau^+) \Big).
\end{aligned}
\end{equation}
With Neumann boundary condition, if any of $e^{--}$, $e^{-}$, $e^{+}$, $e^{++}$ is on the boundary $\partial \mathcal{M}$, we directly set $[[v]]_{c} = 0$.
The triangle that shares edges $e^{--}$ with $\tau^{-}$ is denoted as $\tau^{--}$, while $\tau^{++}$ denotes the triangle that shares edge $e^+$ with $\tau^+$. The two auxiliary curves passing through $(e^{--}, e^{+})$ and $(e^{-}, e^{++})$ are denoted as $c^-$ and $c^+$,  respectively.
We demonstrate all aforementioned descriptions in Fig. \ref{fig:2-formOperator}. Then, we define the 2-form operator $\widetilde{\mathcal{D}}_\mathcal{E}$ as
\begin{equation} \label{eq:2-formOperator}
    \widetilde{\mathcal{D}}_{\mathcal{E}}:V \rightarrow \widetilde{W}, \ \ (\widetilde{\mathcal{D}}_{\mathcal{E}}v)|_{c} =  {[[v]]_{c}},  \  \forall c, \ \mathrm{for} \ v \in V,
\end{equation}
where $\widetilde{W}= \mathbb{R}^{3 \times \mathrm{T}}$.
The $\widetilde{W}$ space is equipped with the following inner product and norm:
\begin{equation}\label{eq:2-formInner}
\!\!(\widetilde{w}^{1},\widetilde{w}^{2})_{\widetilde{W}} \!=\! \sum\limits_{c} \widetilde{w}^{1}|_c \widetilde{w}^{2}|_c \mathrm{len}(c), \ \ \|\widetilde{w}\|_{\widetilde{W}} = \sqrt{(\widetilde{w},\widetilde{w})_{\widetilde{W}}},
\end{equation}
where $\widetilde{w}^{1},\widetilde{w}^{2}, \widetilde{w} \in \widetilde{W}$, and $\mathrm{len}(c)=\frac{1}{4}\big(\mathrm{len}(l^{-}) + 2\mathrm{len}(l) + \mathrm{len}(l^{+}) \big)$ is an approximation of the length of $c$.
$l^+$ is the line segment contained in triangle $\tau^+$ and shares the vertex with $l$, while $l^-$ is the segment contained in $\tau^-$.

\begin{figure}[thb]
  \centering
  \includegraphics[width=0.35\textwidth]{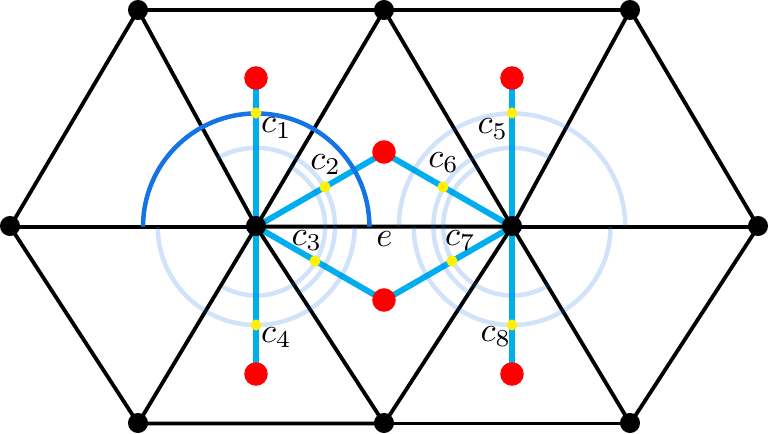}
  \caption{Illustration of the definition of adjoint operator $\widetilde{\mathcal{D}}^{\star}_{\mathcal{E}}
  \widetilde{w}$.
  $B_{2}(e)$ is the set of curves associated with edge $e$, which refers to eight curves.
  The attached lines of curves in $B_{2}(e)$ are also shown.
  \label{fig:joint-2-formOperator}}
\end{figure}

Similarly, the adjoint operator of $\widetilde{\mathcal{D}}_{\mathcal{E}}$, that is $\widetilde{\mathcal{D}}^{\star}_{\mathcal{E}}: \widetilde{W} \rightarrow V$, is given by
\begin{equation}\label{eq:adjoint-2-formOperator}
(\widetilde{\mathcal{D}}^{\star}_{\mathcal{E}}
\widetilde{w})|_{e}= -\frac{1}{\mathrm{len}(e)}\sum\limits_{c \in B_{2}(e)}
\widetilde{w}_c \mathrm{sgn}(e,\tau_c)  \mathrm{len}(c), \ \ \forall e,
\end{equation}
where $B_{2}(e)$ is the set of curves associated with edge $e$ (see Fig. \ref{fig:joint-2-formOperator}), and $\tau_c$ is the triangle satisfying conditions $e \prec \tau_c$ and $\tau_c \in \{\tau^+, \tau^-\}$.
The derivation of $\widetilde{\mathcal{D}}^{\star}_{\mathcal{E}}$ can be found in lemma 2 in Part 1 of the supplementary material.

\textbf{Remark 2}. We give an intuitive interpretation of the 2-form operator $\widetilde{\mathcal{D}}_{\mathcal{E}}$. From \eqref{eq:2-formJump} and \eqref{eq:2-formOperator}, we can see that the 2-form operator w.r.t. $v$ describes the sum of variations of $v$ across edges $(e^{--}, e^+)$ and across edges $(e^{-}, e^{++})$.
This operator can also be seen as an analogue of the second-order operator w.r.t. $u$ (in different directions), which can be expressed as
\begin{equation} \label{eq:2-formOperator-uv}
\begin{aligned}
(\widetilde{\mathcal{D}}_{\mathcal{E}} v)|_{c} = & \Big( v_{e^{--}}\mathrm{sgn}(e^{--},\tau^-) + v_{e^+}\mathrm{sgn}(e^+,\tau^+)  \Big) +  \\
                                    & \Big( v_{e^-}\mathrm{sgn}(e^-,\tau^-) + v_{e^{++}}\mathrm{sgn}(e^{++},\tau^+) \Big) \\
                                   \approx & ( u_{\tau+} + u_{\tau-} - u_{\tau} - u_{\tau--}) + \\
                                    & ( u_{\tau+} + u_{\tau-} - u_{\tau} - u_{\tau++}).
\end{aligned}
\end{equation}
Intuitively, \eqref{eq:2-formOperator-uv} can be used to depict both first-order derivatives $\partial_i v_j + \partial_j v_i $ and second-order directional derivatives $\partial_i \partial_j u + \partial_j \partial_i u$.
We illustrate the discretization of second-order directional derivatives $\partial_i \partial_j u + \partial_j \partial_i u$ in Fig. \ref{fig:illustration}.

Note that with the 1- and 2-form  operators \eqref{eq:1-formOperator} and \eqref{eq:2-formOperator}, the discrete symmetrized gradient operator $\xi : V \rightarrow W$ can be directly approximated.
Space $W=\mathbb{R}^{6 \times \mathrm{T}}$ is a composition of the spaces $\overline{W}$ and $\widetilde{W}$, and it is equipped with the following inner product and norm:
\begin{equation*}
    \begin{aligned}
        (w^{1},w^{2})_{W}&= (\overline{w}^{1},\overline{w}^{2})_{\overline{W}} + (\widetilde{w}^{1},\widetilde{w}^{2})_{\widetilde{W}}, \\
        \|w\|_{W} &= \|\overline{w}\|_{\overline{W}} + \|\widetilde{w}\|_{\widetilde{W}},
\end{aligned}
\end{equation*}
with $\ w^1, w^2, w \in W$, $\overline{w}^1, \overline{w}^2, \overline{w} \in \overline{W}$, and $\widetilde{w}^1, \widetilde{w}^2, \widetilde{w} \in \widetilde{W}$.
Then, the second-order term of TGV can be expressed as
\begin{equation} \label{eq:second-orderOfTGV}
\begin{aligned}
\| \xi(v) \|_{W}&=  \sum \limits_{\tau} \Big( \sum\limits_{i} \| \partial_i v_i \| + \sum\limits_{i,j} \| \partial_i v_j + \partial_j v_i \| \Big) \\
                &=  \| \overline{\mathcal{D}}_{\mathcal{E}} v \|_{\overline{W}} + \| \widetilde{\mathcal{D}}_{\mathcal{E}}  v \|_{\widetilde{W}},
\end{aligned}
\end{equation}
where $i,j = 1, 2, 3$ and $i\neq j$.
Given $u \in U$, with the above definition, we now formulate the discretized TGV as
\begin{equation}\label{eq:discrete-scalar-TGV}
  \mathrm{TGV}(u) = \min \limits_{v \in V} \Big\{ \alpha_1 \|\mathcal{D}_{\mathcal{M}} u - v\|_{V} + \alpha_0 \|\xi(v)\|_{W} \Big\},
\end{equation}
which defines the TGV semi-norm over meshes.

We can extend the TGV semi-norm \eqref{eq:discrete-scalar-TGV} to the vectorial case.
To consider vectorial data, three vectorial spaces $\mathbf{U}$, $\mathbf{V}$, and $\mathbf{W}$ are defined as
$$\mathbf{U} =\underbrace{U \times \cdots \times U}\limits_{\mathfrak{N}},\mathbf{V} =\underbrace{V \times \cdots \times V}\limits_{\mathfrak{N}}, \mathbf{W} =\underbrace{ W \times \cdots \times W}\limits_{\mathfrak{N}},$$
for $\mathfrak{N}$-channel data.
The inner products and norms in $\mathbf{U}$, $\mathbf{V}$, and $\mathbf{W}$ are defined as follows:
$$(\mathbf{u}^{1},\mathbf{u}^{2})_{\mathbf{U}} = \sum\limits_{1\leq i \leq \mathfrak{N}}(u_{i}^{1},u_{i}^{2})_{U},\quad
 \|\mathbf{u}\|_{\mathbf{U}} = \sqrt{(\mathbf{u},\mathbf{u})_{\mathbf{U}}},$$

$$(\mathbf{v}^{1},\mathbf{v}^{2})_{\mathbf{V}} = \sum\limits_{1\leq i \leq \mathfrak{N}}(v_{i}^{1},v_{i}^{2})_{V},\quad
\|\mathbf{v}\|_{\mathbf{V}} = \sqrt{(\mathbf{v},\mathbf{v})_{\mathbf{V}}},$$

$$(\mathbf{w}^{1},\mathbf{w}^{2})_{\mathbf{W}} = \sum\limits_{1\leq i\leq \mathfrak{N}}(w_{i}^{1},w_{i}^{2})_{W},\quad
\|\mathbf{w}\|_{\mathbf{W}} = \sqrt{(\mathbf{w},\mathbf{w})_{\mathbf{W}}},$$
with $\mathbf{u}^{1},\mathbf{u}^{2},\mathbf{u} \in \mathbf{U}$, $\mathbf{v}^{1},\mathbf{v}^{2},\mathbf{v} \in \mathbf{V}$, and $\mathbf{w}^{1}, \mathbf{w}^{2}, \mathbf{w} \in \mathbf{W}$.
Thus, all the aforementioned discrete operators can be evaluated channel by channel, and the vectorial TGV semi-norm is then defined as
\begin{equation}\label{eq:discrete-vectorial-TGV}
  \mathbf{TGV}(\mathbf{u}) = \min \limits_{\mathbf{v} \in \mathbf{V}} \Big\{ \alpha_1 \|\mathcal{D}_{\mathcal{M}} \mathbf{u} - \mathbf{v}\|_{\mathbf{V}} + \alpha_0 \|\xi(\mathbf{v})\|_{\mathbf{W}} \Big\}.
\end{equation}

\subsection {Differences between TV, HO and TGV}
There are some existing works highly related to our discretized TGV.
It is necessary to discuss the differences between our discretized TGV, total variation (TV) in \cite{Zhang15Variational}, and high-order variation (HO) in \cite{Liu2019ANovel}.
Given a signal $u \in V$, Zhang et al. \cite{Zhang15Variational} defined the discretized TV as
\begin{equation}\label{eq:TV-seminorm}
  \mathrm{TV}(u) = \|\mathcal{D}_{\mathcal{M}} u \|_{V},
\end{equation}
which describes the first-order variations over mesh edges.
TV \eqref{eq:TV-seminorm} works exceptionally well in preserving sharp features, but produces staircase artifacts in smooth regions.

To overcome the staircase artifacts of the TV regularizer, Liu et al. \cite{Liu2019ANovel} introduced a second-order difference operator to discretize second-order derivatives,
and propose second-order variation as follows:
\begin{equation}\label{eq:HO-seminorm}
  \mathrm{HO}(u) = \sum_{l} \|2 u_{\tau} - u_{\tau^+} - u_{\tau^-}\| \mathrm{len}(l).
\end{equation}
HO \eqref{eq:HO-seminorm} recovers smooth regions well, but
blurs sharp features in case of large noise.

Next, we discuss the differences between the second-order term \eqref{eq:second-orderOfTGV} of TGV and HO \eqref{eq:HO-seminorm}.
From \eqref{eq:1-formOperator-uv}, we can see $\mathrm{HO}(u) \approx \| \overline{\mathcal{D}}_{\mathcal{E}} v \|_{\overline{W}}$.
Therefore, HO \eqref{eq:HO-seminorm} can be seen as an analogue of the second-order derivative in the same direction ($\partial_i \partial_i u$).
In other words, the HO regularizer minimizes only the second-order variations in the same direction.
In contrast, the minimization of the second-order term of TGV attempts to simultaneously minimize the second-order variations in the same direction ($\partial_i \partial_i u$) as well as those in different directions ($\partial_i \partial_j u + \partial_j \partial_i u$).

As mentioned earlier, TV is more effective than HO in preserving sharp features, while HO handles smooth regions better than TV.
Until now, it has been challenging to have one regularizer simultaneously preserve sharp features in some parts of the mesh and recover smooth regions in some other parts.
To address this problem, we propose the discretized TGV, which automatically balances the first- and second-order terms via the auxiliary variable $v$.
In consequence, it combines the advantages from both TV and HO, and manages to overcome their weakness.
See Section \ref{sec:6} for more detailed comparisons.
\section{Mesh Denoising using Vectorial TGV} \label{sec:5}
In this section, we first propose a vectorial TGV based normal filter,
and then design an algorithm to effectively solve the optimization problem.
After that, we reconstruct vertex positions based on the optimized face normals.

\subsection{Vectorial TGV based Normal Filter} \label{sec:5.1}
Given a noisy mesh, we denote its face normal field as $\mathbf{N}^{in}$.
In order to remove noise in $\mathbf{N}^{in}$ using vectorial TGV \eqref{eq:discrete-vectorial-TGV}, we formulate our normal filter as the following problem,
\begin{equation} \label{eq:TGV-NormalFilteringModel}
\begin{aligned}
\min \limits_{{\mathbf{N}},\mathbf{v}}
\! \Big\{
\frac{\beta}{2} & \left\|\mathbf{N} \! -\! \mathbf{N}^{in}\right\|^2_{\mathbf{U}} \!+\!
  \alpha_1 \! \sum \limits_{e} w_e \|(\mathcal{D}_{\mathcal{M}} \mathbf{N} \!-\! \mathbf{v})|_{e}\| \mathrm{len}(e) \\
+ &\alpha_0 \|\xi(\mathbf{v})\|_{\mathbf{W}}\Big \}, \ \ \textmd{s.t.} \  \left\|\mathbf{N}_\tau\right\|=1, \forall \tau,
\end{aligned}
\end{equation}
where $\| \xi(\mathbf{v}) \|_{\mathbf{W}}= \| \overline{\mathcal{D}}_{\mathcal{E}} \mathbf{v} \|_{\overline{\mathbf{W}}} + \| \widetilde{\mathcal{D}}_{\mathcal{E}}  \mathbf{v} \|_{\widetilde{\mathbf{W}}}$.
Note that $\mathbf{N} \in \mathbf{U}$ and $\mathbf{U}$ denotes 3-channel $U$.
Weight $w_e$ is given by
\begin{equation}\label{eq:TGV-weight}
w_e = \exp\biggl(-\frac{\|\mathbf{N}_{e,1}-\mathbf{N}_{e,2}\|^2}{2{\sigma_e}^2}\biggr),
\end{equation}
where ${\mathbf{N}}_{e,1}$ and ${\mathbf{N}}_{e,2}$ are normals of the triangles sharing edge $e$, and $\sigma_e$ is a user-specified parameter.
$w_e$ is expected to be large when $\|\mathbf{N}_{e,1}-\mathbf{N}_{e,2}\|$, the modulus of the first-order normal difference defined on $e$, is small and vice versa.
Thus, it results in large weights for smooth regions, and small weights for sharp features, and therefore allows the proposed filter \eqref{eq:TGV-NormalFilteringModel} to smooth non-features regions while preserving sharp features.

The vectorial TGV \eqref{eq:discrete-vectorial-TGV} (applied to the face normal field) can produce satisfactory denoising results in most cases.
However, it may oversmooth sharp features for some meshes with large noise.
Thus, we propose the dynamic weighting ($w_e$) in our vectorial TGV based normal filter \eqref{eq:TGV-NormalFilteringModel}.
These weights are updated in each iteration, which enhance the sparsity of the original vectorial TGV \eqref{eq:discrete-vectorial-TGV} for improved sharp feature reconstruction.
Essentially, these dynamically adjusted weights penalize smooth regions more than sharp features, which allows the lower-than-$\ell_1$-sparsity effect to be achieved \cite{avron2010}.

\subsection{Numerical Optimization}
Due to the vectorial $\ell_1$ semi-norm involved, problem \eqref{eq:TGV-NormalFilteringModel} has a non-differentiable objective which makes it difficult to solve.
Here, we use variable-splitting and augmented Lagrange method (ALM) to solve \eqref{eq:TGV-NormalFilteringModel}, which has achieved great success in solving $\ell_1$ related problems \cite{Wu2010Augmented,Boyd2011}.

By introducing new variables $\mathbf{P}$, $\overline{\mathbf{Q}}$, and $\widetilde{\mathbf{Q}}$, we reformulate \eqref{eq:TGV-NormalFilteringModel} as a constrained optimization problem,
\begin{equation*}
\begin{aligned}
\min \limits_{\mathbf{N}, \mathbf{v}, \mathbf{P}, \overline{\mathbf{Q}}, \widetilde{\mathbf{Q}}}
& \Big\{ \frac{\beta}{2} \left\|\textbf{N}-\textbf{N}^{in}\right\|^2_{\mathbf{U}} \!+\!
  \alpha_1 \sum \limits_{e} w_e \|\mathbf{P}_{e}\| \mathrm{len}(e) \\
      & + \alpha_0 \|\overline{\mathbf{Q}}\|_{\overline{\mathbf{W}}} + \alpha_0 \|\widetilde{\mathbf{Q}}\|_{\widetilde{\mathbf{W}}} + \Psi(\mathbf{N})
\Big\}, \\
\textmd{s.t.}& \ \ \mathbf{P} = \mathcal{D}_{\mathcal{M}} \mathbf{N} - \mathbf{v}, \ \overline{\mathbf{Q}}=\overline{\mathcal{D}}_{\mathcal{E}} \mathbf{v}, \ \widetilde{\mathbf{Q}}=\widetilde{\mathcal{D}}_{\mathcal{E}} \mathbf{v},
\end{aligned}
\end{equation*}
where
\begin{eqnarray*}
\Psi (\mathbf{N})= \left
\{ \begin{array}{rl}
0,         & \mathrm{if} \ \ \| \mathbf{N}_{\tau}\| =1, \  \forall \tau, \\
+\infty,   & \mathrm{otherwise}.
\end{array}
\right.
\end{eqnarray*}
To solve the above constrained optimization problem, we introduce the augmented Lagrangian function as follows,
\begin{equation*}\label{eq:TGV-ALGfunction}
\begin{aligned}
 & \mathcal{L}(\mathbf{N}, \mathbf{v}, \mathbf{P}, \overline{\mathbf{Q}}, \widetilde{\mathbf{Q}}; \lambda_\mathbf{P}, \lambda_\mathbf{\overline{Q}}, \lambda_\mathbf{\widetilde{Q}})
  = \frac{\beta}{2} \left\|\textbf{N}-\textbf{N}^{in}\right\|^2_{\mathbf{U}} \\
  &+ \alpha_1 \sum \limits_{e} w_e \|\mathbf{P}_e\| \mathrm{len}(e)
  + \alpha_0 \|\overline{\mathbf{Q}}\|_{\overline{\mathbf{W}}} + \alpha_0 \|\widetilde{\mathbf{Q}}\|_{\widetilde{\mathbf{W}}} + \Psi(\mathbf{N}) \\
  &+ {(\lambda_\mathbf{P}, \mathbf{P} - (\mathcal{D}_{\mathcal{M}} \mathbf{N} - \mathbf{v}))}_{\mathbf{V}}
  + \frac{r_1}{2} \left\|\mathbf{P}
  - (\mathcal{D}_{\mathcal{M}} \mathbf{N} - \mathbf{v}) \right\|^2_{\mathbf{V}} \\
  &+ {(\lambda_\mathbf{\overline{Q}}, \mathbf{\overline{Q}} - \overline{\mathcal{D}}_{\mathcal{E}} \mathbf{v})}_{\mathbf{\overline{W}}} + \frac{r_0}{2} \| \mathbf{\overline{Q}} - \overline{\mathcal{D}}_{\mathcal{E}} \mathbf{v} \|^2_{\mathbf{\overline{W}}} \\
  &+ {(\lambda_\mathbf{\widetilde{Q}}, \mathbf{\widetilde{Q}} - \widetilde{\mathcal{D}}_{\mathcal{E}} \mathbf{v})}_{\mathbf{\widetilde{W}}} + \frac{r_0}{2} \| \mathbf{\widetilde{Q}} - \widetilde{\mathcal{D}}_{\mathcal{E}} \mathbf{v} \|^2_{\mathbf{\widetilde{W}}},
\end{aligned}
\end{equation*}
where $\lambda_\mathbf{P}$, $\lambda_\mathbf{\overline{Q}}$, and $\lambda_\mathbf{\widetilde{Q}}$ are Lagrange multipliers, $r_1$ and $r_0$ are positive penalty weights.

Then we apply the variable-splitting technique and iteratively update the different set of variables in alternation with the following five subproblems:
\begin{itemize}
  \item The $\mathbf{N}$-subproblem:
  \begin{equation}\label{eq:n-sub}
  \begin{aligned}
    \min\limits_{\mathbf{N}}  \frac{\beta}{2} \|\mathbf{N}
    & - \mathbf{N}^{in}\|^2_{\mathbf{U}}  + \Psi(\mathbf{N}) \\
    & +\frac{r_1}{2} \|  \mathcal{D}_{\mathcal{M}} \textbf{N} - \mathbf{v} - (\mathbf{P} + \frac{\lambda_\mathbf{P}}{r_1})\|^2_{\mathbf{V}};
  \end{aligned}
  \end{equation}

  \item The $\mathbf{v}$-subproblem:
  \begin{equation}\label{eq:v-sub}
  \begin{aligned}
    \min\limits_{\mathbf{v}}
   & \frac{r_0}{2} \|  \mathcal{\overline{D}}_{\mathcal{E}} \mathbf{v} \!-\! (\mathbf{\overline{Q}} + \frac{\lambda_\mathbf{\overline{Q}}}{r_0})\|^2_{\mathbf{\overline{W}}} \!+\!
    \frac{r_0}{2} \|  \mathcal{\widetilde{D}}_{\mathcal{E}} \mathbf{v} \!-\! (\mathbf{\widetilde{Q}} + \frac{\lambda_\mathbf{\widetilde{Q}}}{r_0})\|^2_{\mathbf{\widetilde{W}}} \\
    & + \frac{r_1}{2} \|  \mathcal{D}_{\mathcal{M}} \textbf{N} - \mathbf{v} - (\mathbf{P} + \frac{\lambda_\mathbf{P}}{r_1})\|^2_{\mathbf{V}};
  \end{aligned}
  \end{equation}

  \item The $\mathbf{P}$-subproblem:
  \begin{equation}\label{eq:p-sub}
  \begin{aligned}
    \min\limits_{\mathbf{P}}  \alpha_1 \! \sum\limits_{e} \! w_{e} \! \left\|  \mathbf{P}_{e}\right\|\mathrm{len}(e)
    \!+\! \frac{r_1}{2} \| \mathbf{P} \!-\! (\mathcal{D}_{\mathcal{M}} \mathbf{N} \!-\! \mathbf{v} \!-\! \frac{\lambda_\mathbf{P}}{r_1})\|^2_{\mathbf{V}};
  \end{aligned}
  \end{equation}

  \item The $\mathbf{\overline{Q}}$-subproblem:
  \begin{equation}\label{eq:q1-sub}
  \begin{aligned}
    \min\limits_{\mathbf{\overline{Q}}}  \alpha_0 \!  \left\|  \mathbf{\overline{Q}}\right\|_{\mathbf{\overline{W}}}
    \!+\! \frac{r_0}{2} \| \mathbf{\overline{Q}} - ( \mathcal{\overline{D}}_{\mathcal{E}} \mathbf{v} - \frac{\lambda_{\mathbf{\overline{Q}}}}{r_0}) \|^2_{\mathbf{\overline{W}}};
  \end{aligned}
  \end{equation}

  \item The $\mathbf{\widetilde{Q}}$-subproblem:
  \begin{equation}\label{eq:q2-sub}
  \begin{aligned}
    \min\limits_{\mathbf{\widetilde{Q}}}  \alpha_0 \!  \left\|  \mathbf{\widetilde{Q}}\right\|_{\mathbf{\widetilde{W}}}
    \!+\! \frac{r_0}{2} \| \mathbf{\widetilde{Q}} - ( \mathcal{\widetilde{D}}_{\mathcal{E}} \mathbf{v} - \frac{\lambda_{\mathbf{\widetilde{Q}}}}{r_0}) \|^2_{\mathbf{\widetilde{W}}}.
  \end{aligned}
  \end{equation}
\end{itemize}

The $\mathbf{N}$-subproblem  \eqref{eq:n-sub} is a quadratic optimization problem with the unit normal constraints.
Here, we adopt an approximation strategy to solve this problem.
We first ignore the unit normal constraints and solve the quadratic program, and then project the minimizer onto the unit sphere.
Specifically, we check the first-order optimality condition of \eqref{eq:n-sub}, and obtain the following Euler-Lagrange equation
\begin{equation}\label{eq:n-solution}
   \beta \mathbf{N} - r_1 \mathcal{D}^{\star}_{\mathcal{M}}\mathcal{D}_{\mathcal{M}} \mathbf{N} = \beta \mathbf{N}^{in} - \mathcal{D}^{\star}_{\mathcal{M}}\big(\lambda_{\mathbf{P}} + r_1 (\mathbf{P}+\mathbf{v})\big).
\end{equation}
Plugging in the first-order operator \eqref{firstOrderOperator} and its adjoint operator \eqref{adjoint-firstOrderOperator}, we can rewrite the above equation as a sparse and positive semidefinite linear system, which can be solved by efficient sparse linear solvers, such as TAUCS and Intel Math Kernel Library (MKL).

The $\mathbf{v}$-subproblem \eqref{eq:v-sub} is also a quadratic program, whose Euler-Lagrange equation is given as
\begin{equation}\label{eq:v-solution}
\begin{aligned}
  r_1 \mathbf{v} \!-\! r_0 \overline{\mathcal{D}}^{\star}_{\mathcal{E}} \mathcal{\overline{D}}_{\mathcal{E}} \mathbf{v} \!  -\! & r_0  \widetilde{\mathcal{D}}^{\star}_{\mathcal{E}}  \mathcal{\widetilde{D}}_{\mathcal{E}} \mathbf{v} = -\lambda_{\mathbf{P}} - r_1 (\mathbf{P}- \mathcal{D}_{\mathcal{M}} \mathbf{N}) \\
  &- \overline{\mathcal{D}}^{\star}_{\mathcal{E}}(\lambda_{\overline{Q}} \!+\! r_0 \overline{Q}) - \widetilde{\mathcal{D}}^{\star}_{\mathcal{E}}(\lambda_{\widetilde{Q}} \!+\! r_0 \widetilde{Q}).
\end{aligned}
\end{equation}
Plugging the 1- and 2-form operators \eqref{eq:1-formOperator} and \eqref{eq:2-formOperator} and their adjoint operators \eqref{eq:adjoint-1-formOperator} and \eqref{eq:adjoint-2-formOperator}, we can rewrite the above equation as a sparse linear system, which again can be solved by linear solvers.

The $\mathbf{P}$-subproblem \eqref{eq:p-sub} is solved directly as it can be spatially decoupled, where the minimization problem for each edge is solved separately.
For each $\mathbf{P}_e$, we have the following simplified problem:
\begin{equation*}
      \min\limits_{\mathbf{P}_e}  \alpha_1  w_{e} \! \left\|  \mathbf{P}_{e}\right\|
    + \frac{r_1}{2} \| \mathbf{P}_e - \big((\mathcal{D}_{\mathcal{M}} \mathbf{N}) |_{e} - \mathbf{v}_e - \frac{\lambda_{\mathbf{P}_e}}{r_1}\big)\|^2,
\end{equation*}
which has a closed form solution:
\begin{equation}\label{eq:p-solution}
\mathbf{P}_{e} = \mathrm{Shrink}(\alpha_1 w_e, \ r_1, \ (\mathcal{D}_{\mathcal{M}} \mathbf{N}) |_{e} \!-\! \mathbf{v}_e \!-\! \frac{\lambda_{\mathbf{P}_e}}{r_1}),
\end{equation}
with the soft shrinkage operator defined as:
\begin{equation*}
\mathrm{Shrink}(x,y,z)=\max(0, 1 - \frac{x}{y\|z\|})z.
\end{equation*}

The $\mathbf{\overline{Q}}$-subproblem \eqref{eq:q1-sub} is solved for each line independently.
For each $\mathbf{\overline{Q}}_l$, we solve the following problem:
\begin{equation*}
      \min\limits_{\mathbf{\overline{Q}}_l}  \alpha_0 \!  \left\|  \mathbf{\overline{Q}}_l \right\|
    \!+\! \frac{r_0}{2} \| \mathbf{\overline{Q}}_l - \big( (\mathcal{\overline{D}}_{\mathcal{E}} \mathbf{v})|_{l} - \frac{\lambda_{\mathbf{\overline{Q}}_l}}{r_0}\big) \|^2,
\end{equation*}
whose closed form solution is:
\begin{equation}\label{eq:q1-solution}
\mathbf{\overline{Q}}_{l} = \mathrm{Shrink}(\alpha_0, \ r_0, \ (\mathcal{\overline{D}}_{\mathcal{E}} \mathbf{v})|_{l} - \frac{\lambda_{\mathbf{\overline{Q}}_l}}{r_0}).
\end{equation}

Similarly, the $\mathbf{\widetilde{Q}}$-subproblem \eqref{eq:q2-sub} can be separated into the following problem for each curve $\mathbf{\widetilde{Q}}_c$:
\begin{equation*}
      \min\limits_{\mathbf{\widetilde{Q}}_c}  \alpha_0 \!  \left\|  \mathbf{\widetilde{Q}}_c \right\|
    \!+\! \frac{r_0}{2} \| \mathbf{\widetilde{Q}}_c - \big( (\mathcal{\widetilde{D}}_{\mathcal{E}} \mathbf{v})|_{c} - \frac{\lambda_{\mathbf{\widetilde{Q}}_c}}{r_0}\big) \|^2,
\end{equation*}
which has a closed form solution:
\begin{equation}\label{eq:q2-solution}
\mathbf{\widetilde{Q}}_{c} = \mathrm{Shrink}(\alpha_0, \ r_0, \ (\mathcal{\widetilde{D}}_{\mathcal{E}} \mathbf{v})|_{c} - \frac{\lambda_{\mathbf{\widetilde{Q}}_c}}{r_0}).
\end{equation}

\begin{figure}[thb]
  \centering
  \includegraphics[width=0.48\textwidth]{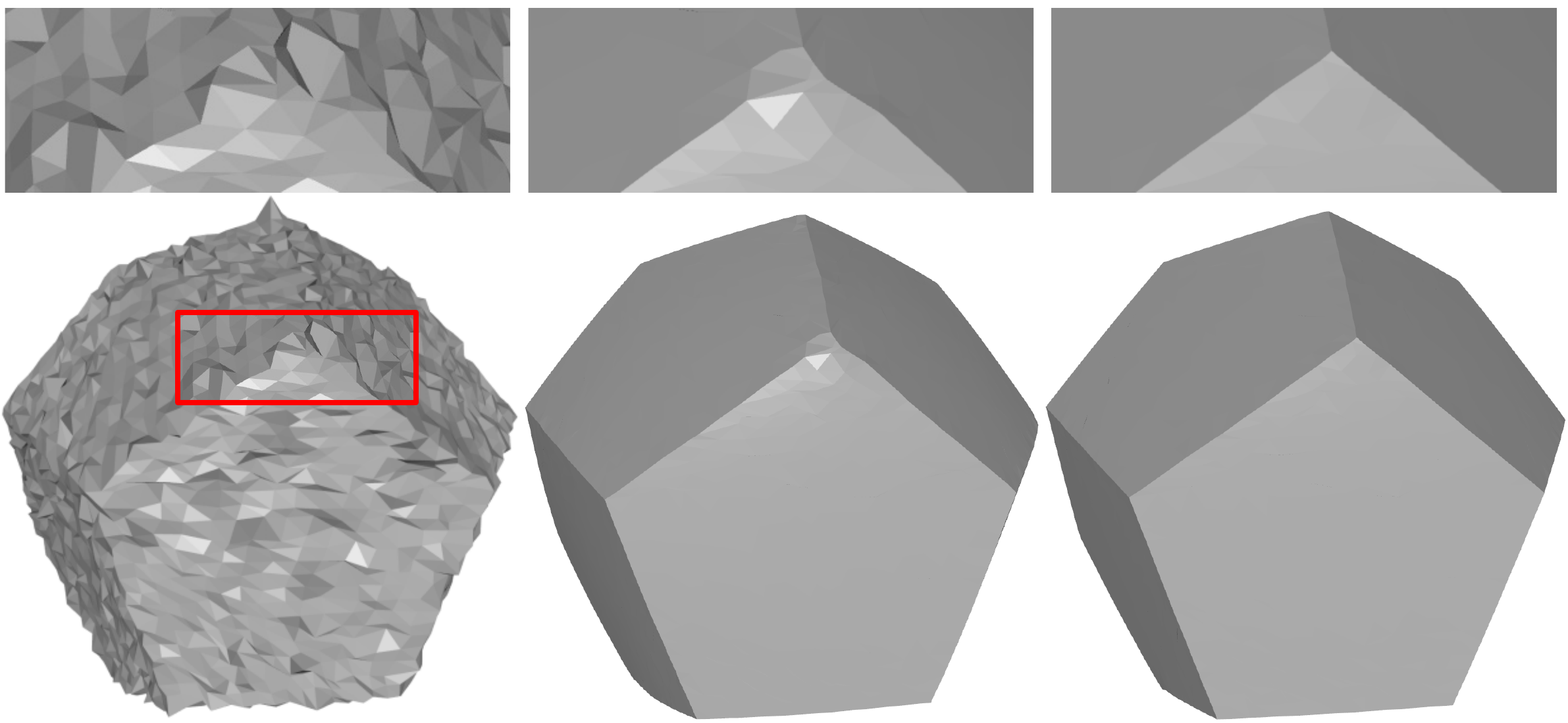}
  \caption{Denoising results of Dodecahedron (corrupted with $\sigma =0.3 \bar{l}_e$, where $\sigma$ is standard deviation of Gaussian noise and $\bar{l}_e$ is mean edge length). From left to right: noisy mesh, denoising results produced by vectorial TGV normal filter \eqref{eq:TGV-NormalFilteringModel} without and with dynamic weights, respectively.
  \label{fig:weight}}
\end{figure}

In summary, the full procedure for the TGV normal filter \eqref{eq:TGV-NormalFilteringModel} is sketched in Algorithm \ref{alg:normalFiltering}.
Based on variable-splitting and ALM, this algorithm iteratively solves the above five subproblems and updates the Lagrange multipliers. The algorithm terminates when one of the stopping criteria is met.
As mentioned in section \ref{sec:5.1}, the dynamic weights $w_e$ play a key role in recovering sharp features.
As Fig. \ref{fig:weight} shows, without these dynamic weights, some sharp features are blurred in the denoised result.

\begin{algorithm}[hbt]
\caption{ALM for TGV normal filtering \eqref{eq:TGV-NormalFilteringModel}}
\label{alg:normalFiltering}
\textbf{Initialization:} $\mathbf{N}^{-1} = \mathbf{v}^{-1} = \mathbf{P}^{-1} = \mathbf{\overline{Q}}^{-1}=\mathbf{\widetilde{Q}}^{-1}=0, \lambda_\mathbf{P}^{0} = \lambda_\mathbf{\overline{Q}}^{0} = \lambda_\mathbf{\widetilde{Q}}^{0}=0, k=0$\;
\Repeat{$\|\mathbf{N}^{k}-\mathbf{N}^{k-1}\|_{\mathbf{U}}^2 < 1e-10$ $\mathrm{or}$ $k \geq 100$}
{
    \SetAlgoNoLine
     1. fix $(\mathbf{v}^{k-1}, \mathbf{P}^{k-1}, \lambda^k_{\mathbf{P}})$, solve $\mathbf{N}^k$ by \eqref{eq:n-solution}\; \quad normalize $\mathbf{N}^{k}$\;
     2. fix \!$(\mathbf{N}^k, \mathbf{P}^{k-1} \!\!,\! \mathbf{\overline{Q}}^{k-1} \!\!,\! \mathbf{\widetilde{Q}}^{k-1} \!\!,\! \lambda^k_\mathbf{P}, \lambda^k_{\mathbf{\overline{Q}}}, \lambda^k_{\mathbf{\widetilde{Q}}})$, solve $\mathbf{v}^k$ by \eqref{eq:v-solution}\;
     3. fix $(\mathbf{N}^k, \mathbf{v}^k, \lambda^k_\mathbf{P})$, solve $\mathbf{P}^k$ by \eqref{eq:p-solution}\;
     4. fix $(\mathbf{v}^k, \lambda^k_{\mathbf{\overline{Q}}})$, solve $\mathbf{\overline{Q}}^k$ by \eqref{eq:q1-solution}\;
     5. fix $(\mathbf{v}^k, \lambda^k_{\mathbf{\widetilde{Q}}})$, solve $\mathbf{\widetilde{Q}}^k$ by \eqref{eq:q2-solution}\;
     6. update Lagrange multipliers\\
     \quad $\lambda_\mathbf{P}^{k+1} =  \lambda_\mathbf{P}^{k}+r_1\big(\mathbf{P}^k- (\mathcal{D}_{\mathcal{M}} \mathbf{N}^k-\mathbf{v}^k) \big) $\;
    \quad $\lambda_\mathbf{\overline{Q}}^{k+1} = \lambda_\mathbf{\overline{Q}}^{k}+r_0(\mathbf{\overline{Q}}^k - \mathcal{\overline{D}}_{\mathcal{E}} \mathbf{v}^k ) $\;
    \quad $\lambda_\mathbf{\widetilde{Q}}^{k+1} =  \lambda_\mathbf{\widetilde{Q}}^{k}+r_0(\mathbf{\widetilde{Q}}^k - \mathcal{\widetilde{D}}_{\mathcal{E}} \mathbf{v}^k ) $\;
    7. update weights $w_e$ using \eqref{eq:TGV-weight} \;
    8. Increment $k$: $k=k+1$\;
}
\Return $\mathbf{N}^k$.
\end{algorithm}

\subsection{Vertex Updating Scheme}
After smoothing the normal field via the proposed TGV-based normal filter, vertex positions should be updated to match the filtered normals.
To avoid the triangle orientation ambiguity problem in the traditional vertex updating scheme \cite{Sun07Fast}, we reconstruct the mesh using the vertex updating scheme proposed by Zhang et al. \cite{Zhang19Static}.
We empirically fix the iteration number as 30 in our experiments, which allows producing satisfactory results.
We refer the interested reader to the work \cite{Zhang19Static} for more details.

\section{Experiments and Discussions} \label{sec:6}
We test the proposed denoising method on a variety of meshes including CAD, non-CAD, and scanned data.
The tested meshes are corrupted by either synthetic or raw noise.
The synthetic noise is generated by a zero-mean Gaussian function with mean edge length ($\bar{l}_e$) as the standard deviation ($\sigma$).
We present visual and numerical comparisons between the proposed method (TGV) and the state-of-the-art, including the total variation filter (TV) \cite{Zhang15Variational}, the high-order filter (HO) \cite{Liu2019ANovel}, $\ell_0$ minimization (L0) \cite{He13Mesh}, the bilateral filter (BF) \cite{Zheng11Bilateral}, the non-local low-rank filter (NLLR) \cite{Li2018NonLocal}, and the cascaded filter (CNR) \cite{Wang2016Mesh}.
We implemented TV, HO, L0, and BF according to the literature in C++.
For NLLR, we execute the code kindly provided by the authors of \cite{Li2018NonLocal} to produce the results.
For CNR, we directly use the trained neural networks kindly provided by the authors \cite{Wang2016Mesh} to generate the results.
We carefully tune the parameters of each competing methods so that satisfactory results are produced.
All the methods are performed on a laptop with an Intel i7 dual core 2.6 GHz processor and 16 GB RAM. All the meshes are rendered in flat-shading to emphasize faceting effect.
To promote reproducibility, we release our executable program and data in the GitHub page \footnote{https://github.com/LabZhengLiu/MeshTGV}.

\begin{figure*}[htb]
    \centering
    \subfloat[Noisy]{\label{fig:parameter-alpha1-a}\includegraphics[width=0.18\textwidth]{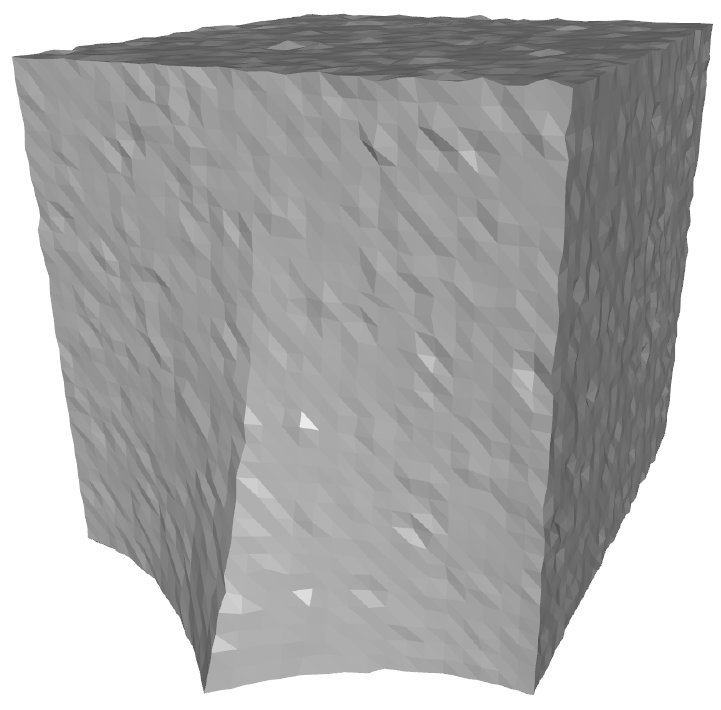}}
    \subfloat[ $\alpha_1=0.05$]{\label{fig:parameter-alpha1-b}\includegraphics[width=0.18\textwidth]{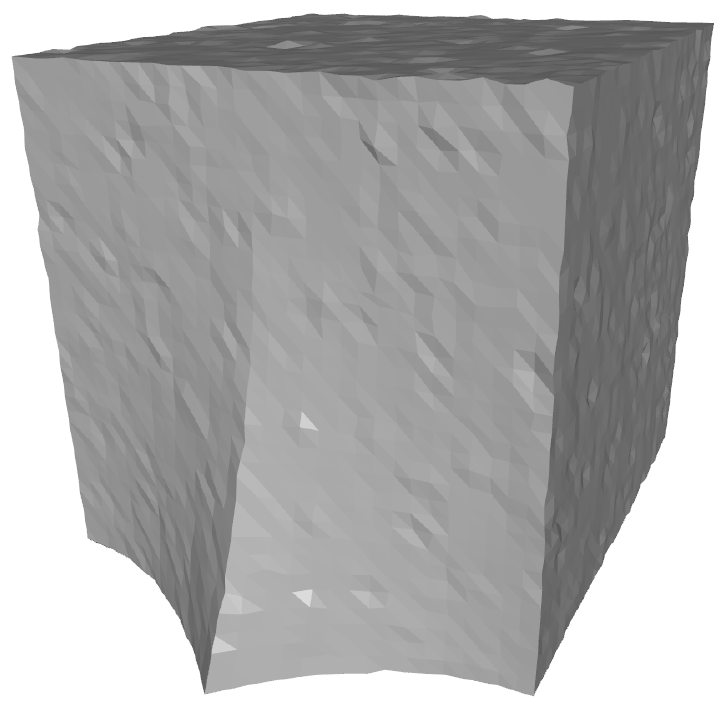}}
    \subfloat[$\alpha_1=0.4$]{\label{fig:parameter-alpha1-c}\includegraphics[width=0.18\textwidth]{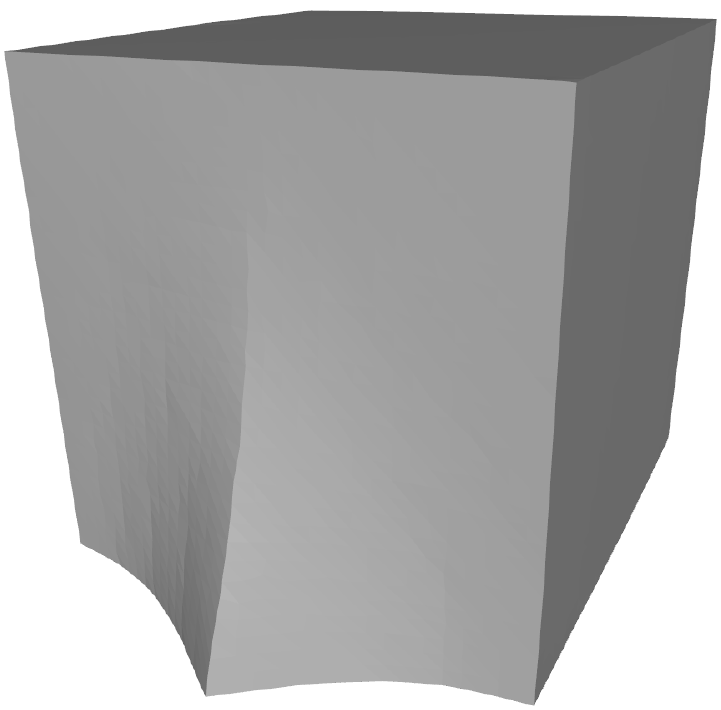}}
    \subfloat[$\alpha_1=0.7$]{\label{fig:parameter-alpha1-d}\includegraphics[width=0.18\textwidth]{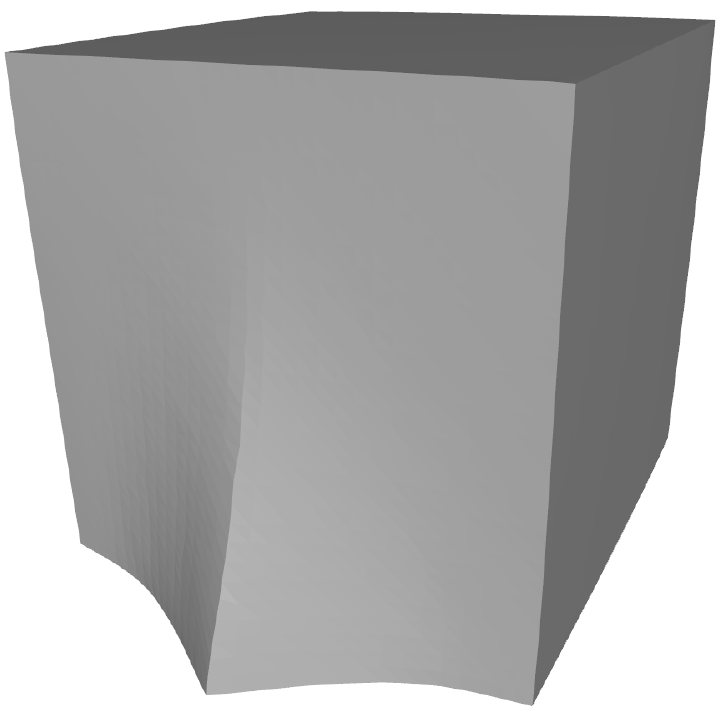}}
    \subfloat[$\alpha_1=2.0$]{\label{fig:parameter-alpha1-e}\includegraphics[width=0.18\textwidth]{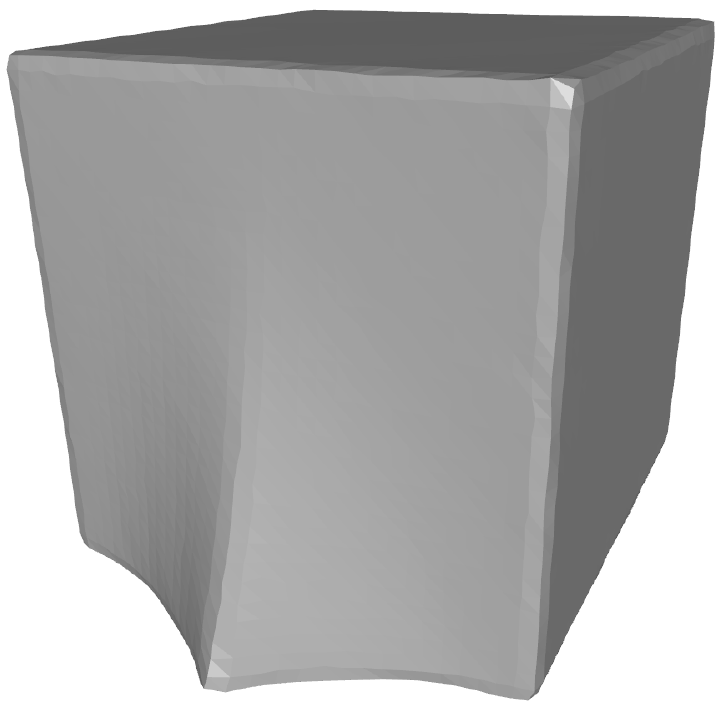}}
    \caption{Denoising results for varying $\alpha_1$ with fixed $\alpha_0$ and $\beta$. From left to right: noisy mesh (corrupted with $\sigma=0.1 \bar{l}_e$), and results with increasing $\alpha_1$.}
    \label{fig:parameter-alpha1}
\end{figure*}

\begin{figure*}[htb]
    \centering
    \subfloat[Noisy]{\label{fig:parameter-alpha0-a}\includegraphics[width=0.17\textwidth]{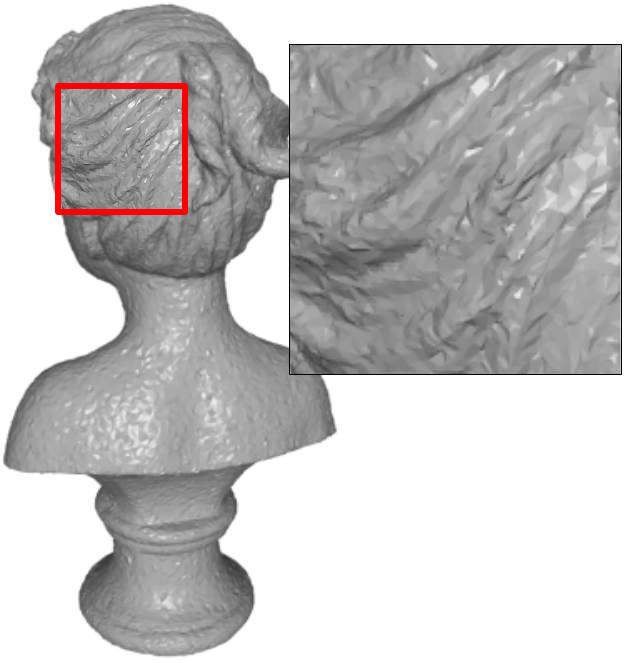}}
    \subfloat[ $\alpha_0=0.01$]{\label{fig:parameter-alpha0-b}\includegraphics[width=0.17\textwidth]{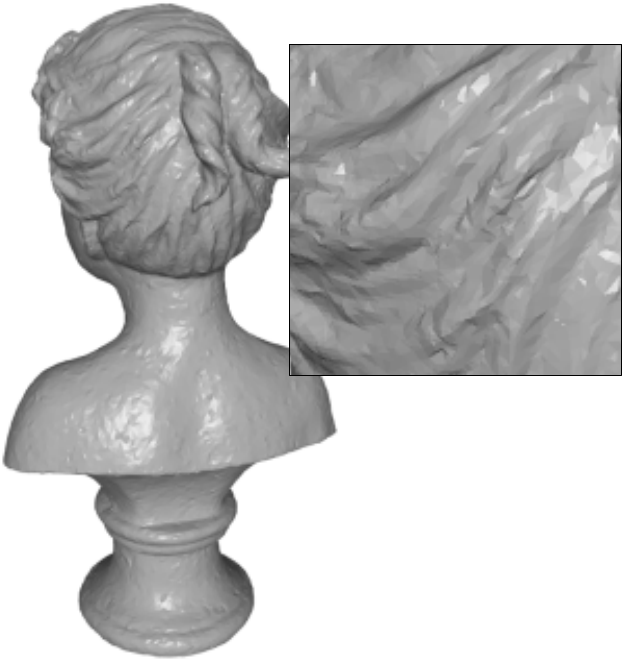}}
    \subfloat[$\alpha_0=0.03$]{\label{fig:parameter-alpha0-c}\includegraphics[width=0.17\textwidth]{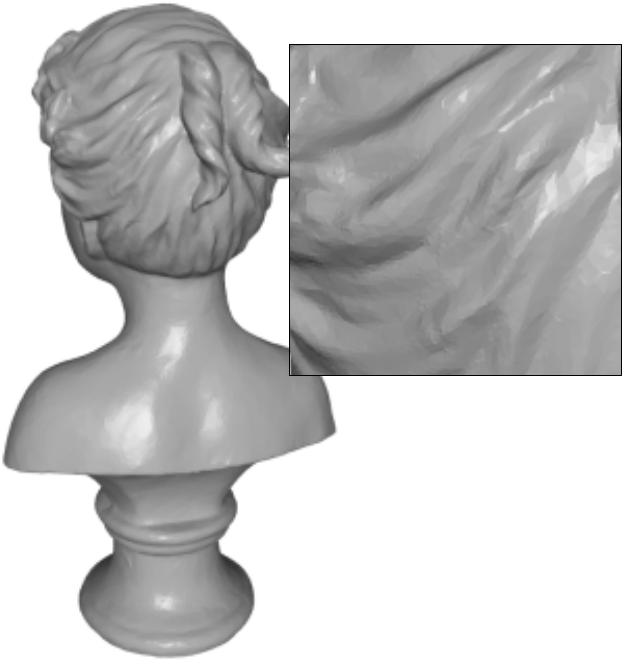}}
    \subfloat[$\alpha_0=0.05$]{\label{fig:parameter-alpha0-d}\includegraphics[width=0.17\textwidth]{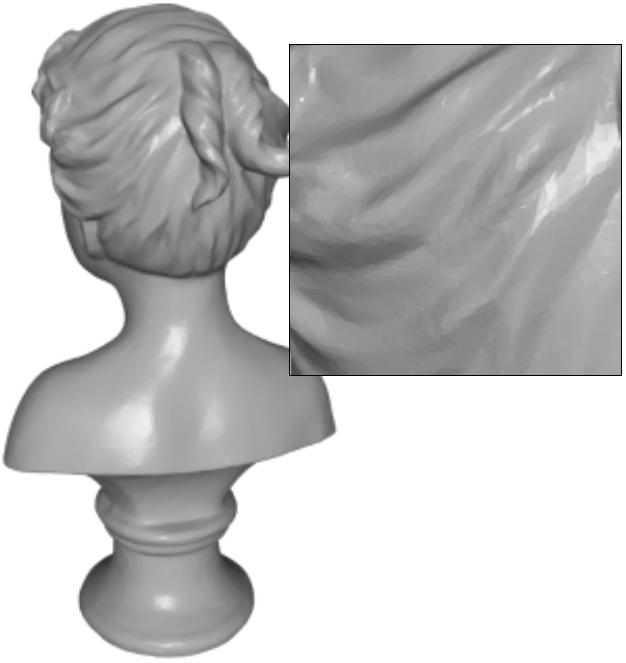}}
    \subfloat[$\alpha_0=1.0$]{\label{fig:parameter-alpha0-e}\includegraphics[width=0.17\textwidth]{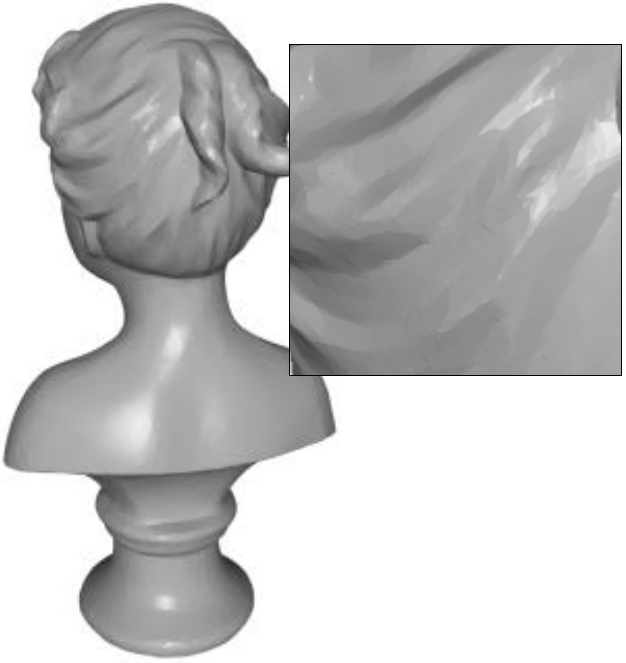}}
    \caption{Denoising results for varying $\alpha_0$ with fixed $\alpha_1$ and $\beta$. From left to right: noisy mesh (corrupted with $\sigma=0.15 \bar{l}_e$), and results with increasing $\alpha_0$.}
    \label{fig:parameter-alpha0}
\end{figure*}

\subsection{Parameters Setting}
\begin{figure}[thb]
  \centering
  \subfloat[Input (Result)]{\label{fig:gradientAndv-a}\includegraphics[width=0.148\textwidth]{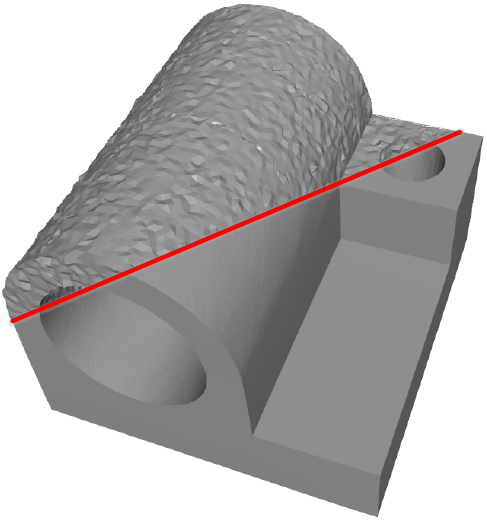}}
  \subfloat[$\mathcal{D}_{\mathcal{M}} \mathbf{N}$]{\label{fig:gradientAndv-b}\includegraphics[width=0.148\textwidth]{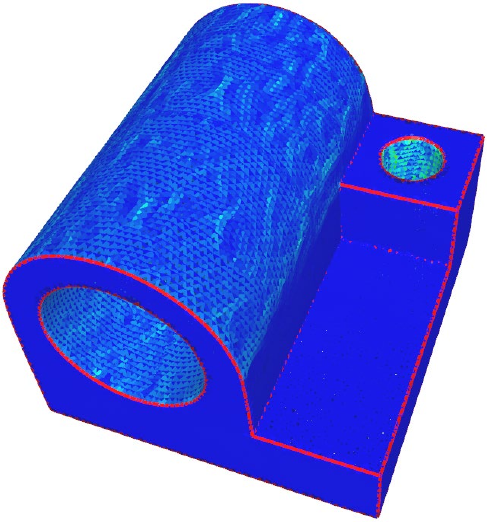}}
  \subfloat[$\mathbf{v}$]{\label{fig:gradientAndv-c}\includegraphics[width=0.155\textwidth]{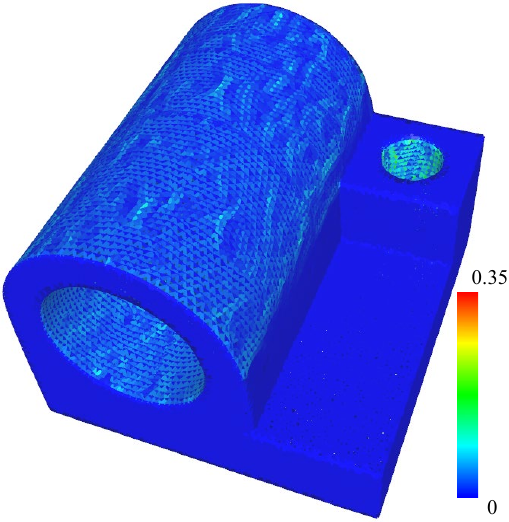}}
  \caption{(a) Noisy input (Denoising result). (b) $\mathcal{D}_{\mathcal{M}} \mathbf{N}$ of the result visualized using color coding. (c) The visualization of $\mathbf{v}$ of the result.
  \label{fig:gradientAndv}}
\end{figure}

Our TGV filter \eqref{eq:TGV-NormalFilteringModel} has three parameters, i.e., $\alpha_1$, $\alpha_0$, and $\beta$, which balance the first-, second-order, and fidelity terms of \eqref{eq:TGV-NormalFilteringModel}.
When the parameters are chosen properly, on one hand, in smooth regions we have $\mathbf{v} \approx \mathcal{D}_{\mathcal{M}} \mathbf{N}$, which results in the first-order term
being close to vanishing.
Thus, the minimization of \eqref{eq:TGV-NormalFilteringModel} is mainly controlled by the second-order term 
in these smooth regions; see the corresponding regions in Figs. \ref{fig:gradientAndv-b} and \ref{fig:gradientAndv-c}.
On the other hand, in regions near sharp features, we have $\mathbf{v} \approx 0$, which causes the second-order term to be close to vanishing.
Thus, the minimization mainly depends on the first-order term in these regions; see the corresponding regions in Figs. \ref{fig:gradientAndv-b} and \ref{fig:gradientAndv-c}.

Parameter $\alpha_1$ controls the impact of the first-order term in \eqref{eq:TGV-NormalFilteringModel}.
For each noisy mesh, these exist a range ([0.5, 3.0]) for $\alpha_1$, that leads to promising results.
This indicates that our method is insensitive to the perturbation of $\alpha_1$; see Figs. \ref{fig:parameter-alpha1-c} and \ref{fig:parameter-alpha1-d}.
If $\alpha_1$ is too small, the first-order term gets practically ignored causing $\mathbf{v} \approx 0$
over the whole mesh, which in turn causes the second-order term being close to vanishing.
Therefore, the TGV regularizer fails, leading to residual noise in the result; see Fig. \ref{fig:parameter-alpha1-b}.
If $\alpha_1$ is too large, in regions near sharp features, the first-order term tends to have $\mathbf{v} \approx \mathcal{D}_{\mathcal{M}} \mathbf{N}$, and the minimization of \eqref{eq:TGV-NormalFilteringModel} will be controlled by the second-order term in these regions, which may smooth sharp features; see Fig. \ref{fig:parameter-alpha1-e}.

Parameter $\alpha_0$ influences the effect of the second-order term in \eqref{eq:TGV-NormalFilteringModel}.
Similar to $\alpha_1$, for each noisy mesh, there exist a range ([0.05, 1]) for $\alpha_0$ that can produce satisfactory results; see Figs. \ref{fig:parameter-alpha0-c} and \ref{fig:parameter-alpha0-d}.
Underweighting the second-order term leads to residual noise in the result; see Fig. \ref{fig:parameter-alpha0-b}.
In contrast, overweighting the second-order term will penalize smooth regions as well as fine features and therefore oversmooth the details; see Fig. \ref{fig:parameter-alpha0-e}.

Parameter $\beta$ controls the degree of denoising procedure.
It is empirically fixed as $100$ for CAD and scanned surfaces usually, and is fixed as $1000$ for non-CAD surfaces.

\begin{figure*}[htb]
    \centering
    \subfloat[Noisy]{\label{block-a}\includegraphics[width=0.12\textwidth]{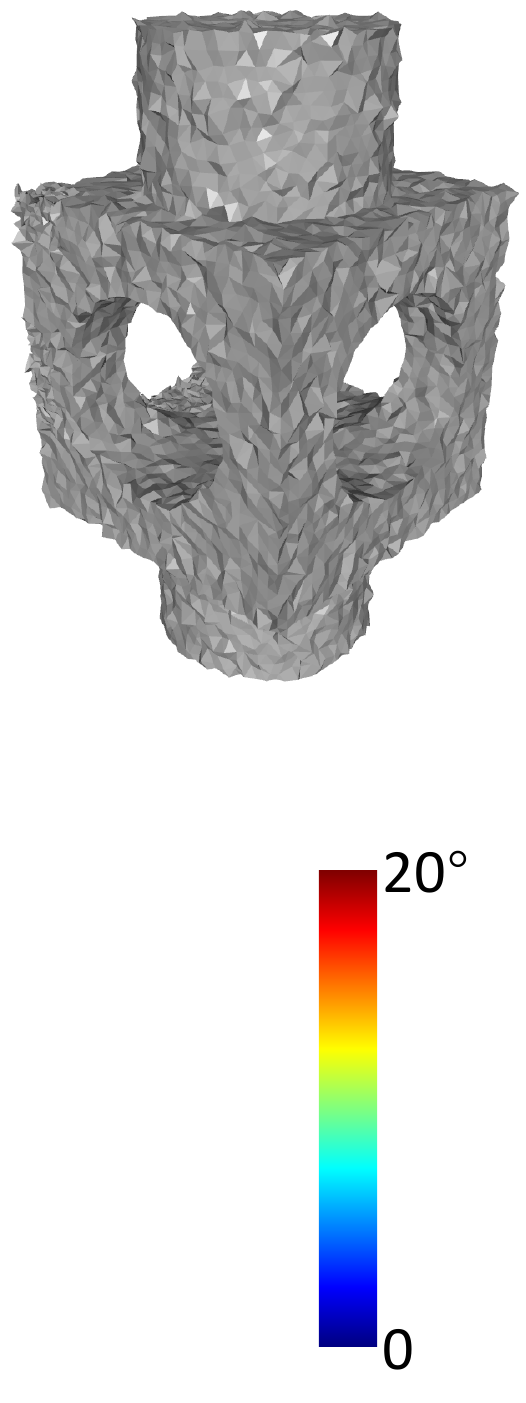}}
    \subfloat[TV]{\label{block-b}\includegraphics[width=0.12\textwidth]{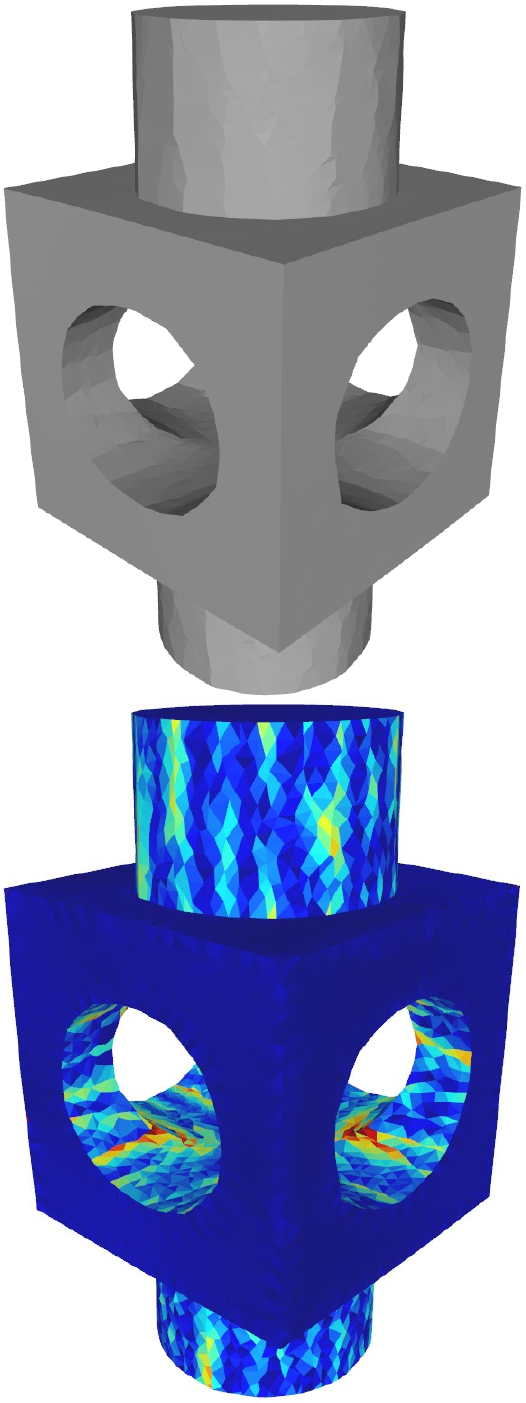}}
    \subfloat[HO]{\label{block-c}\includegraphics[width=0.12\textwidth]{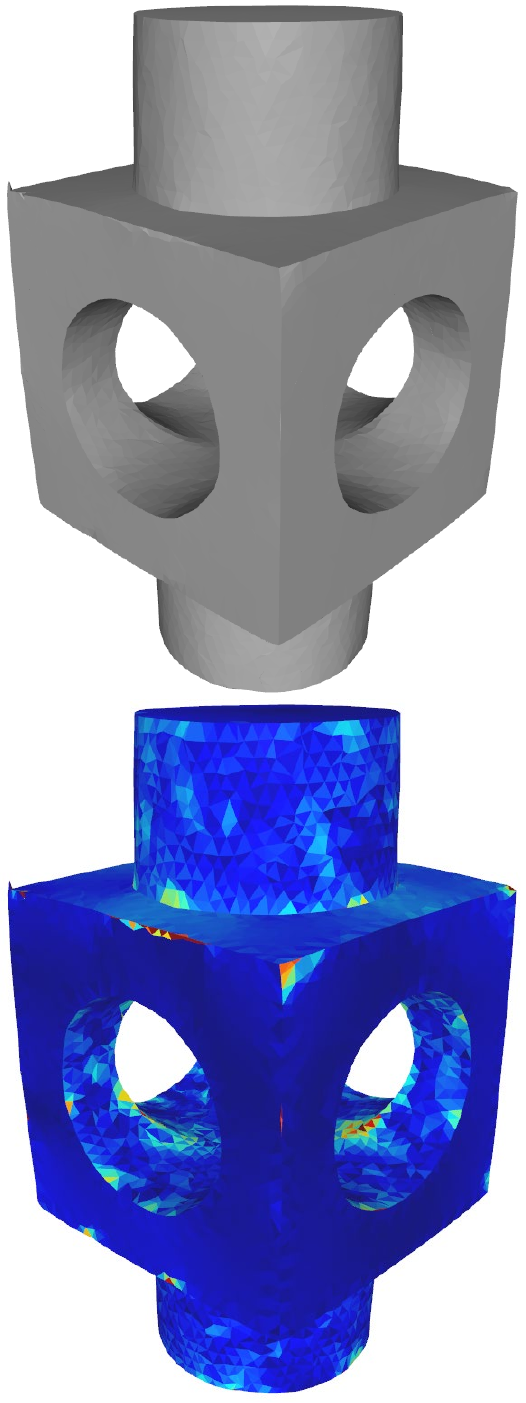}}
    \subfloat[L0]{\label{block-d}\includegraphics[width=0.12\textwidth]{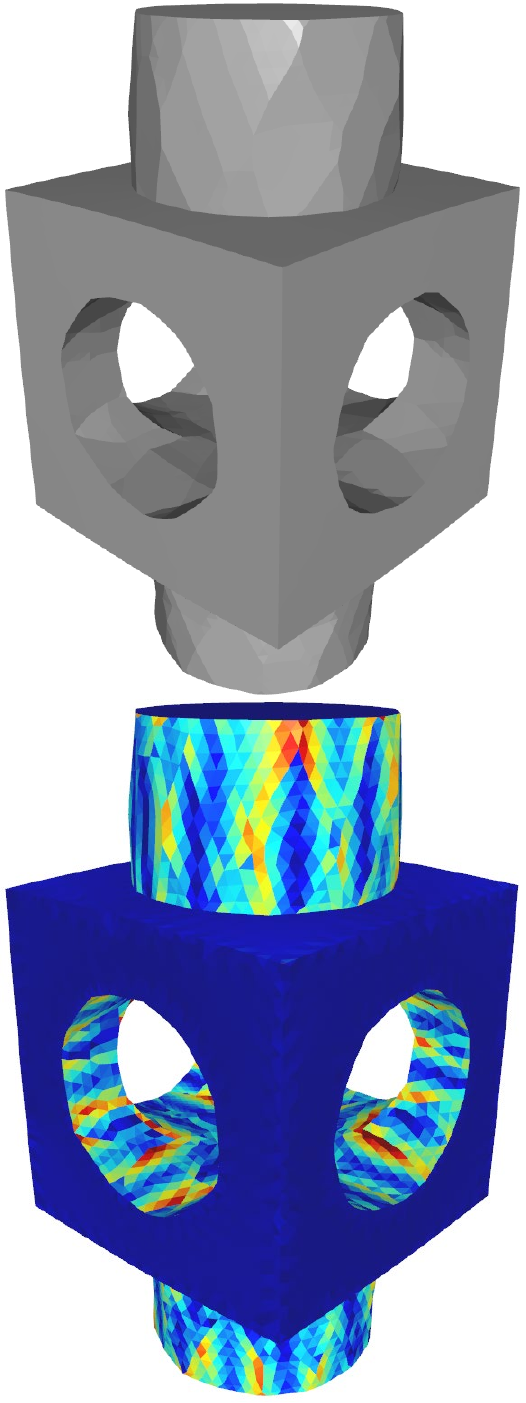}}
    \subfloat[BF]{\label{block-e}\includegraphics[width=0.12\textwidth]{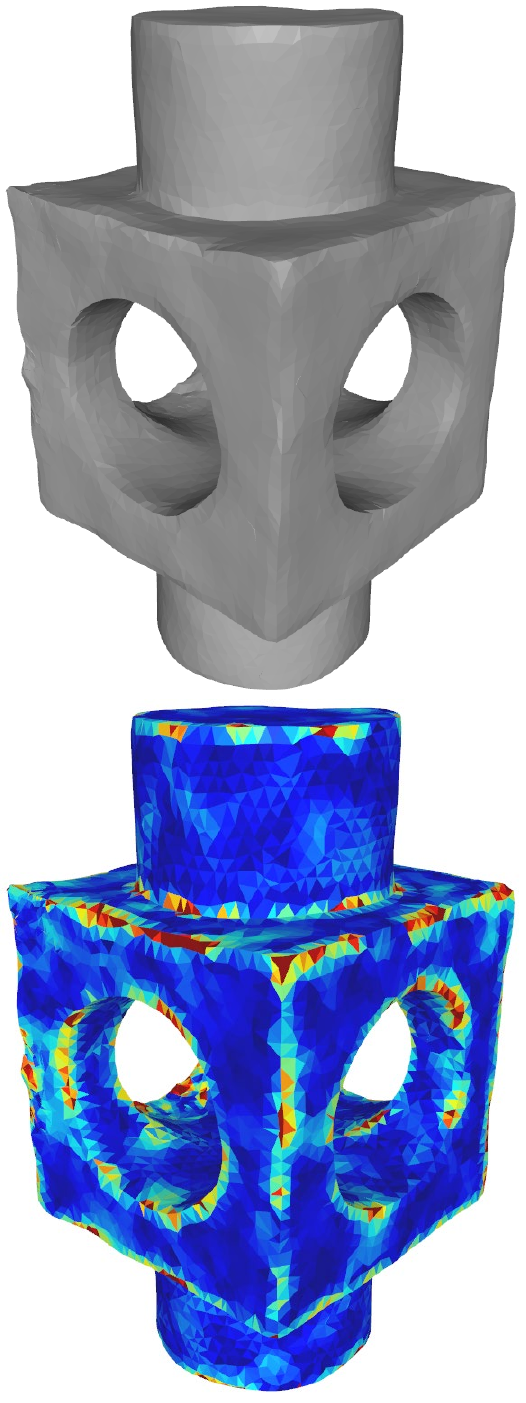}}
    \subfloat[NLLR]{\label{block-f}\includegraphics[width=0.12\textwidth]{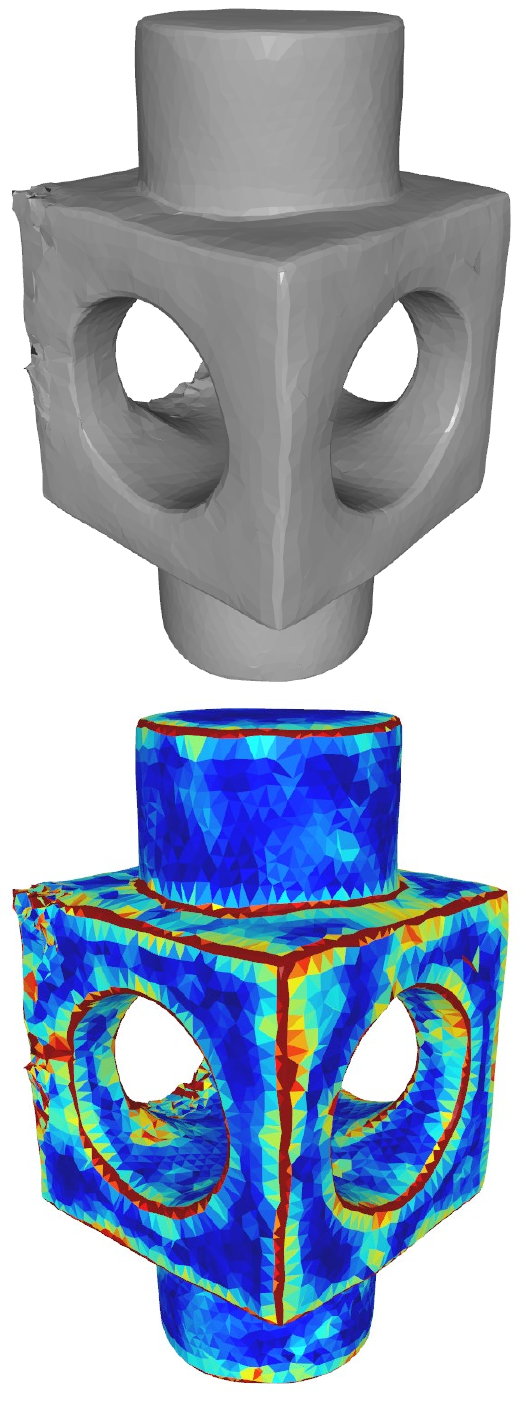}}
    \subfloat[CNR]{\label{block-g}\includegraphics[width=0.12\textwidth]{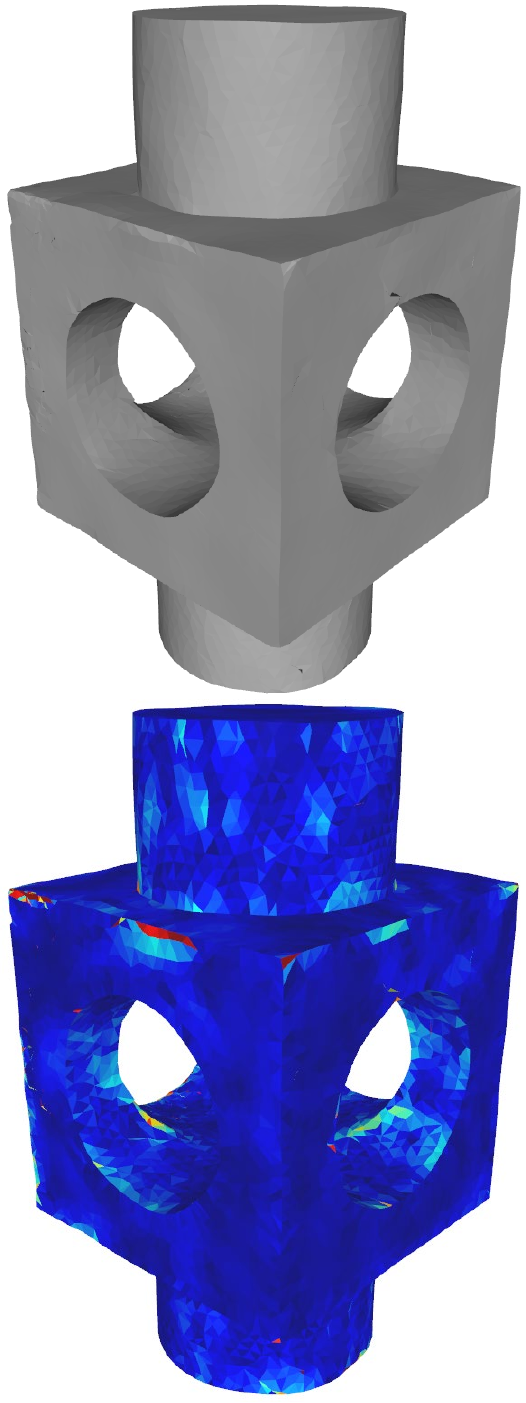}}
    \subfloat[Ours]{\label{block-h}\includegraphics[width=0.12\textwidth]{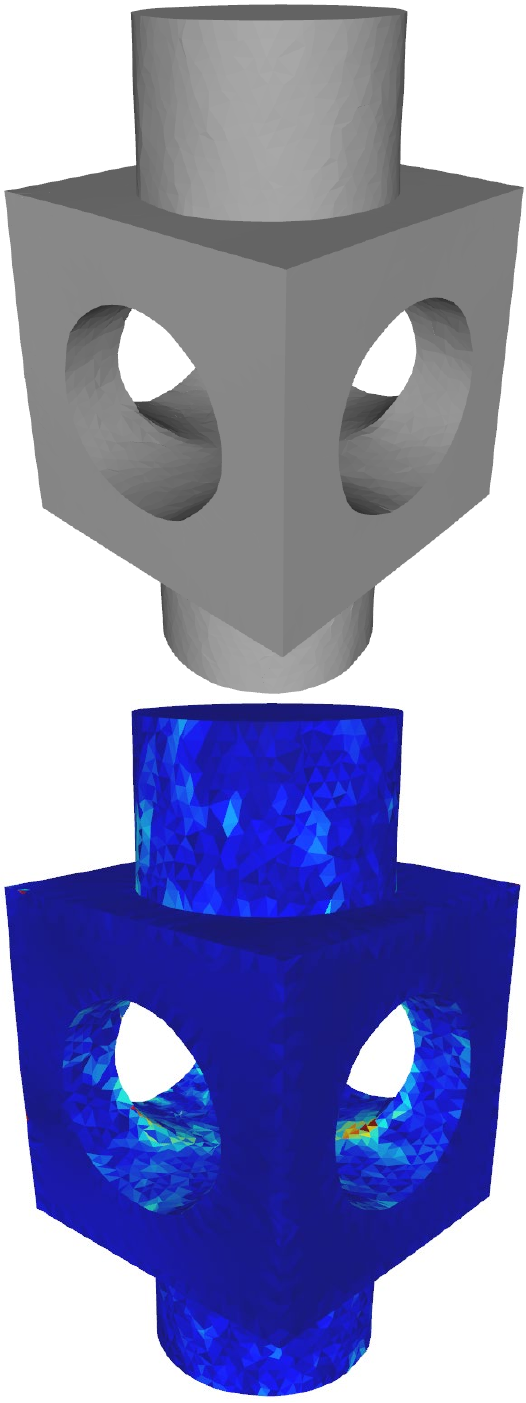}}
    \caption{Comparison of denoising results of Block, corrupted with $\sigma=0.35 \bar{l}_e$.
    The second row visualizes the corresponding error maps, using the angular difference between face normals of denoised meshes and ground truth meshes. }
    \label{fig:block}
\end{figure*}

\begin{figure*}[htb]
    \centering
    \subfloat[Noisy]{\label{fandisk-a}\includegraphics[width=0.125\textwidth]{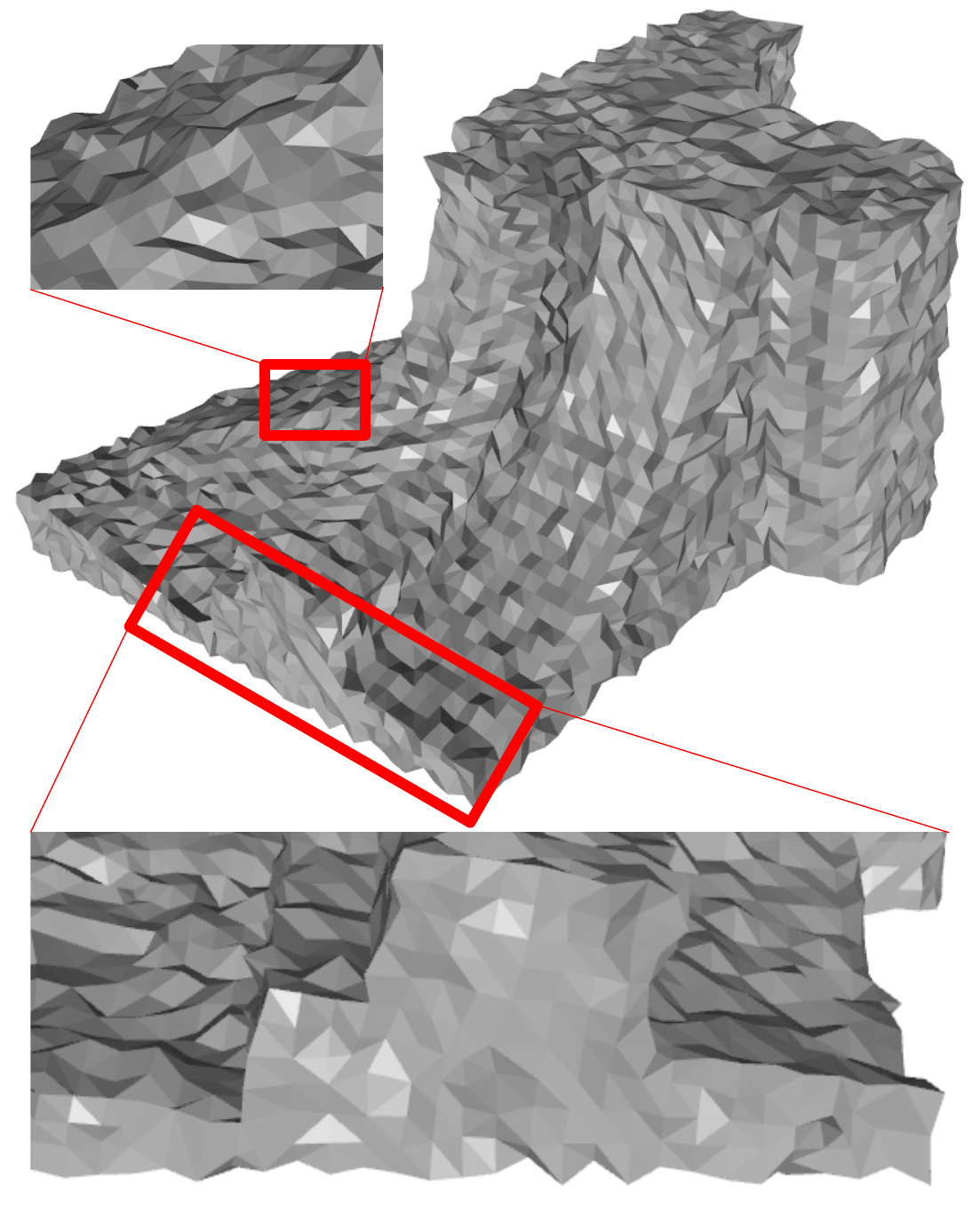}}
    \subfloat[TV]{\label{fandisk-b}\includegraphics[width=0.125\textwidth]{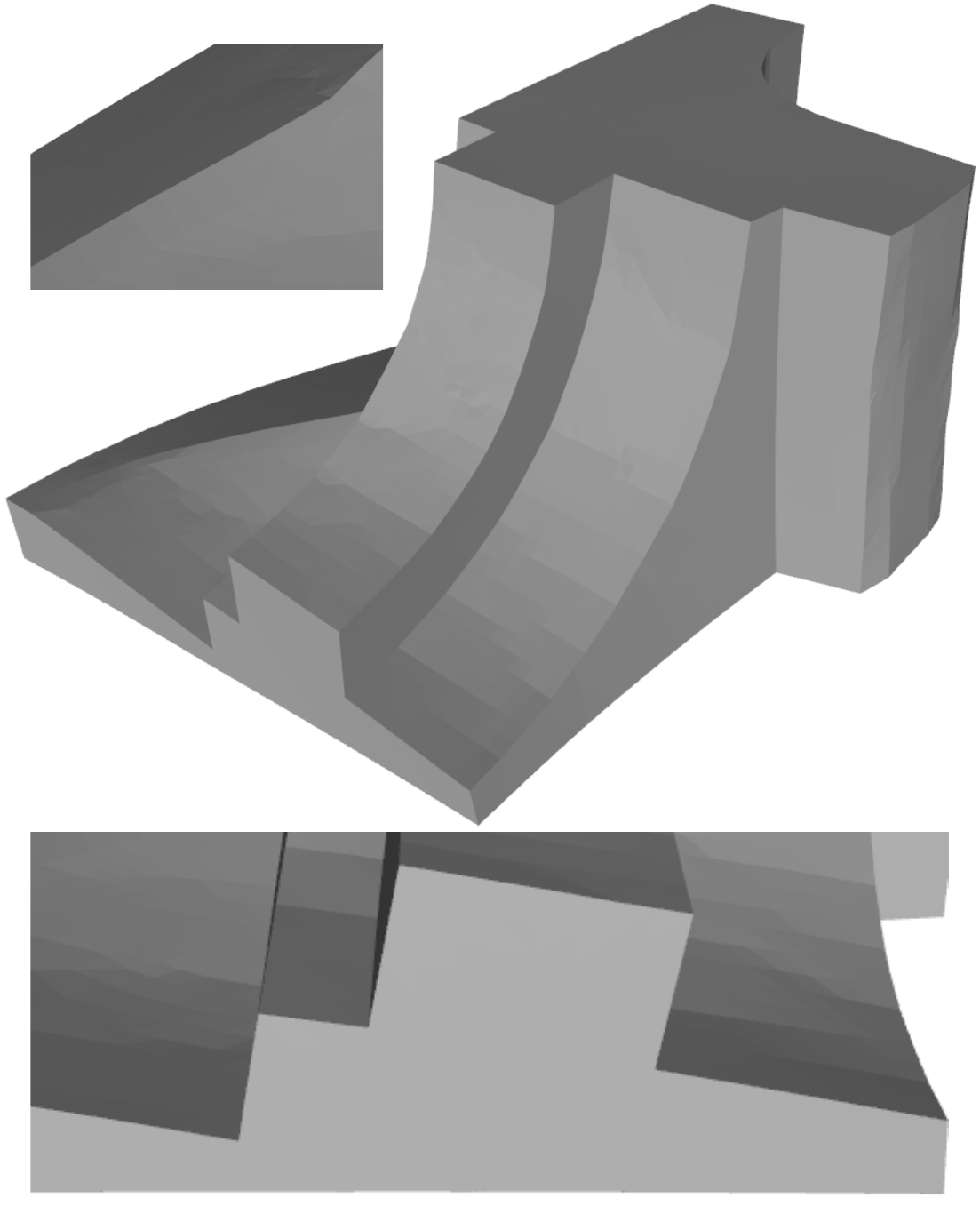}}
    \subfloat[HO]{\label{fandisk-c}\includegraphics[width=0.125\textwidth]{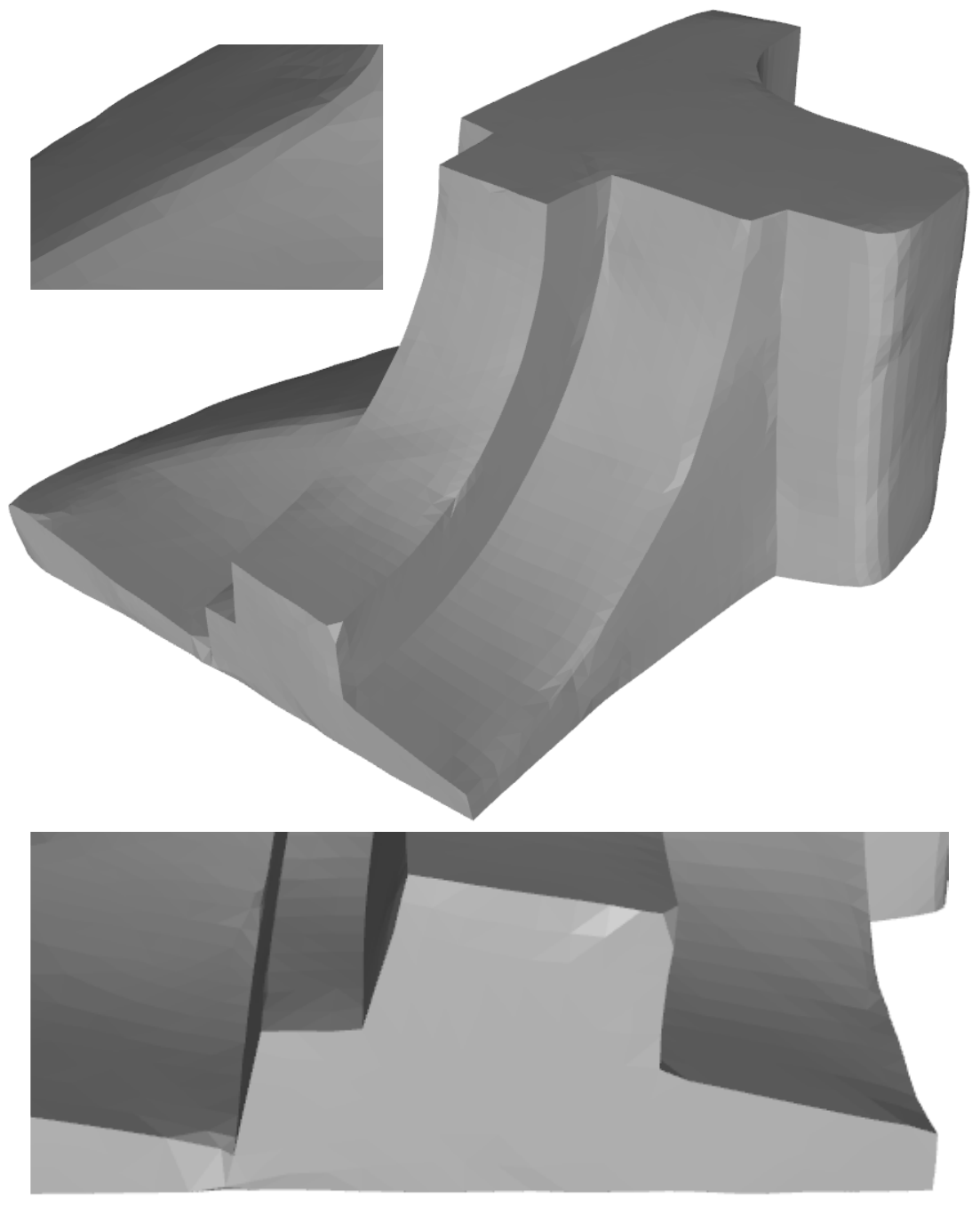}}
    \subfloat[L0]{\label{fandisk-d}\includegraphics[width=0.125\textwidth]{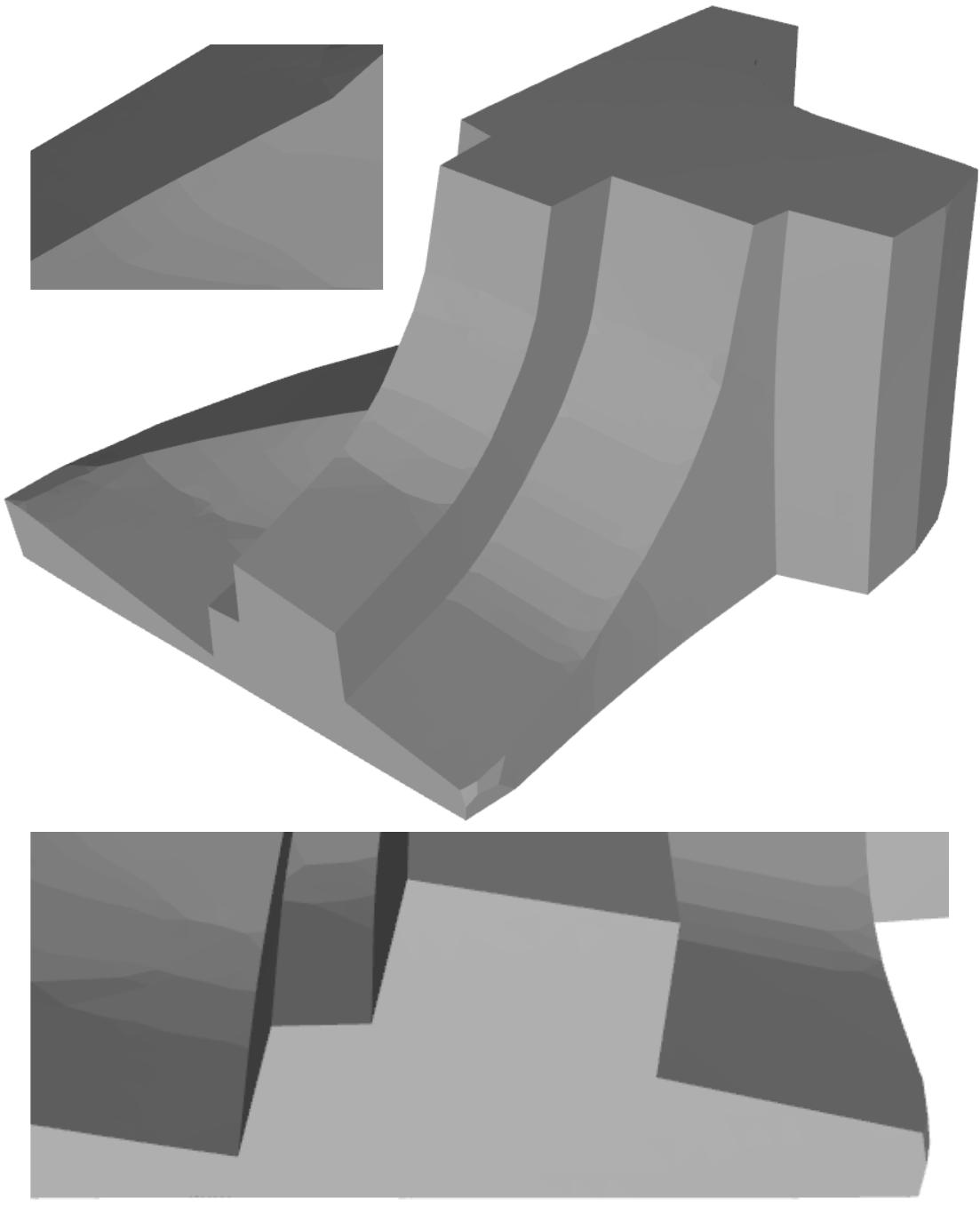}}
    \subfloat[BF]{\label{fandisk-e}\includegraphics[width=0.125\textwidth]{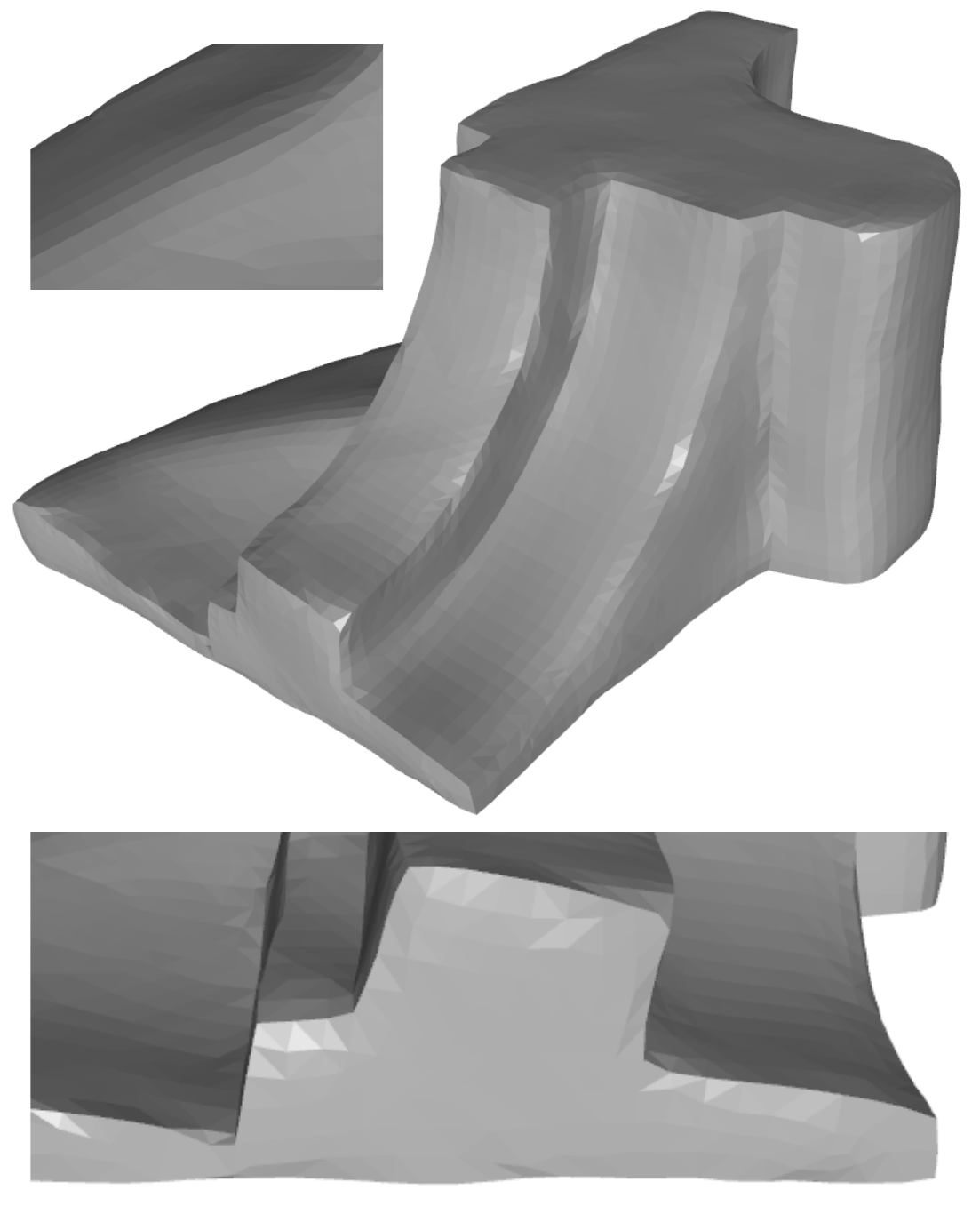}}
    \subfloat[NLLR]{\label{fandisk-f}\includegraphics[width=0.125\textwidth]{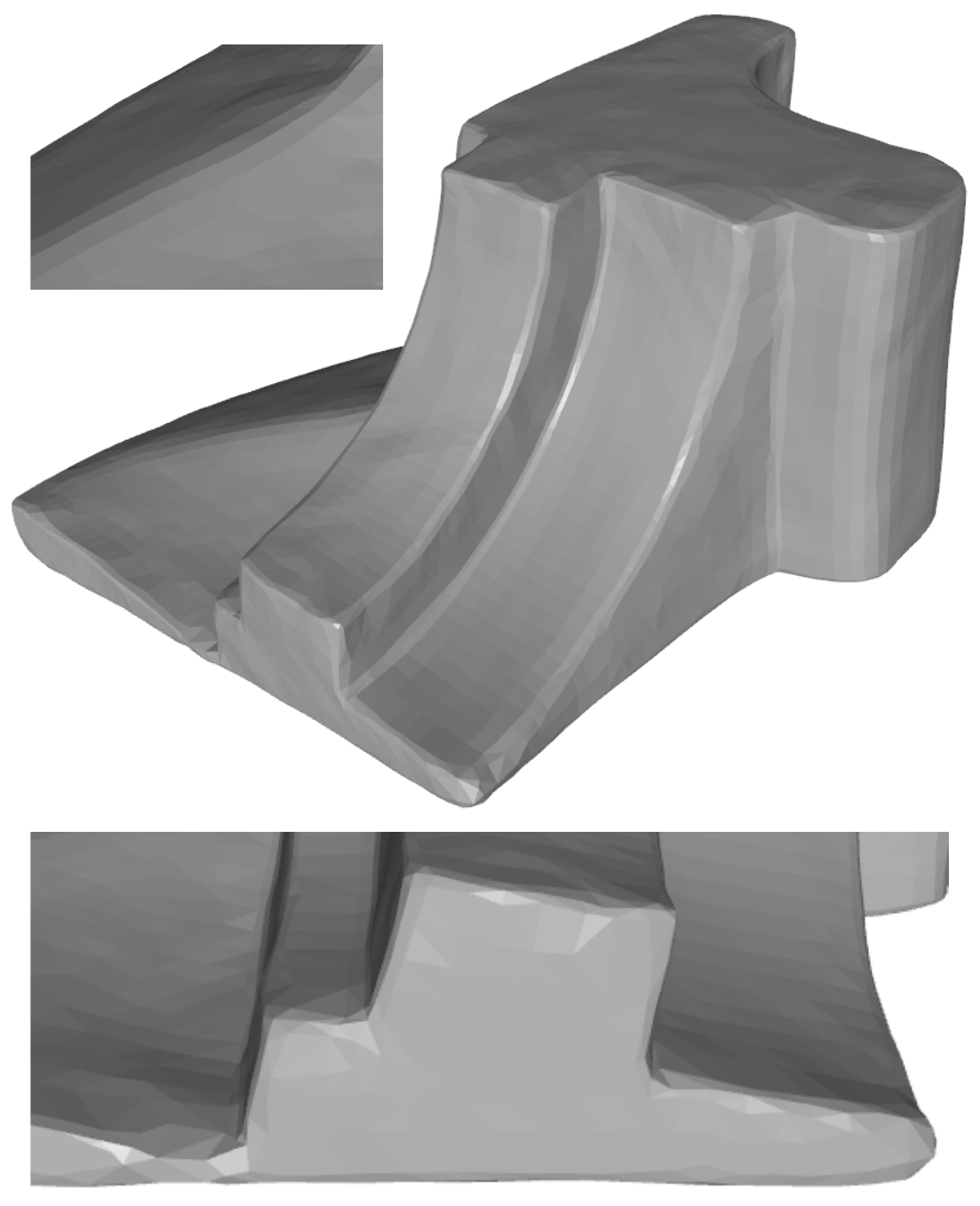}}
    \subfloat[CNR]{\label{fandisk-g}\includegraphics[width=0.125\textwidth]{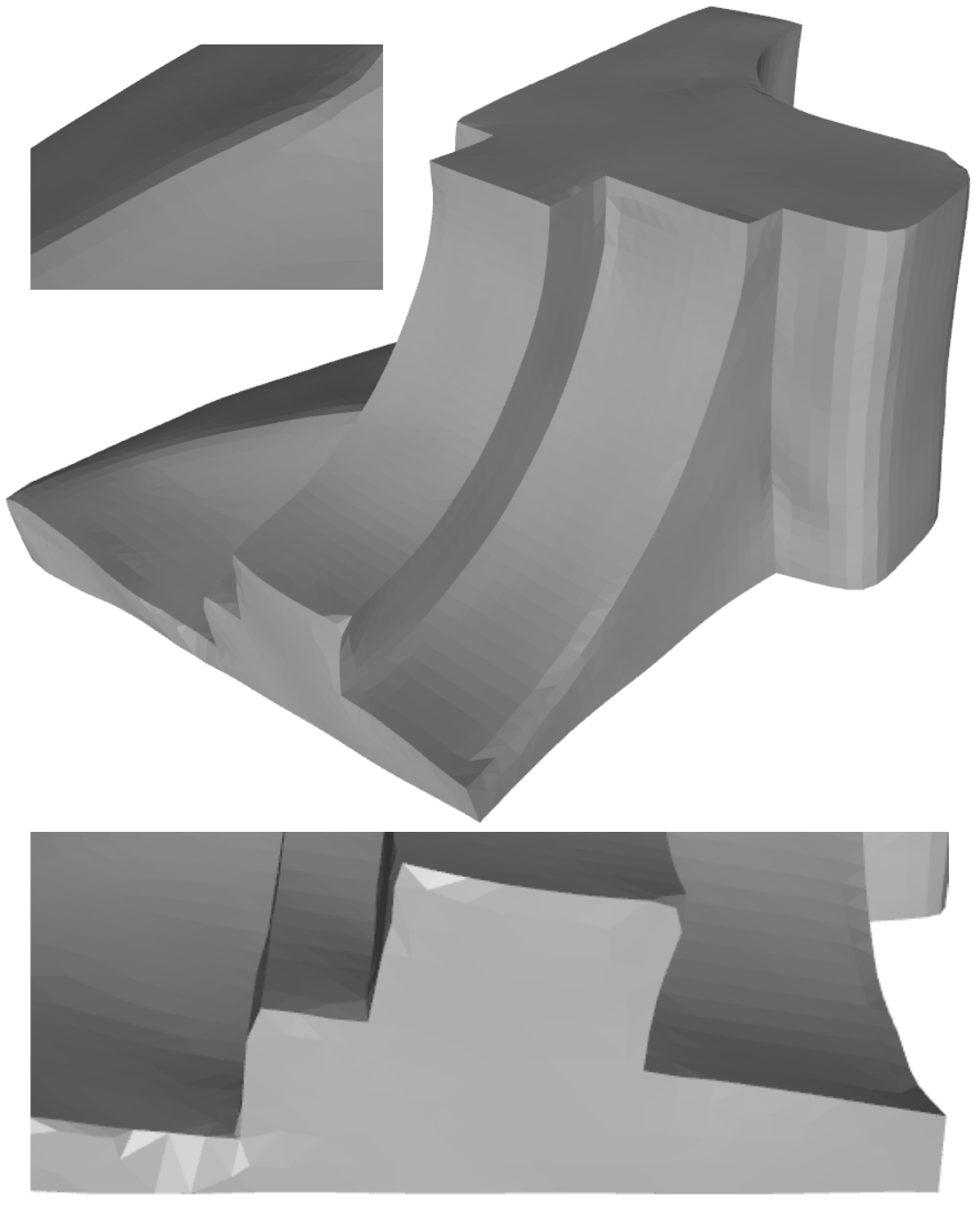}}
    \subfloat[Ours]{\label{fandisk-h}\includegraphics[width=0.125\textwidth]{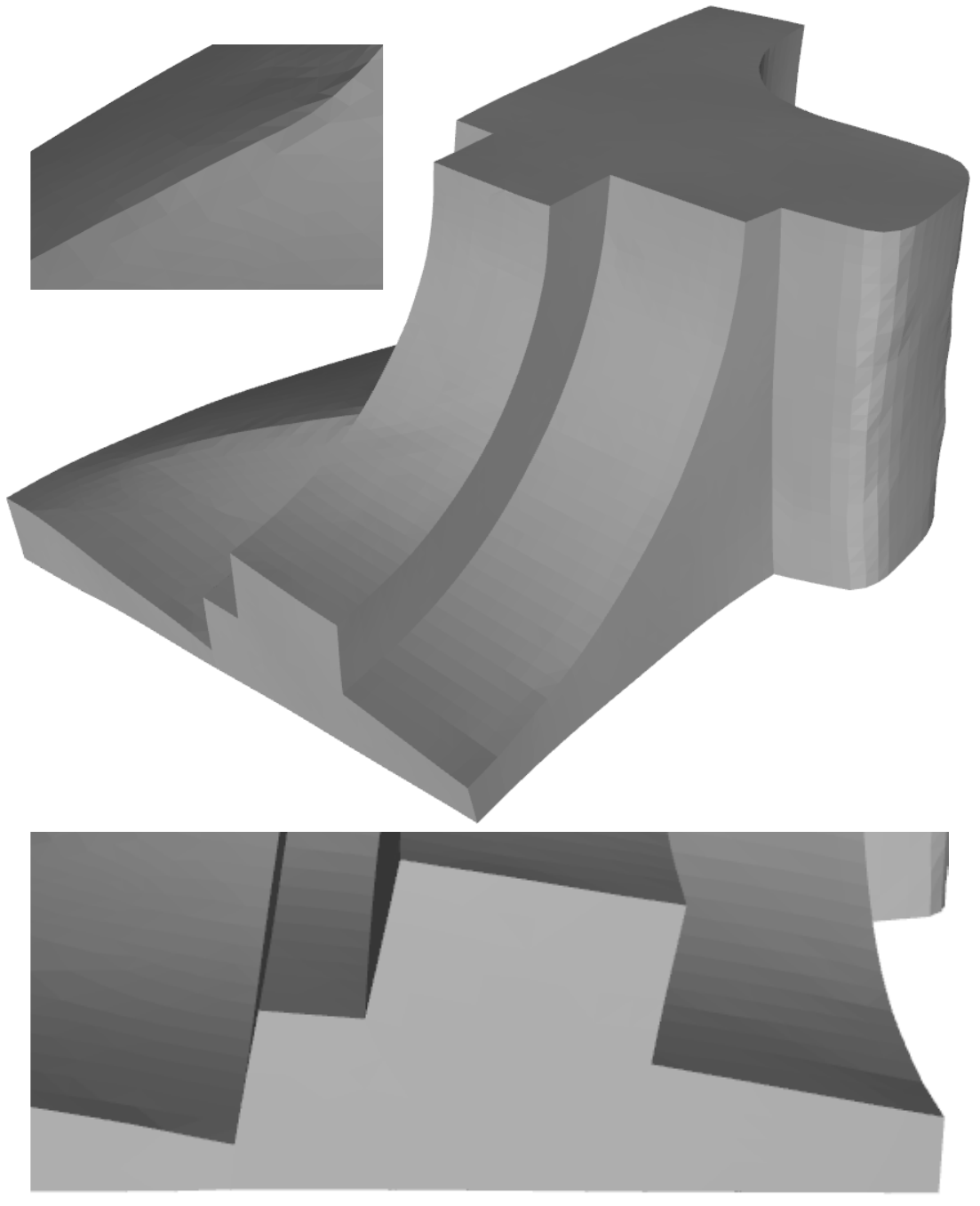}}
    \caption{Comparison of denoising results of Fandisk, corrupted with $\sigma=0.25 \bar{l}_e$.
    }
    \label{fig:fandisk}
\end{figure*}

\subsection{Qualitative Performance}
\textbf{Denoise CAD surfaces}. In Fig. \ref{fig:block}, we present the denoising results on a CAD surface containing both sharp features and smooth regions.
It can be seen that, except for BF and NLLR, all the other testing methods preserve sharp features to some extent.
As both geometric features and noise belong to high frequency information, BF and NLLR cannot distinguish them, especially for sharp features, and as a result, some features are treated as noise and get blurred; see Figs. \ref{block-e} and \ref{block-f}.
CNR, the learning-based method, performs well, however, it induces artifacts near sharp features; see Fig.  \ref{block-g}.
We observe that,  sparse optimization based methods, including TV, HO, TGV, and L0, preserve sharp features more accurately.
However, due to its higher sparsity requirement,  L0 flattens some smooth regions and induces false features in smooth regions sometimes, as Fig. \ref{block-d} shows.
In contrast, TGV is free from these artifacts in smooth regions,
which makes it an significant improvement over TV; see Figs. \ref{block-b} and \ref{block-h}.
Compared to HO, TGV recovers sharp features and flat regions more accurately; see Fig. \ref{block-c}.
For each testing method, we also visualize the error map for the normals, where the error is defined as the angular difference between the filtered normals and the ground truth.
The normals produced by our method are noticeably closer to the ground truth; see the second row of Fig. \ref{fig:block}.

In Fig. \ref{fig:fandisk}, we compare the results for a CAD surface including sharp features and shallow edge.
Again, BF and NLLR blur sharp features in varying degrees, while L0 flattens smooth regions and produces false features; see Figs. \ref{fandisk-e}, \ref{fandisk-f}, and \ref{fandisk-d}.
As TV applies only to the first-order information, it suffers from staircase artifacts in smoothly curved regions; see Fig. \ref{fandisk-b}.
HO recovers smooth regions more accurately than TV.
However, as HO only uses high-order information, it bends straight-line edges and blurs shallow features; see the zoomed-in view of Fig. \ref{fandisk-c}.
Hence TV and HO are effective in preserving either sharp features or smooth regions, but not both.
In contrast, TGV combines the advantages from both methods and accurately recovers both sharp features and smooth regions; see Fig. \ref{fandisk-h}.
Visual comparisons for this example show the superior performance of TGV
in simultaneously preserving features and recovering smooth regions.

\begin{figure*}[htb]
    \centering
    \subfloat[Noisy]{\label{lucy-a}\includegraphics[width=0.125\textwidth]{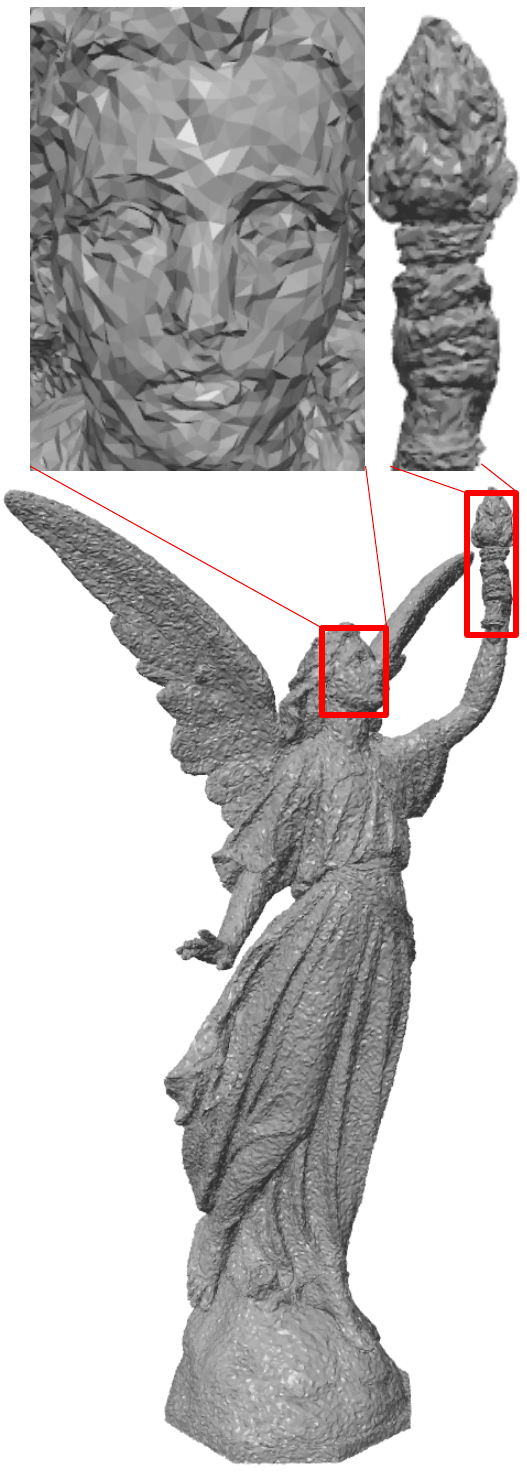}}
    \subfloat[TV]{\label{lucy-b}\includegraphics[width=0.125\textwidth]{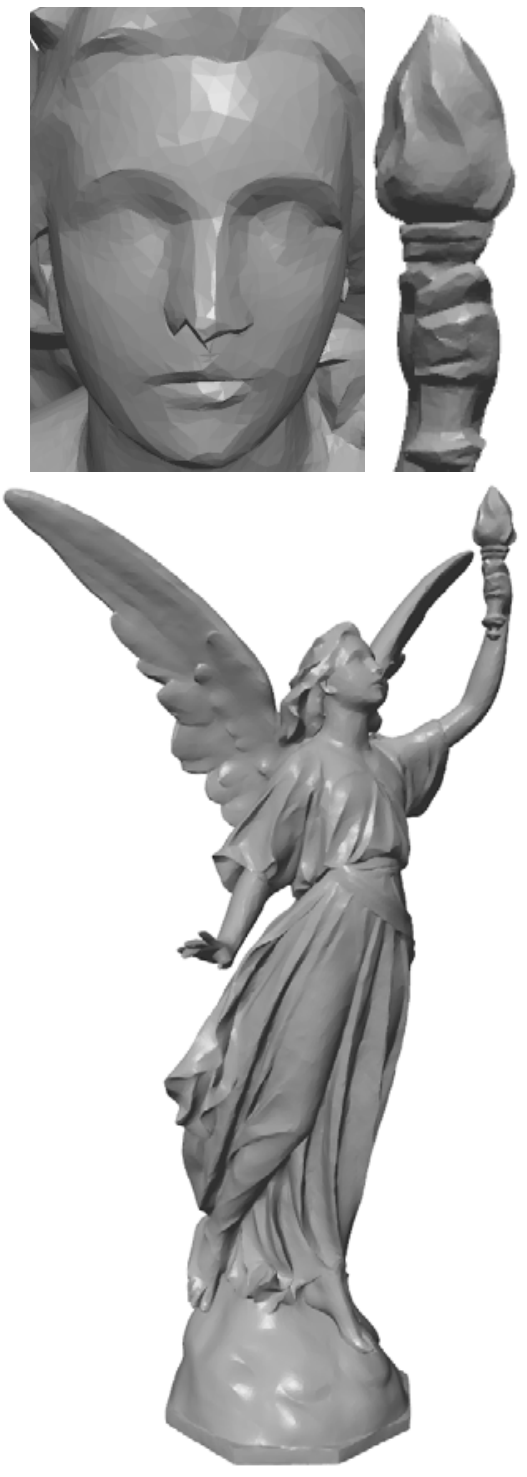}}
    \subfloat[HO]{\label{lucy-c}\includegraphics[width=0.125\textwidth]{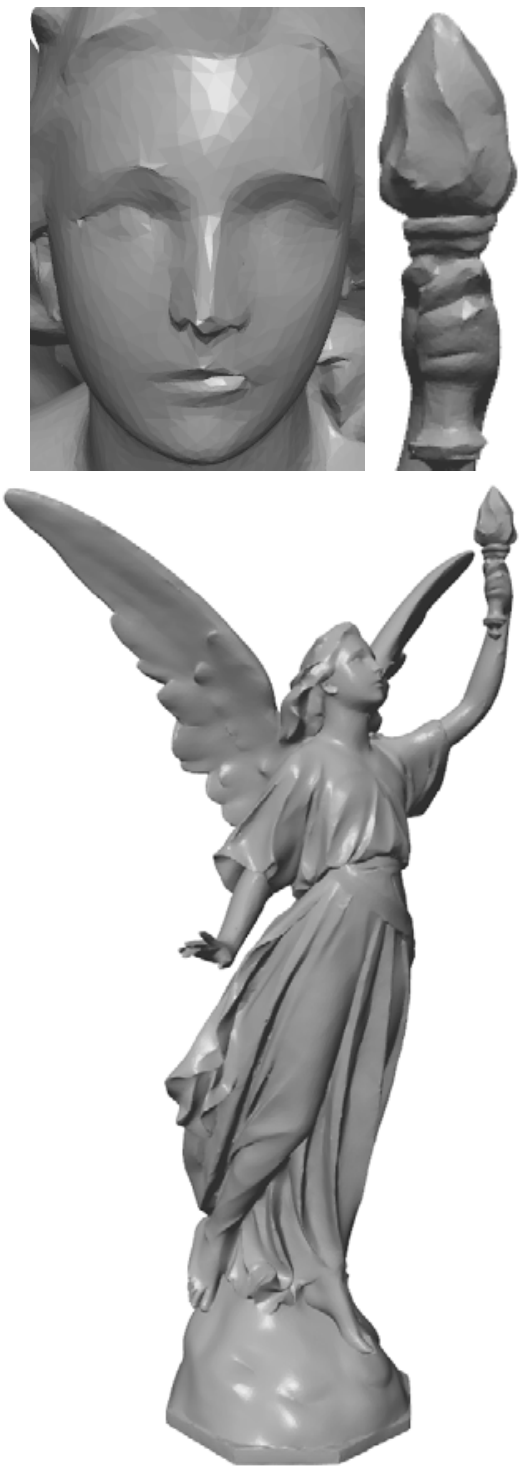}}
    \subfloat[L0]{\label{lucy-d}\includegraphics[width=0.125\textwidth]{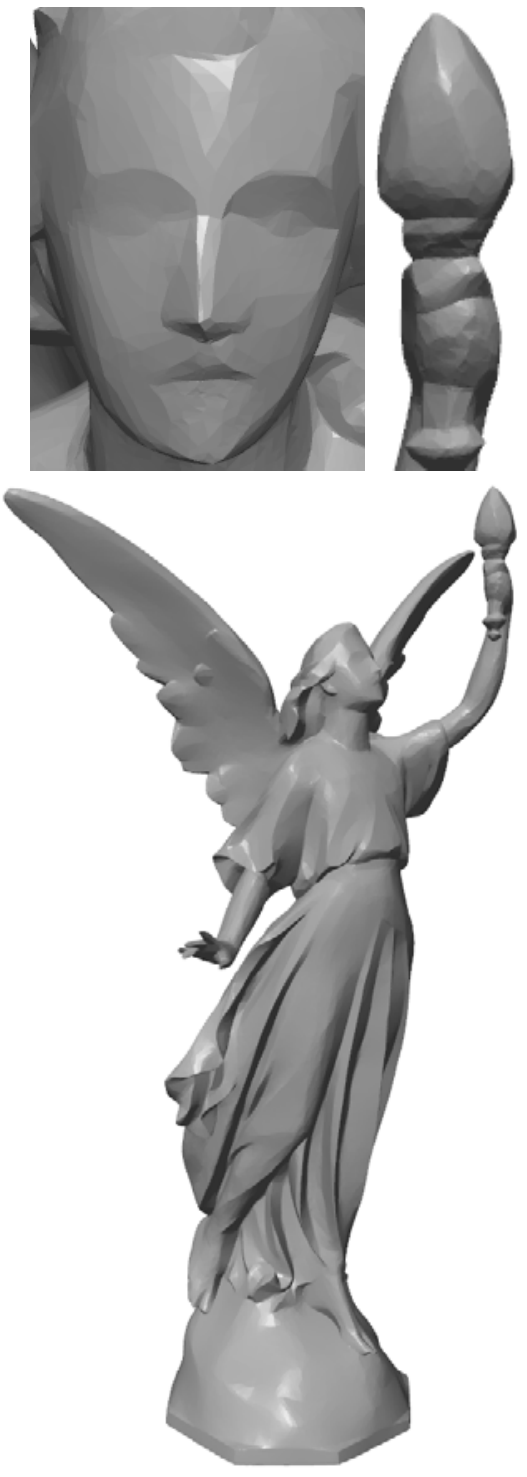}}
    \subfloat[BF]{\label{lucy-e}\includegraphics[width=0.125\textwidth]{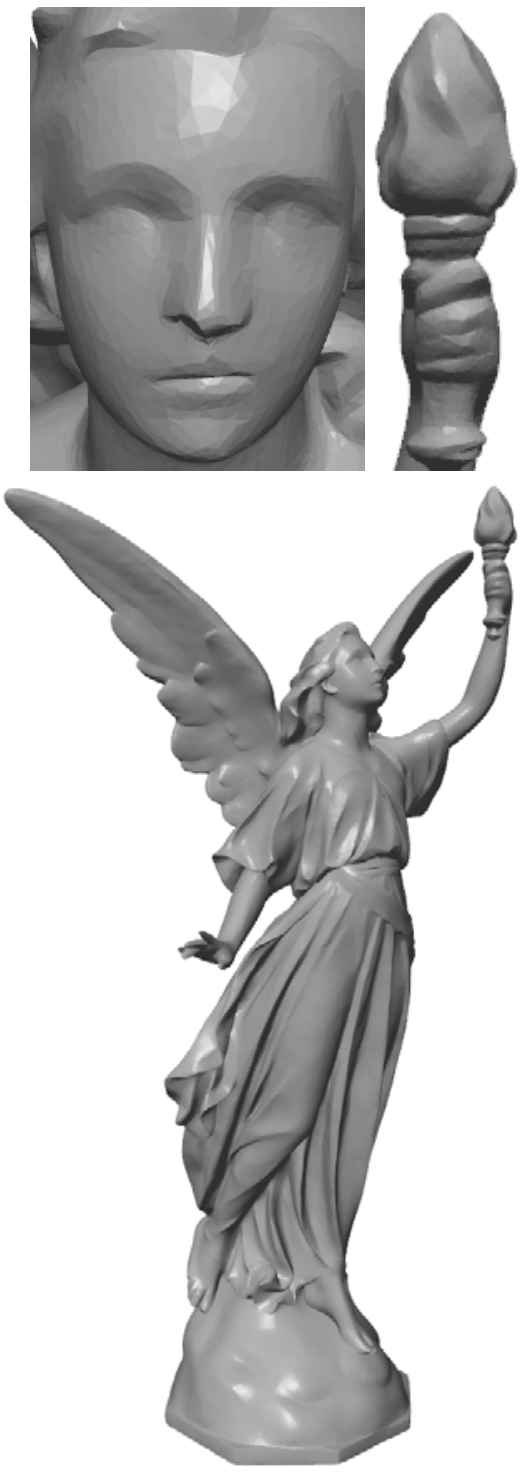}}
    \subfloat[NLLR]{\label{lucy-f}\includegraphics[width=0.125\textwidth]{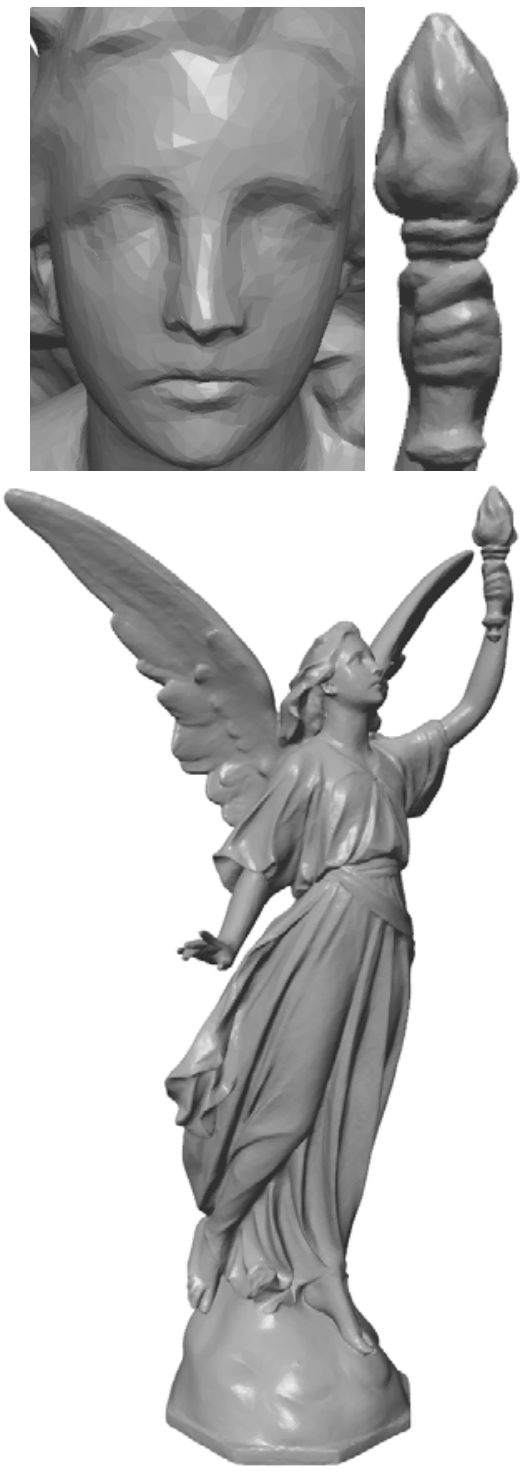}}
    \subfloat[CNR]{\label{lucy-g}\includegraphics[width=0.125\textwidth]{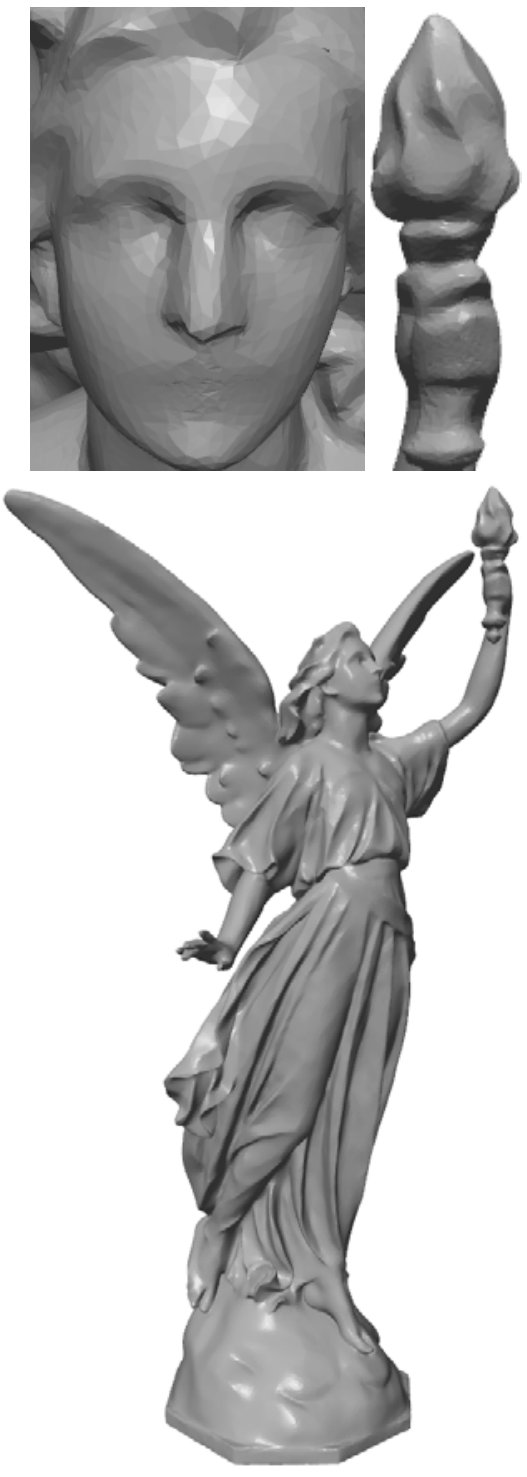}}
    \subfloat[Ours]{\label{lucy-h}\includegraphics[width=0.125\textwidth]{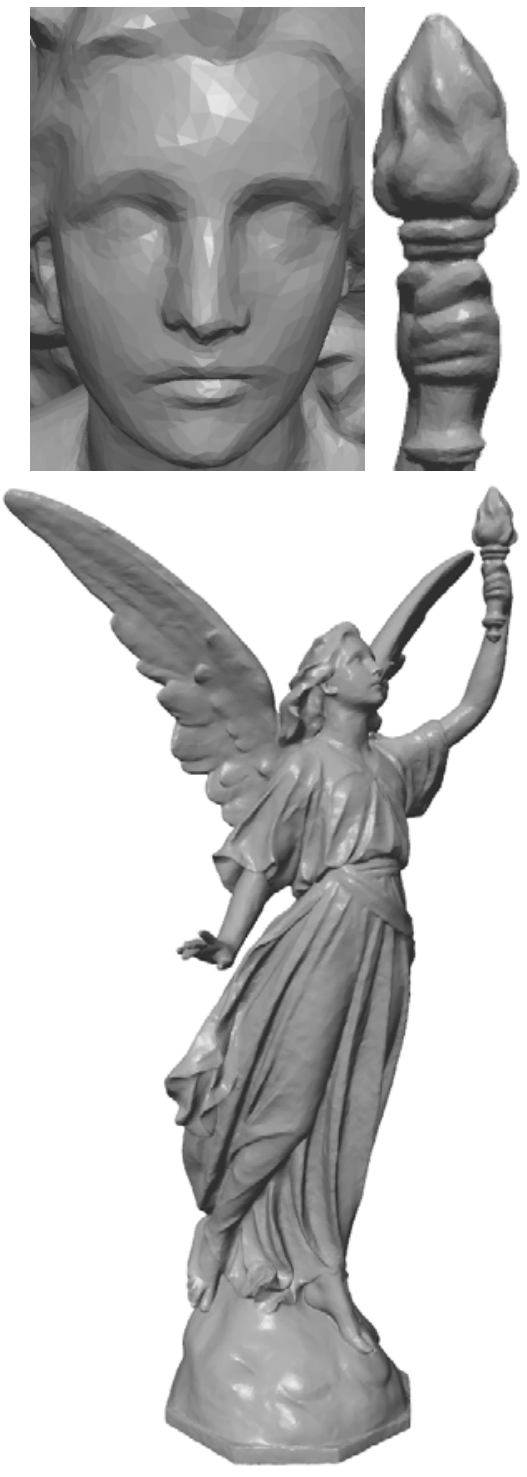}}
    \caption{Comparison of denoising results of Lucy, corrupted with $\sigma=0.2 \bar{l}_e$.
    }
    \label{fig:lucy}
\end{figure*}

\begin{figure*}[htb]
    \centering
    \subfloat[Noisy]{\label{gargoyle-a}\includegraphics[width=0.125\textwidth]{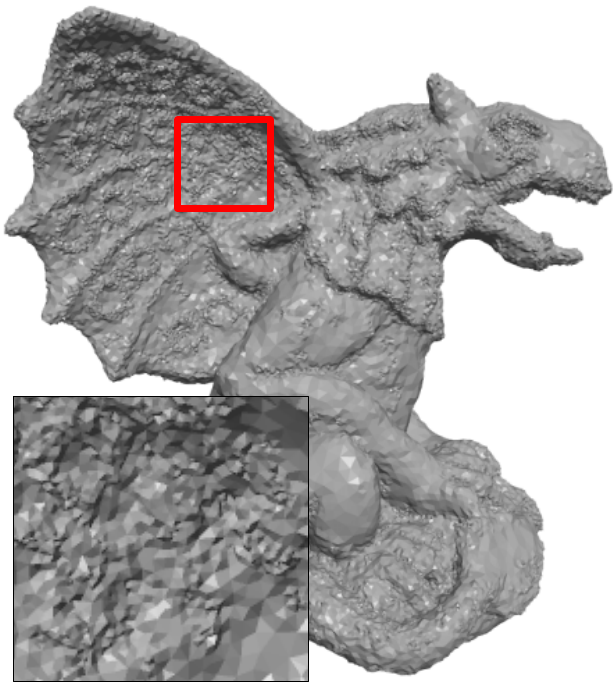}}
    \subfloat[TV]{\label{gargoyle-b}\includegraphics[width=0.125\textwidth]{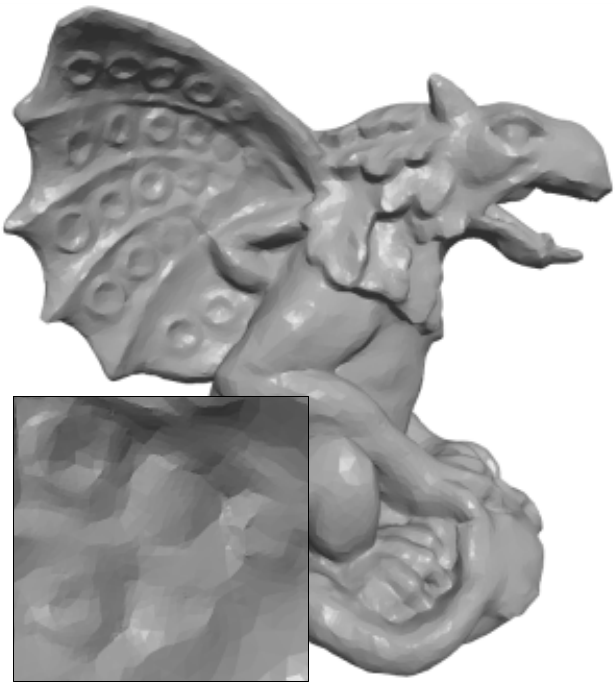}}
    \subfloat[HO]{\label{gargoyle-c}\includegraphics[width=0.125\textwidth]{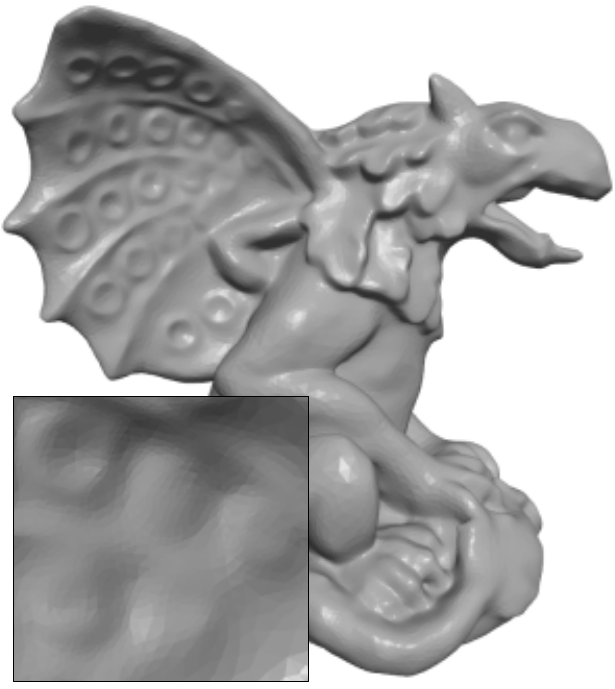}}
    \subfloat[L0]{\label{gargoyle-d}\includegraphics[width=0.125\textwidth]{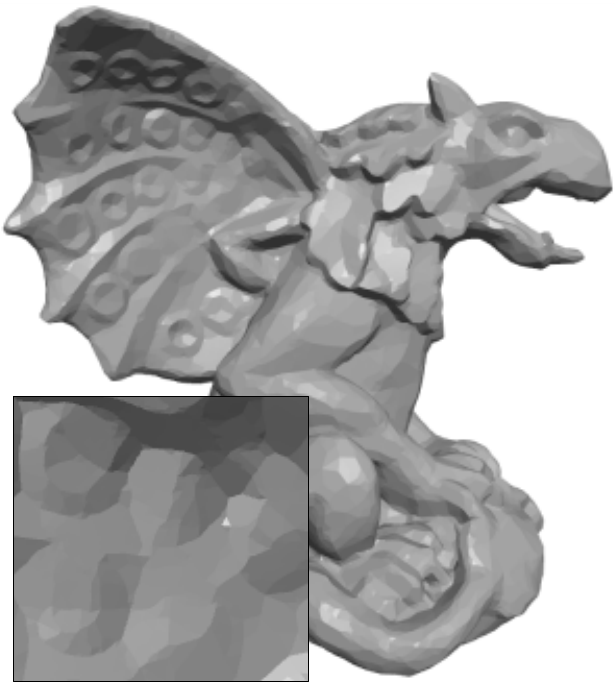}}
    \subfloat[BF]{\label{gargoyle-e}\includegraphics[width=0.125\textwidth]{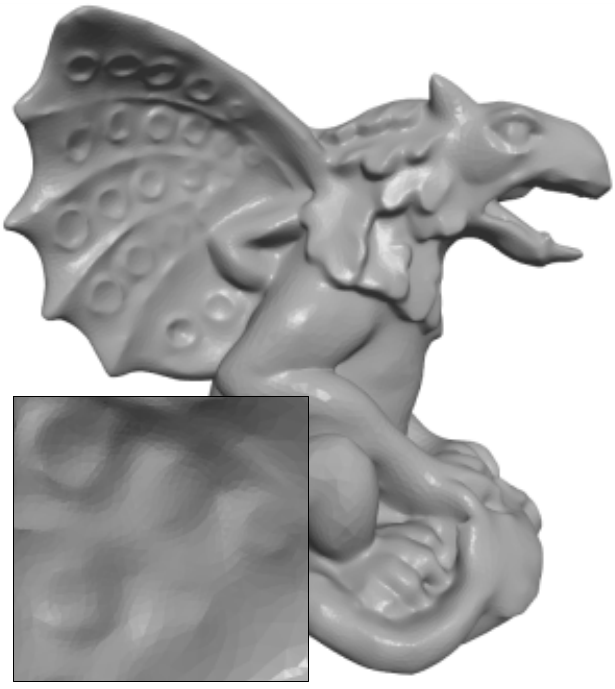}}
    \subfloat[NLLR]{\label{gargoyle-f}\includegraphics[width=0.125\textwidth]{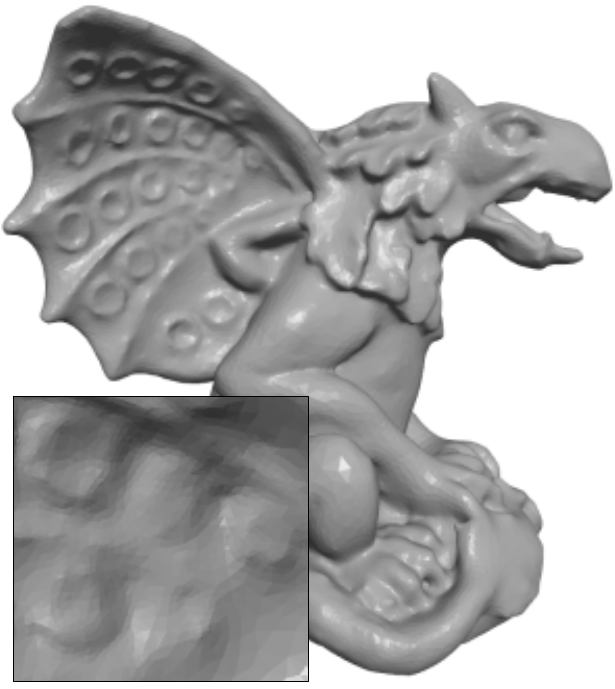}}
    \subfloat[CNR]{\label{gargoyle-g}\includegraphics[width=0.125\textwidth]{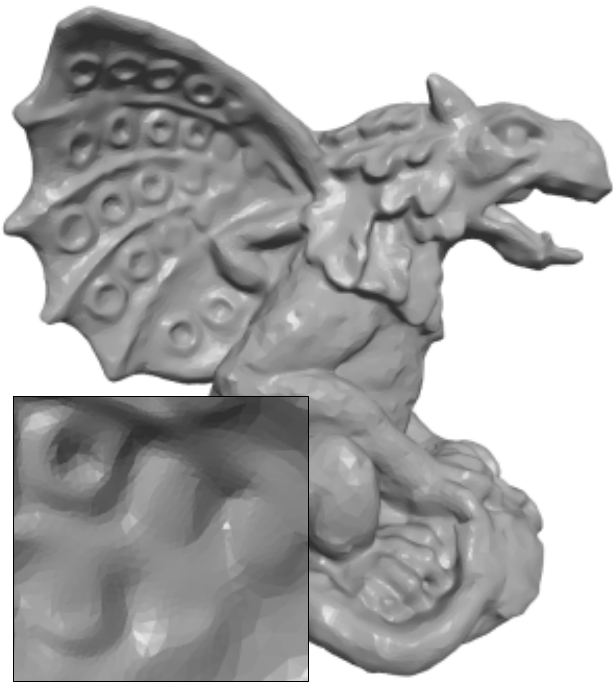}}
    \subfloat[Ours]{\label{gargoyle-h}\includegraphics[width=0.125\textwidth]{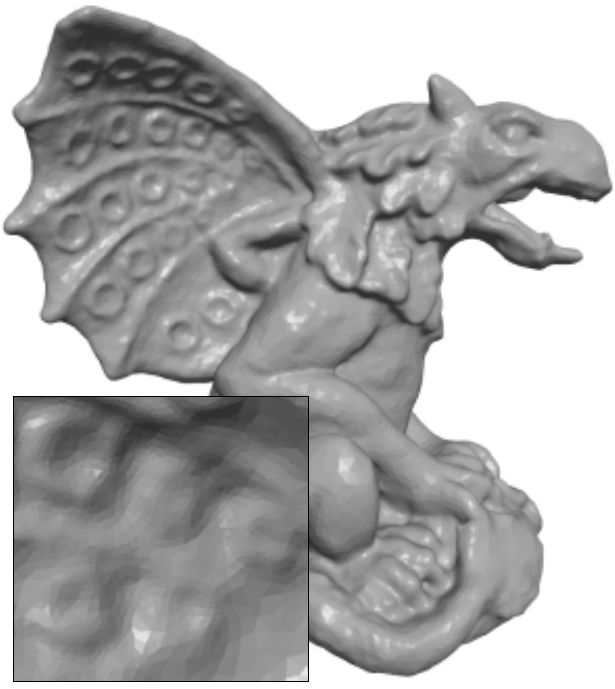}}
    \caption{Comparison of denoising results for Gargoyle, corrupted with $\sigma=0.25 \bar{l}_e$.
    }
    \label{fig:gargoyle}
\end{figure*}

\textbf{Denoise non-CAD surfaces}. Fig. \ref{fig:lucy} shows a comparison on a non-CAD surface with rich geometric features.
As expected, TV over-smoothes some small-scale features, and exhibits slight staircase artifacts.
L0 makes this situation even worse by transforming smooth regions into piecewise constant ones while over sharpening medium-scale features; see Fig. \ref{lucy-d}.
HO and CNR are effective in preserving medium-scale features; see the torch in Figs. \ref{lucy-c} and \ref{lucy-g}.
But they may smooth small-scale features and fine details; see the hand regions in Figs. \ref{lucy-c} and \ref{lucy-g}.
In contrast, NLLR and our method TGV produce visually compelling results. 
Furthermore, from Table \ref{tab:error}, we can see our method achieves higher numerical accuracy than NLLR.
Therefore, our method produces appealing results with geometric features recovered better than all competing methods.

Fig. \ref{fig:gargoyle} compares the results on a non-CAD surface containing multi-scale features.
As expected, comparing to the other methods, NLLR, CNR, and our method TGV recover different levels of features in a better manner.
NLLR may retain some extra noise in the result, while CNR slightly blurs small-scale features.
In contrast, TGV produces visually the best result with most geometric features well preserved.

Overall, for non-CAD meshes, our method TGV generates satisfactory results with features recovered better, and at the same time it prevents introducing additional artifacts (e.g., staircase artifacts, over-smoothing, over-sharpening effects, extra noise).

\begin{figure*}[]
    \centering
    \subfloat[Noisy]{\label{vase-a}\includegraphics[width=0.125\textwidth]{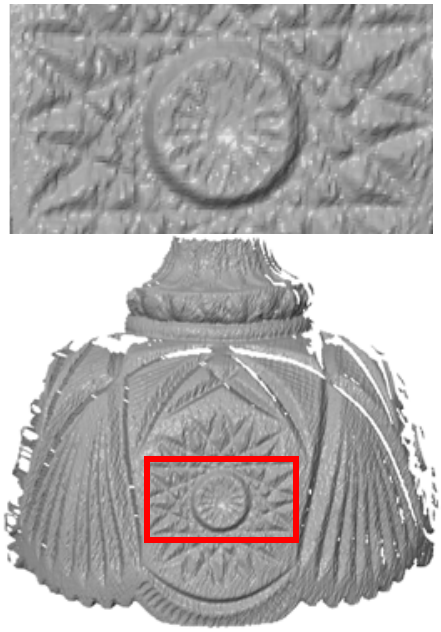}}
    \subfloat[TV]{\label{vase-b}\includegraphics[width=0.125\textwidth]{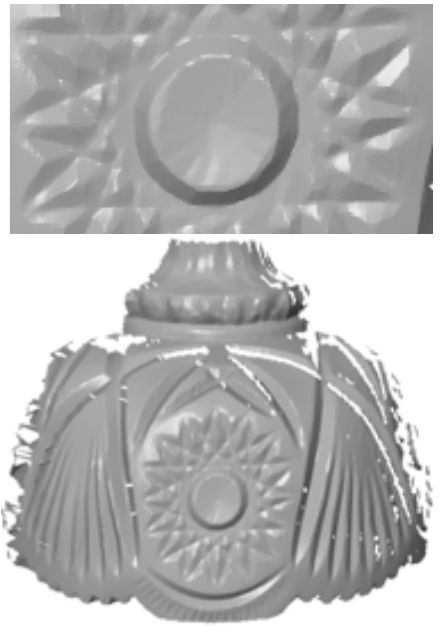}}
    \subfloat[HO]{\label{vase-c}\includegraphics[width=0.125\textwidth]{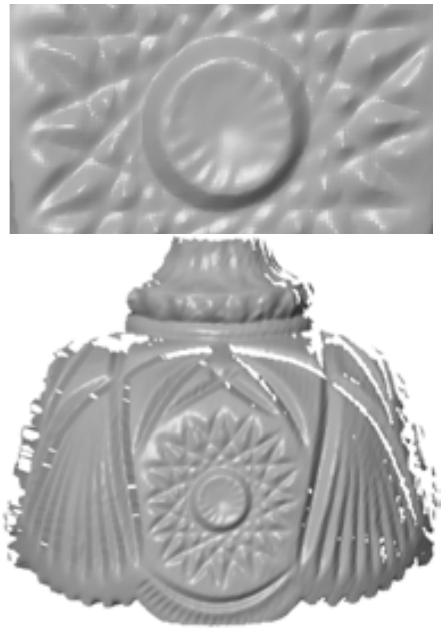}}
    \subfloat[L0]{\label{vase-d}\includegraphics[width=0.125\textwidth]{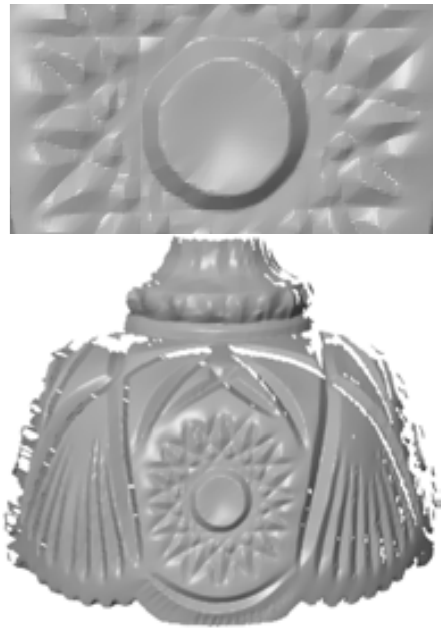}}
    \subfloat[BF]{\label{vase-e}\includegraphics[width=0.125\textwidth]{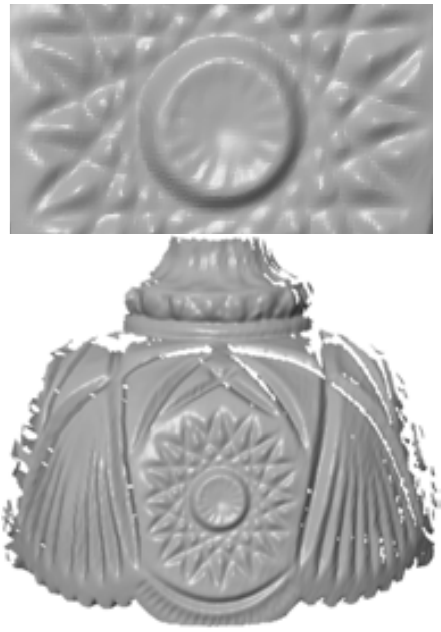}}
    \subfloat[NLLR]{\label{vase-f}\includegraphics[width=0.125\textwidth]{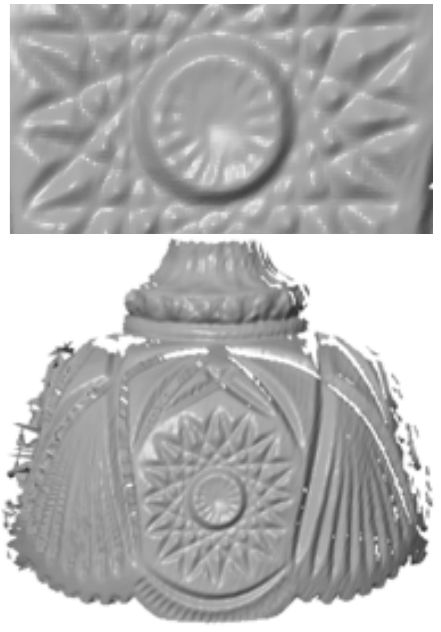}}
    \subfloat[CNR]{\label{vase-g}\includegraphics[width=0.125\textwidth]{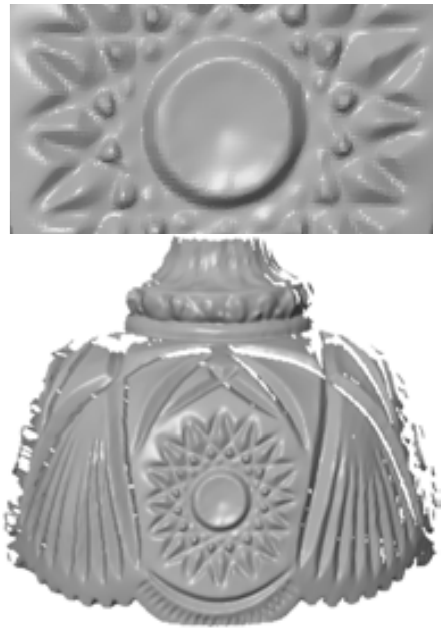}}
    \subfloat[Ours]{\label{vase-h}\includegraphics[width=0.125\textwidth]{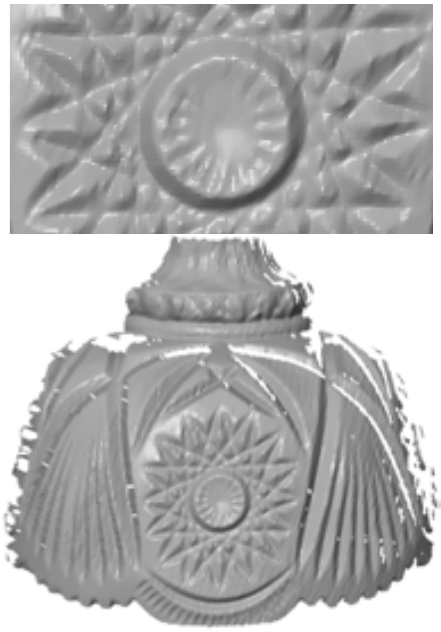}}
    \caption{Comparison of denoising results for scanned data acquired by a laser scanner. }
    \label{fig:vase}
\end{figure*}

\begin{figure*}[htp]
    \centering
    \subfloat[Noisy]{\label{kinect-a}\includegraphics[width=0.12\textwidth]{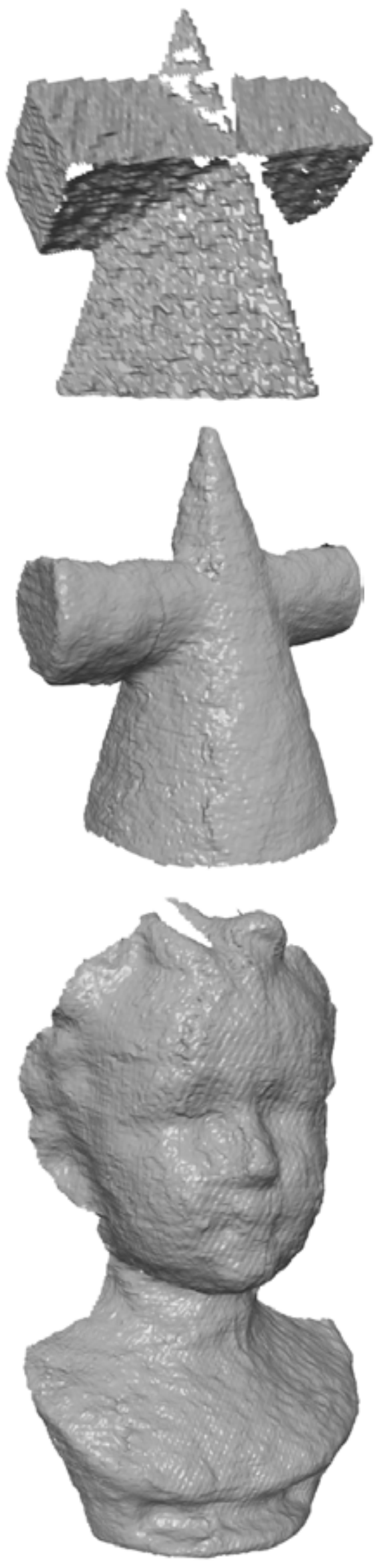}}
    \subfloat[TV]{\label{kinect-b}\includegraphics[width=0.12\textwidth]{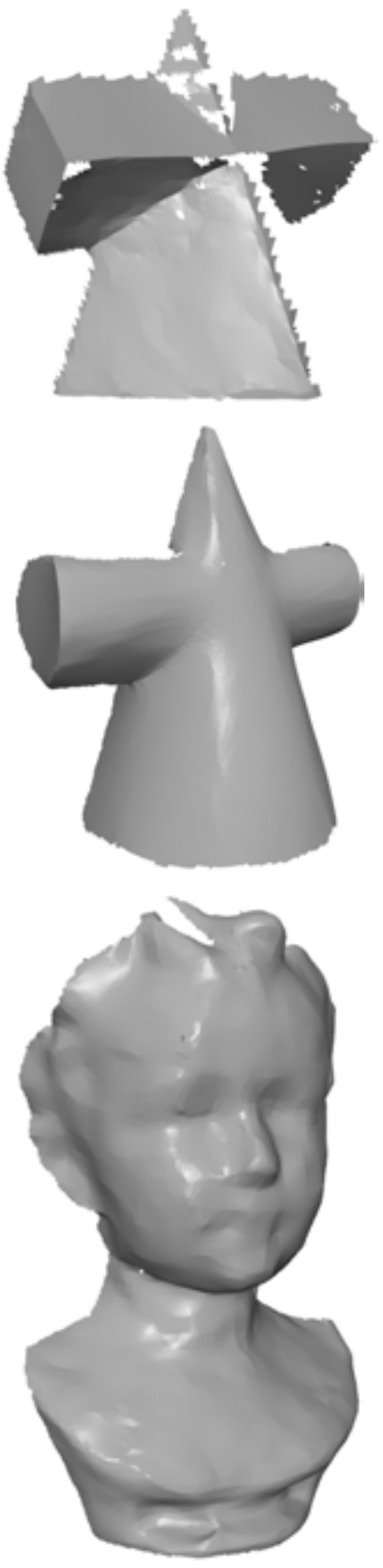}}
    \subfloat[HO]{\label{kinect-c}\includegraphics[width=0.12\textwidth]{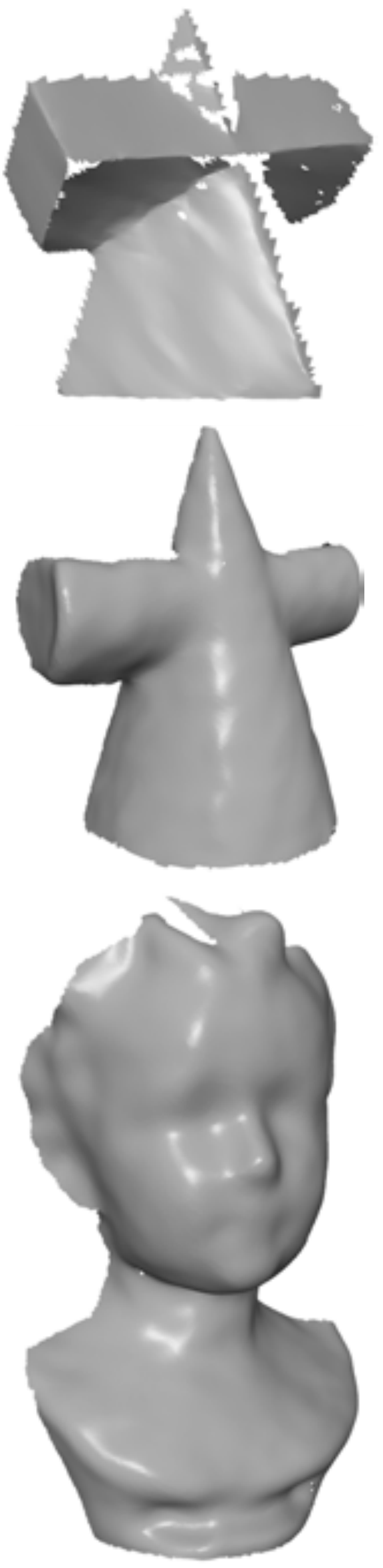}}
    \subfloat[L0]{\label{kinect-d}\includegraphics[width=0.12\textwidth]{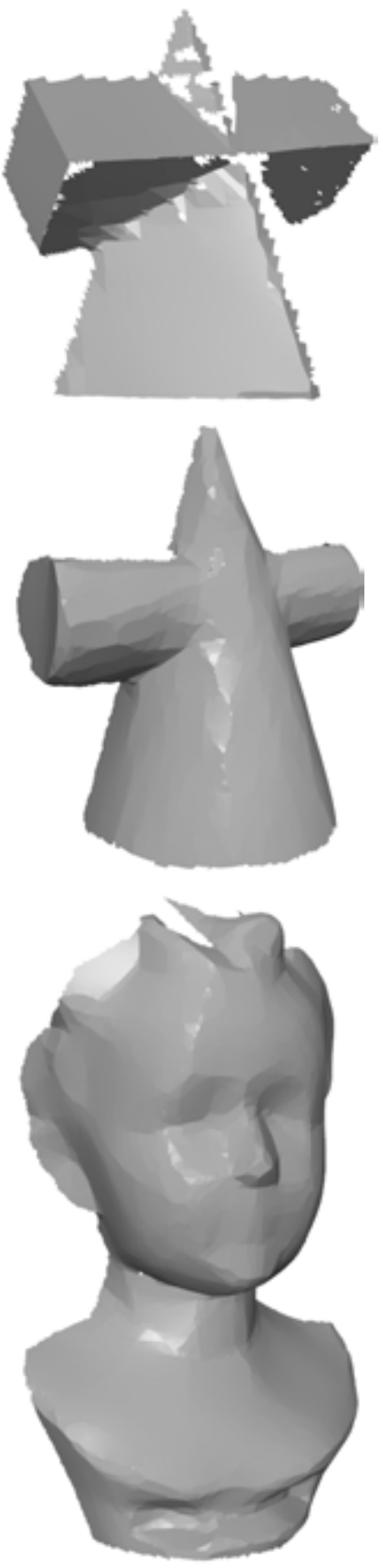}}
    \subfloat[BF]{\label{kinect-e}\includegraphics[width=0.12\textwidth]{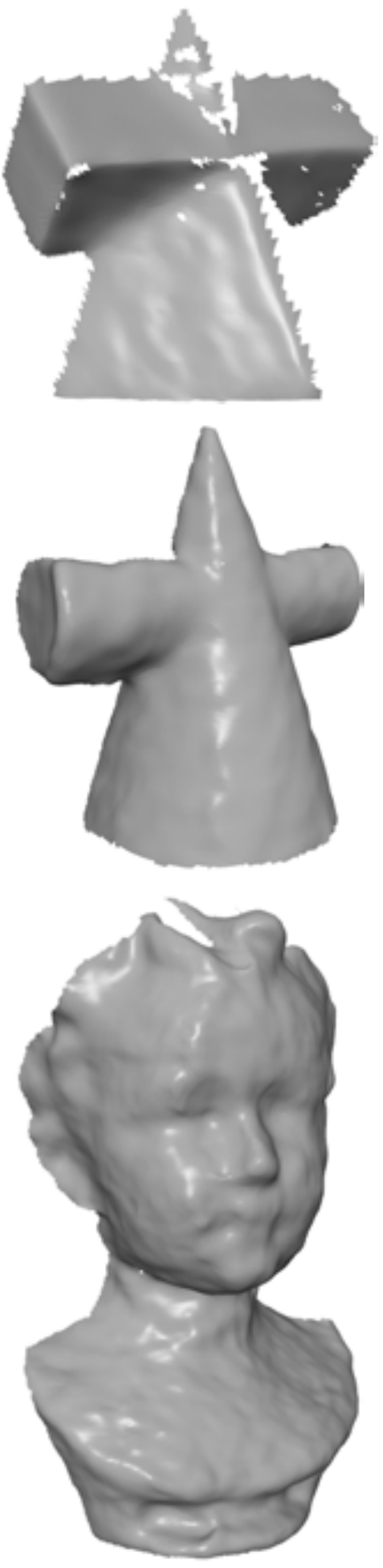}}
    \subfloat[NLLR]{\label{kinect-f}\includegraphics[width=0.12\textwidth]{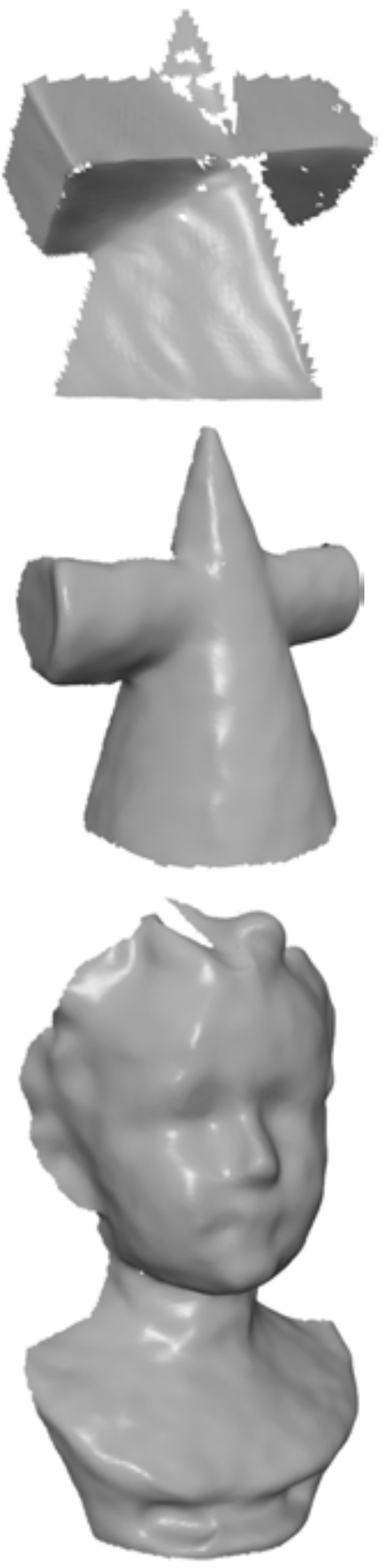}}
    \subfloat[CNR]{\label{kinect-g}\includegraphics[width=0.12\textwidth]{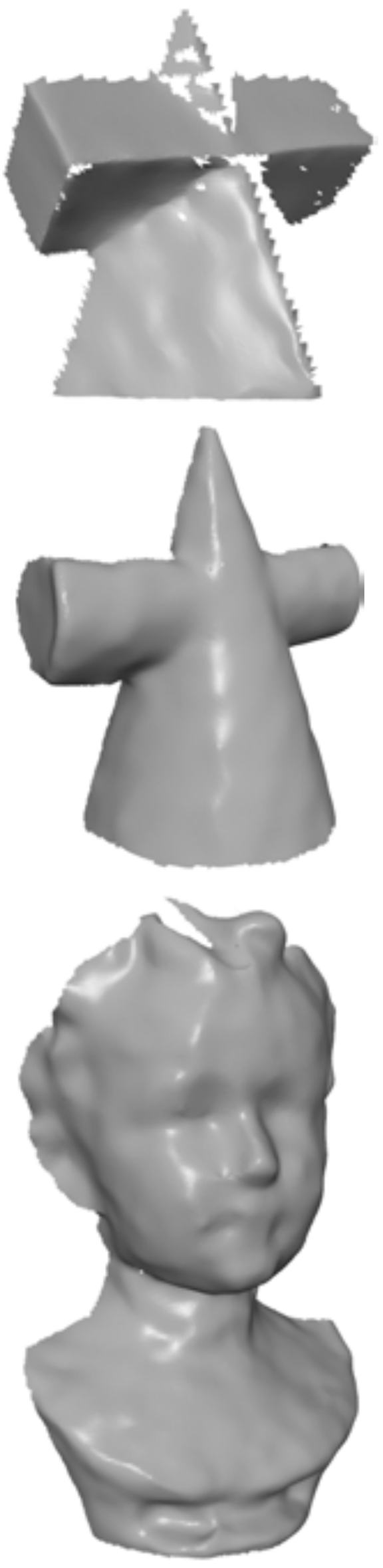}}
    \subfloat[Ours]{\label{kinect-h}\includegraphics[width=0.12\textwidth]{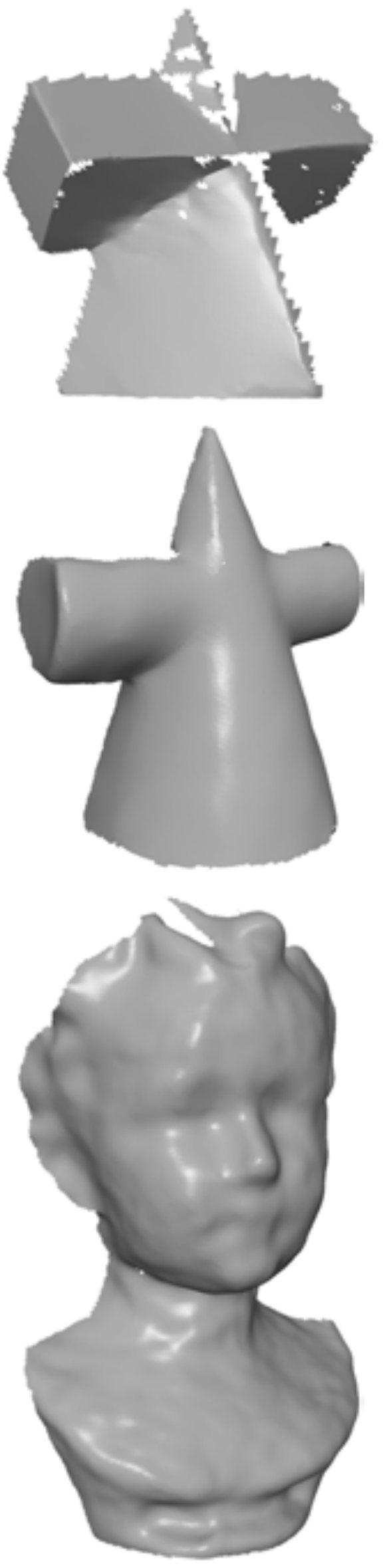}}
    \caption{Comparison of denoising results for scanned data acquired by Kinect.}
    \label{fig:kinectData}
\end{figure*}

\textbf{Denoise scanning data}.  We also compare the different methods on scanned data, where the noise pattern is unknown.
Fig. \ref{fig:vase} shows the results for data acquired by a laser scanner.
First, TV, L0, and CNR over-smooth fine details while sharpen some features, which makes the results look less natural (see the zoomed-in views of Figs. \ref{vase-b}, \ref{vase-d}, and \ref{vase-g}).
In contrast, HO and BF blur details to varying degrees.
In this example, the results produced by NLLR and our method TGV look more natural and compelling than those produced by the other methods; see Figs. \ref{vase-f} and \ref{vase-h}.
Our result is free of visible artifacts and almost does not lose any features from the underlying surface.

In Fig. \ref{fig:kinectData}, we examine the performance of our method on meshes acquired by the Kinect sensor.
These scanned meshes are provided by Wang et al. \cite{Wang2016Mesh}.
As can be seen, all the competing methods remove noise effectively, except for BF which cannot distinguish features and noise clearly.
TV and L0 produce staircase artifacts in smooth regions and sharpen curved features; see Figs. \ref{kinect-b} and \ref{kinect-d}. This phenomenon is more severe for L0.
Although HO does a good job in smooth regions, it slightly blurs small-scale features.
NLLR and CNR yield visually excellent results, although they induce some small bumps in smooth regions.
In contrast, our method outperforms the other methods in preserving features and recovering smooth regions, while preventing visible artifacts.

Overall, in all the meshes being tested, our results present visually cleaner geometric features without noticeable artifacts.
Hence, our method has succeeded in simultaneously recovering smoothly curved regions and preserving sharp features, even in the presence of heavy noise.

\begin{figure*}[htp]
    \centering
    \includegraphics[width=0.95\textwidth]{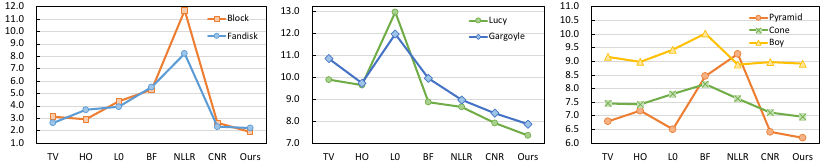}
    \caption{Error plots of mean angular difference ($\theta$) of the results in Figs. \ref{fig:block}, \ref{fig:fandisk}, \ref{fig:lucy}, \ref{fig:gargoyle}, and \ref{fig:kinectData}, for all competing methods.}
    \label{fig:thetaCurves}
\end{figure*}

\begin{table*}[htp]
    \centering    \footnotesize
    \caption{Quantitative evaluation of the results in Figs. \ref{fig:block}, \ref{fig:fandisk}, \ref{fig:lucy}, \ref{fig:gargoyle}, and \ref{fig:kinectData}. For each result, we list mean angular difference $\theta$ (in degrees), vertex-based mesh-to-mesh error $E_v \ (\times 10^{-2})$ and the execution time (in seconds).}

    \begin{tabular}{lccccccccccccccccccccc}
    \toprule

    \footnotesize Mesh &
    \multicolumn{3}{c}{\footnotesize TV}&
    \multicolumn{3}{c}{\footnotesize HO}&
    \multicolumn{3}{c}{\footnotesize L0}&
    \multicolumn{3}{c}{\footnotesize BF}&
    \multicolumn{3}{c}{\footnotesize NLLR}&
    \multicolumn{3}{c}{\footnotesize CNR}&
    \multicolumn{3}{c}{\footnotesize TGV}
    \\
    \midrule
    \footnotesize Block &
    \multicolumn{3}{c|}{3.12, 2.01; 1.44} &
    \multicolumn{3}{c|}{2.90, 1.70; 2.03} &
    \multicolumn{3}{c|}{4.35, 2.15; 13.3} &
    \multicolumn{3}{c|}{5.30, 1.80; 1.07} &
    \multicolumn{3}{c|}{11.7, 2.44; 10.4} &
    \multicolumn{3}{c|}{2.63, \textbf{0.86}; 0.71} &
    \multicolumn{3}{c}{\textbf{1.88}, 0.95; 6.28}\\

    \footnotesize Fandisk &
    \multicolumn{3}{c|}{2.62, 2.39; 0.88} &
    \multicolumn{3}{c|}{3.67, 1.67; 1.88} &
    \multicolumn{3}{c|}{3.92, 2.17; 6.73} &
    \multicolumn{3}{c|}{5.51, 1.62; 0.61} &
    \multicolumn{3}{c|}{8.21, 1.57; 3.64} &
    \multicolumn{3}{c|}{2.31, 1.47; 0.55} &
    \multicolumn{3}{c}{\textbf{2.20}, \textbf{1.20}; 3.08}\\

    \footnotesize Lucy &
    \multicolumn{3}{c|}{9.88, 0.38; 30.9} &
    \multicolumn{3}{c|}{9.63, 0.61; 60.1} &
    \multicolumn{3}{c|}{13.0, 0.72; 87.6} &
    \multicolumn{3}{c|}{8.86, 0.50; 8.80} &
    \multicolumn{3}{c|}{8.64, 0.36; 67.5} &
    \multicolumn{3}{c|}{7.91, 0.30; 10.5} &
    \multicolumn{3}{c}{\textbf{7.34}, \textbf{0.26}; 66.6}\\

    \footnotesize Gargoyle &
    \multicolumn{3}{c|}{10.8, 0.59; 13.7} &
    \multicolumn{3}{c|}{9.72, 0.67; 20.3} &
    \multicolumn{3}{c|}{12.0, 0.63; 55.5} &
    \multicolumn{3}{c|}{9.94, 2.22; 4.70} &
    \multicolumn{3}{c|}{8.96, 1.58; 30.5} &
    \multicolumn{3}{c|}{8.36, 0.77; 6.51} &
    \multicolumn{3}{c}{\textbf{7.86}, \textbf{0.54}; 49.8}\\

    \footnotesize Pyramid &
    \multicolumn{3}{c|}{6.79, 4.69; 1.28} &
    \multicolumn{3}{c|}{7.18, 4.23; 1.93} &
    \multicolumn{3}{c|}{6.50, 3.47; 8.03} &
    \multicolumn{3}{c|}{8.45, 4.45; 0.77} &
    \multicolumn{3}{c|}{9.27, 3.41; 104.2} &
    \multicolumn{3}{c|}{6.40, \textbf{3.38}; 0.76} &
    \multicolumn{3}{c}{\textbf{6.19}, 4.52; 4.14}\\

    \footnotesize Cone &
    \multicolumn{3}{c|}{7.45, 3.98; 4.48} &
    \multicolumn{3}{c|}{7.41, 3.33; 7.91} &
    \multicolumn{3}{c|}{7.80, 3.45; 47.7} &
    \multicolumn{3}{c|}{8.16, 2.96; 3.59} &
    \multicolumn{3}{c|}{7.63, 3.04; 312.8} &
    \multicolumn{3}{c|}{7.11, \textbf{2.78}; 1.82} &
    \multicolumn{3}{c}{\textbf{6.96}, 3.57; 30.1}\\

    \footnotesize Boy &
    \multicolumn{3}{c|}{9.16, 6.08; 10.2} &
    \multicolumn{3}{c|}{8.98, \textbf{5.66}; 17.3} &
    \multicolumn{3}{c|}{9.42, 6.04; 88.5} &
    \multicolumn{3}{c|}{10.0, 6.19; 12.3} &
    \multicolumn{3}{c|}{\textbf{8.88}, 5.97; 409.1} &
    \multicolumn{3}{c|}{8.97, 6.07; 4.93} &
    \multicolumn{3}{c}{8.91, 5.94; 55.8}\\

\bottomrule
\end{tabular}
\label{tab:error}
\end{table*}

\begin{table}[htp]\footnotesize
    \centering
    \caption{Mesh sizes for the surfaces in Table \ref{tab:error}}
    \begin{tabular}{ccccccccc}
    \toprule
      \!\!\!\!Mesh\!\!\! & \!\!\! Block\!\!\!  & \!\!\! Fandisk\!\!\!  & \!\!\! Lucy\!\!\!  & \!\!\! Gargoyle\!\!\!  & \!\!\! Pyramid\!\!\!  & \!\!\! Cone\!\!\!  & \!\!\! Boy\!\!\! \\
    \midrule
      \!\!\!\!$|V|$\!\!\! & \!\! 8.8K\!\!  & \!\! 6.5K\!\!  & \!\! 149.3K\!\!  & \!\! 85.6K\!\!  & \!\! 6.6K\!\!  & \!\! 31.2K\!\!  & \!\! 76.9K\!\!\!  \\
      \!\!\!\!$|F|$\!\!\! & \!\! 17.6K\!\!  & \!\! 12.9K\!\!  & \!\! 298.5K\!\!  & \!\! 171.1K\!\!  & \!\! 12.6K\!\!  & \!\! 61.3K\!\!  & \!\! 152.2K\!\!\! \\
    \bottomrule
\end{tabular} \label{tab:meshSize}
\end{table}

\subsection{Quantitative Evaluation}
To quantitatively evaluate the quality of the denoising results, we adopt the mean angular difference, abbreviated as $\theta$, as an error metric.
This error metric is widely used in recent work \cite{Wang2016Mesh}, \cite{Li2018NonLocal}, \cite{Li2020DNF}.
It measures the mean angular difference ($\theta$) of normals between the clean mesh and the denoised result.
For fair comparison, we compute $\theta$ after the filtering step for each testing method (except L0).
Table \ref{tab:error} lists the error metric $\theta$  for all competing methods, and Fig. \ref{fig:thetaCurves} further shows $\theta$ as polyline plots.
The mesh sizes of the tested surfaces in Table \ref{tab:error} are listed in Table \ref{tab:meshSize}.
As we can see from Table \ref{tab:error}, our method TGV produces highly competitive results.
More specifically, for CAD surfaces, our method outperforms the other methods in comparison in the sense that the $\theta$ values are significantly smaller; see the first column of Fig. \ref{fig:thetaCurves}.
That is consistent with the visual comparisons in Figs. \ref{fig:block} and \ref{fig:fandisk}.
For non-CAD surfaces, it is not surprising that our method gives $\theta$ values lower than the competing methods, indicating that our results are more faithful to the ground truth. Visually, results from our method and NLLR look almost identical in Fig. \ref{fig:lucy}.
For scanned data, NLLR exhibits slightly better performance than our method in the Boy example, even though the results look almost identical.
However, in the other two examples (Cone and Pyramid), our method shows better performance in terms of $\theta$ values; see the third column of Fig. \ref{fig:thetaCurves}.

To further evaluate the vertex deviation from the ground truth, we use the vertex-based Hausdorff distance \cite{Zhang15Variational,Wei2018Mesh} to measure the position error between the denoised mesh and the ground truth.
The results are listed in Table. \ref{tab:error}.
As can be seen, for CAD and non-CAD surfaces, our method TGV outperforms the other methods in most cases.
For scanned data, CNR shows better performance.

Overall, the quantitative comparisons show that our method is more effective in recovering shape details, including sharp features, multi-scale features, and smooth regions, from the noisy input, leading to the least amount of error in most cases, in comparison to the other methods.
It is worth noting that our method performs favorably on all types of the tested meshes (CAD, non-CAD, and scanned meshes) rather than having a peak performance on specific types.

\textbf{Computational time}. We list the execution time of each method in Table \ref{tab:error}.
As we can see, CNR is the fastest method, thanks to its pre-trained neural networks.
BF is slower than CNR, but significantly faster than the other methods.
L0 is the slowest for synthetic meshes, while NLLR is the slowest for scanned data.
Our method takes more execution time than TV and HO.
Furthermore, we adopt the conjugate gradient (CG) method to iteratively solve our two linear systems (one for \textbf{N}-subproblem and the other for \textbf{v}-subproblem), and found that
the runtime can be reduced by decreasing the number of iterations, trading off accuracy.
Furthermore, since the coefficient matrices of the two linear systems stay fixed during the iteration, they can be pre-factorized and thus the runtime of the full algorithm is still acceptable.

Overall, our method produces much better results in terms of visual quality and error metrics in most cases, although it seems to be computationally more intensive, hence using modern GPUs and multi-core CPUs to speed up our method is one further direction.

\begin{figure}[htb]
    \centering
    \subfloat[Noisy]{\label{fig:vsMS-HLO-a}\includegraphics[width=0.1\textwidth]{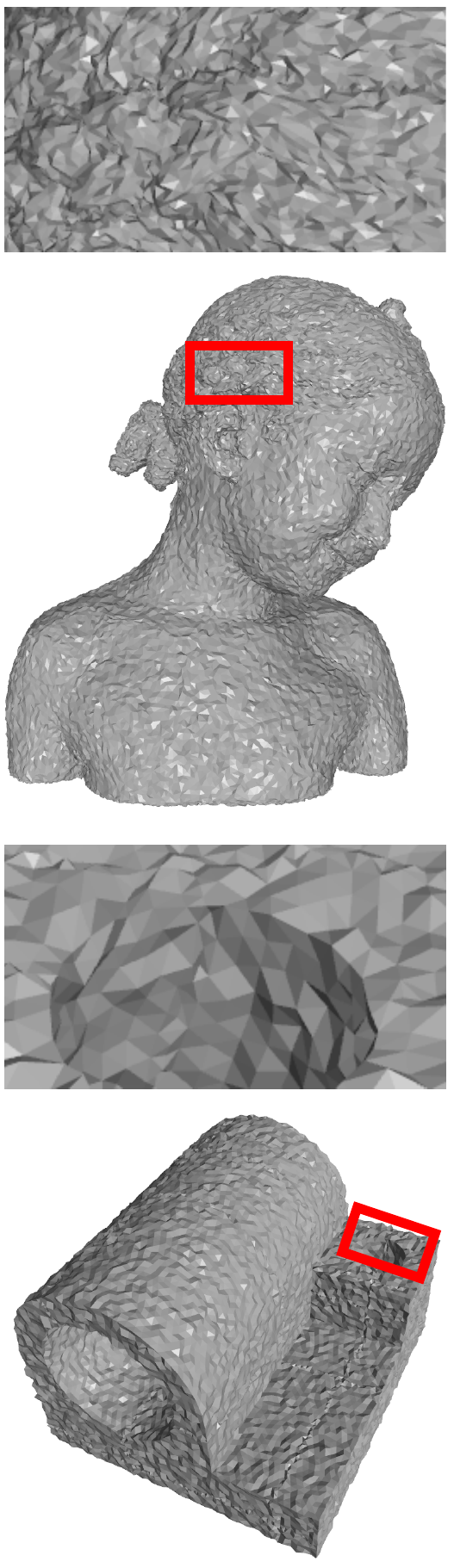}}
    \subfloat[AT]{\label{fig:vsMS-HLO-b}\includegraphics[width=0.1\textwidth]{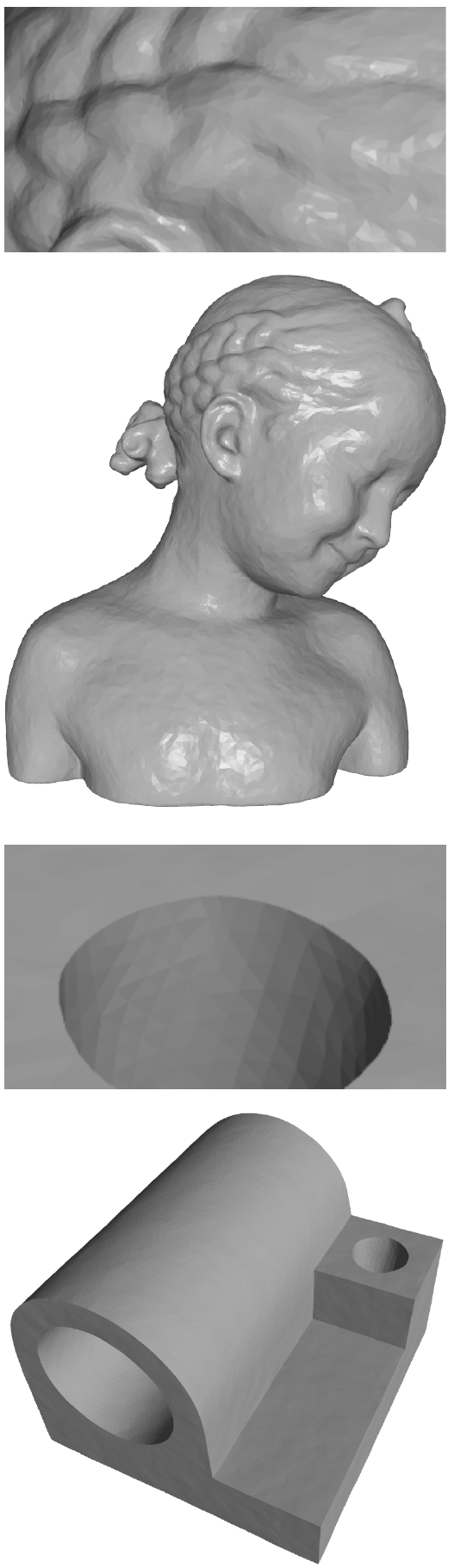}}
    \subfloat[MSTV]{\label{fig:vsMS-HLO-c}\includegraphics[width=0.1\textwidth]{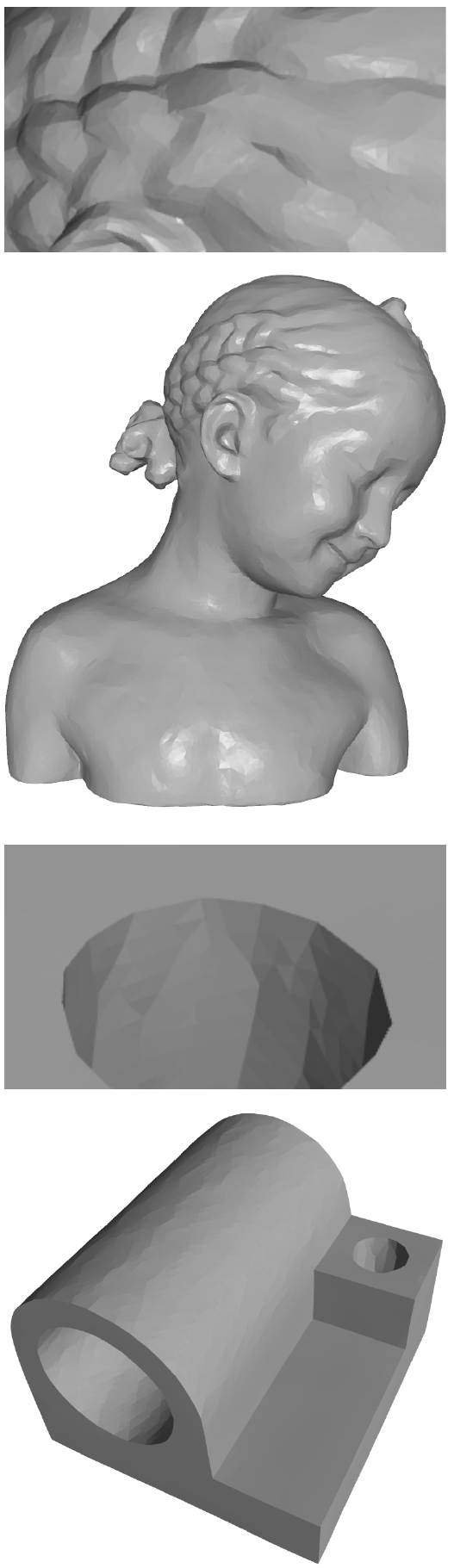}}
    \subfloat[HLO]{\label{fig:vsMS-HLO-d}\includegraphics[width=0.1\textwidth]{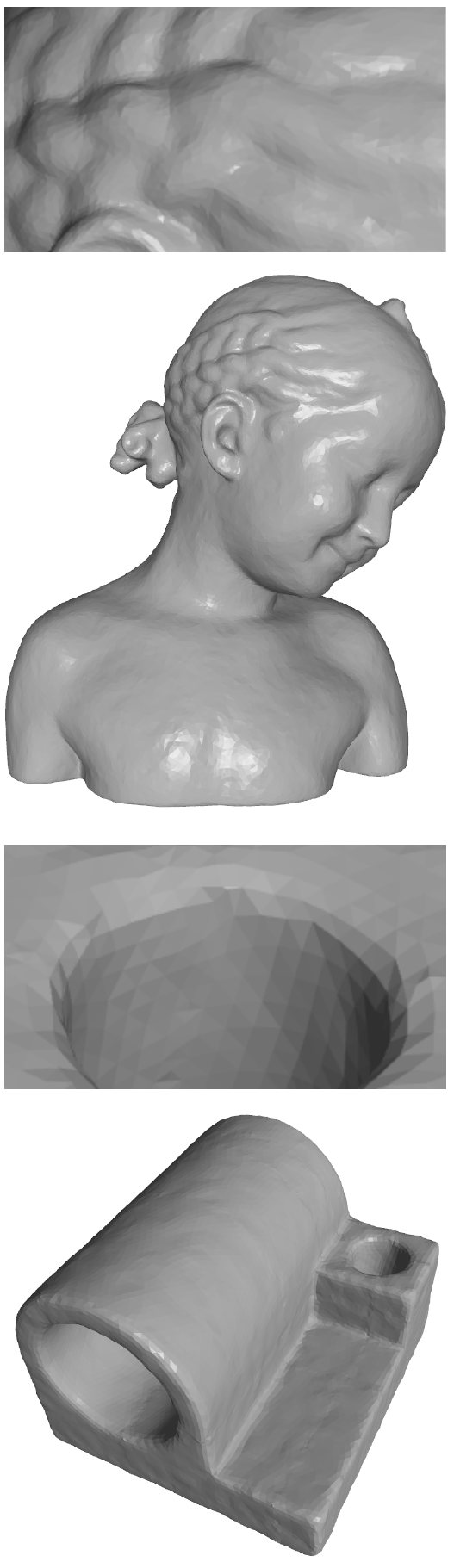}}
    \subfloat[Ours]{\label{fig:vsMS-HLO-e}\includegraphics[width=0.1\textwidth]{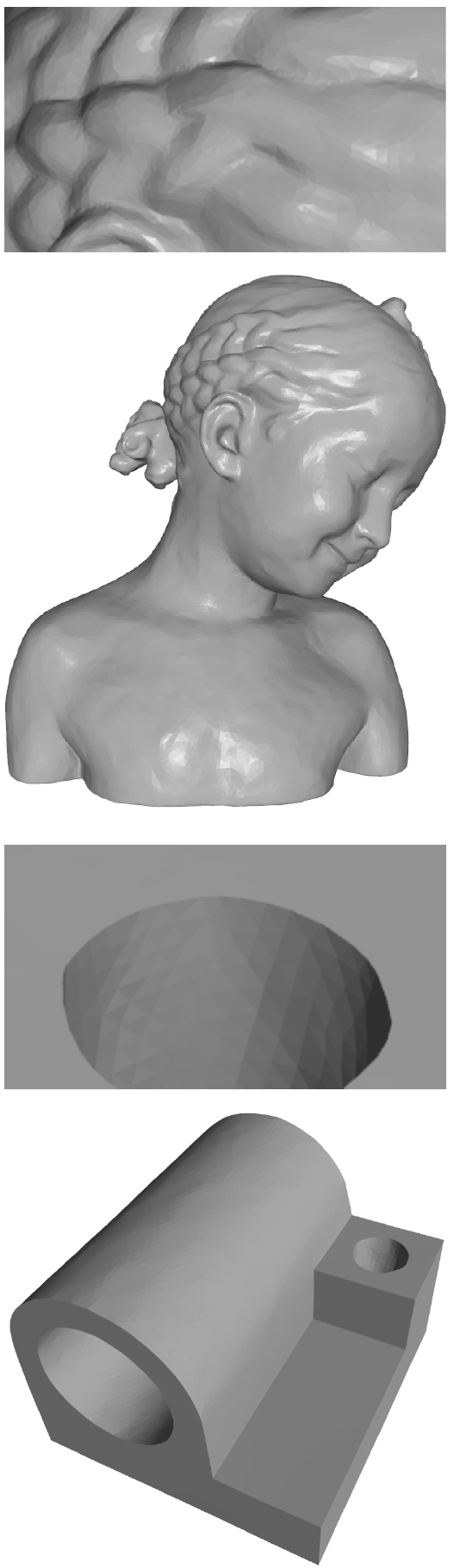}}
    \caption{Comparison of denosing results of Child and Joint, corrupted with $\sigma=0.2 \bar{l}_e$. } \label{fig:vsMS-HLO}
\end{figure}

\begin{table}[!htp]
    \centering
    \setlength{\tabcolsep}{3pt}
    \footnotesize
    \caption{Quantitative evaluation of the results in Fig. \ref{fig:vsMS-HLO} for AT \cite{Liu2020Mesh}, MSTV \cite{Liu2020Mesh}, HLO \cite{pan2020hlo}, and our method TGV. For each result, we list mean angular difference $\theta$ (in degrees), vertex-based mesh-to-mesh error $E_v \ (\times 10^{-2})$ and the execution time (in seconds).}
    \begin{tabular}{lcccccccc}
    \toprule
    \footnotesize Mesh &
    \multicolumn{2}{c}{\footnotesize AT}&
    \multicolumn{2}{c}{\footnotesize MSTV}&
    \multicolumn{2}{c}{\footnotesize HLO} &
    \multicolumn{2}{c}{\footnotesize TGV}\\
    \midrule
    \footnotesize Child& \multicolumn{2}{c|}{7.61,\,0.62;\,3.80} & \multicolumn{2}{c|}{7.22,\,0.55;\,7.06} & \multicolumn{2}{c|}{7.93,\,0.69;\,1.31} & \multicolumn{2}{c}{\textbf{6.29},\,\textbf{0.48};\,25.9}\\
    \footnotesize Joint& \multicolumn{2}{c|}{1.89,\,0.79;\,2.84} & \multicolumn{2}{c|}{1.99,\,0.81;\,2.75} &
    \multicolumn{2}{c|}{6.06,\,2.15;\,0.34} & \multicolumn{2}{c}{\textbf{1.45},\,\textbf{0.70};\,12.4}\\
    \bottomrule
    \end{tabular}
    \label{tab:MSHLO}
\end{table}

\subsection{TGV vs. AT, MSTV, and HLO}
To further demonstrate the effectiveness of TGV, we compare it to the Mumford-Shah methods (AT and MSTV) in \cite{Liu2020Mesh} and the Laplacian diffusion method (HLO) in \cite{pan2020hlo}.
As Fig. \ref{fig:vsMS-HLO} shows, for non-CAD meshes containing different levels of features, AT and TGV produce visually better results. MSTV suffers from staircase artifacts, while HLO tends to smooth weak features.
For CAD meshes, all methods, except HLO, recovers sharp features in the tested noise level.
However, AT causes some bumping in flat regions, while MSTV suffers from staircase artifacts in smooth regions.
In contrast, TGV produces visually more compelling results which are free of noticeable artifacts.
As we can see in Table \ref{tab:MSHLO}, for both CAD and non-CAD meshes, the TGV results show the lowest error values, indicating that TGV outperforms the other three competing methods (AT, MSTV and HLO) numerically.
Table \ref{tab:MSHLO} also lists the CPU execution time for the four methods.
We can see that HLO is the fastest method, while TGV is the slowest.

\begin{figure*}[htb]
    \centering
    \subfloat[Sharpsphere]{\label{fig:bar-sharpsphere}\includegraphics[width=0.32\textwidth]{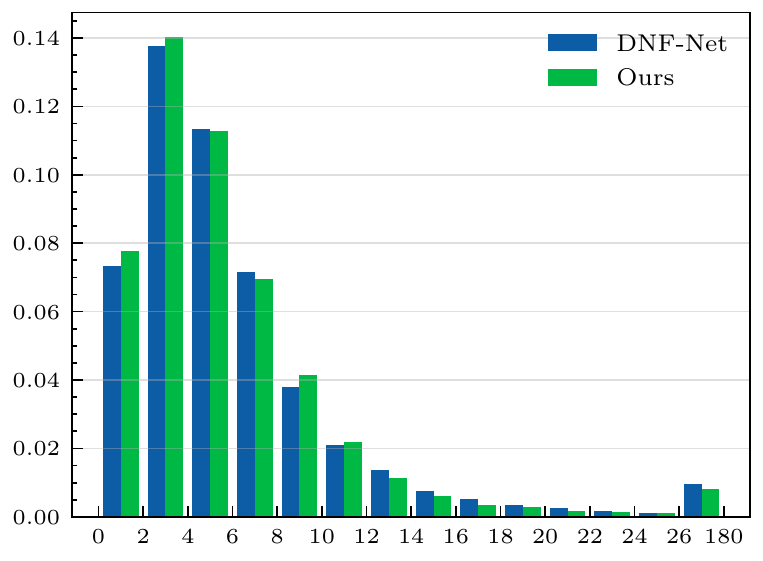}}
    \subfloat[Carter]{\label{fig:bar-carter}\includegraphics[width=0.32\textwidth]{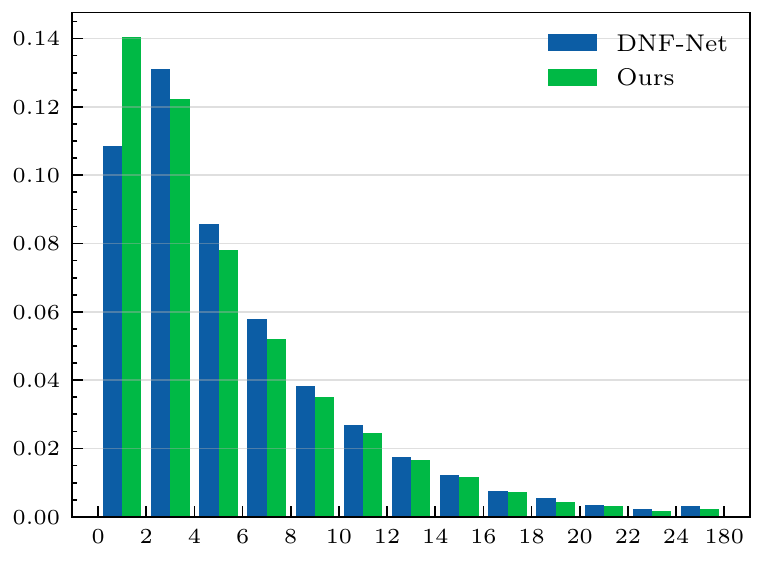}}
    \subfloat[Cone04]{\label{fig:bar-cone}\includegraphics[width=0.32\textwidth]{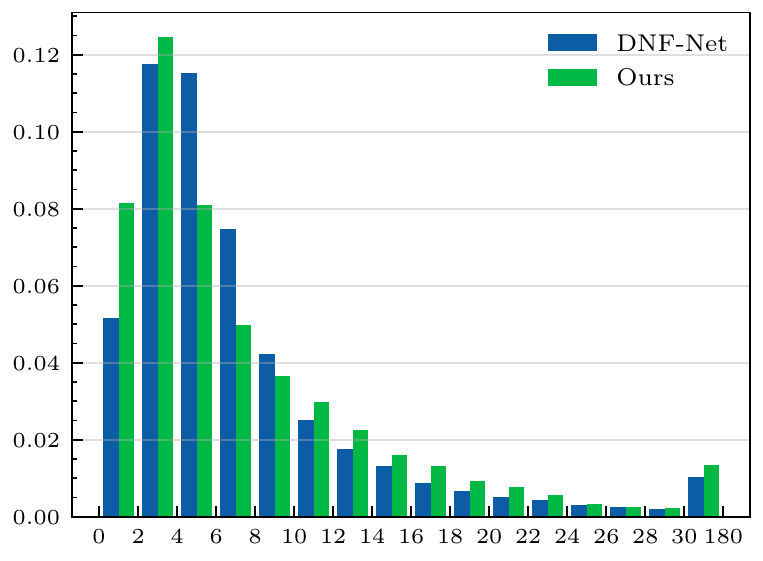}}
    \caption{Histograms of mean angular difference ($\theta$) of the results in Fig. \ref{fig:vsDNFNet}.
    The horizontal axis denotes $\theta$ (in degrees), while the vertical axis denotes the ratio of faces falling in the fixed range of $\theta$. }
    \label{fig:bar}
\end{figure*}

\begin{figure}[htb]
    \centering
    \subfloat[Noisy]{\label{fig:vsDNFNet-a}\includegraphics[width=0.16\textwidth]{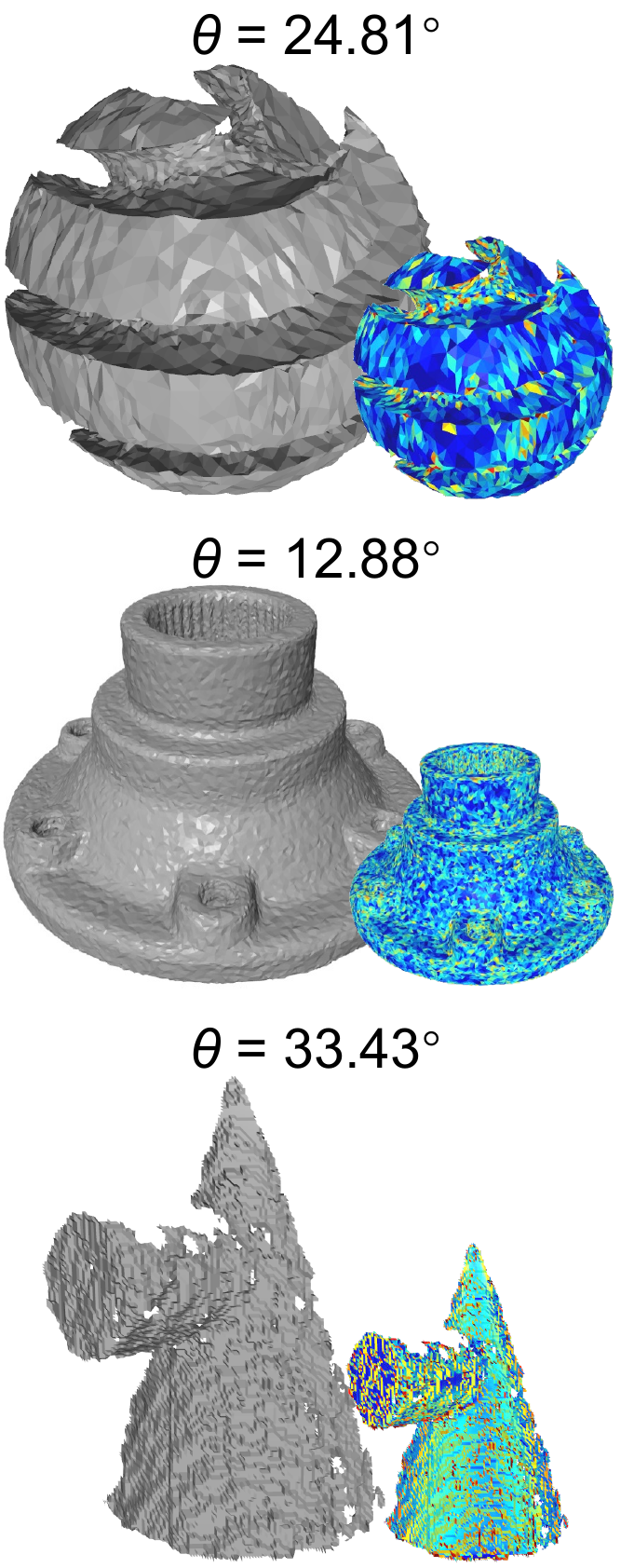}}
    \subfloat[DNF-Net]{\label{fig:vsDNFNet-b}\includegraphics[width=0.16\textwidth]{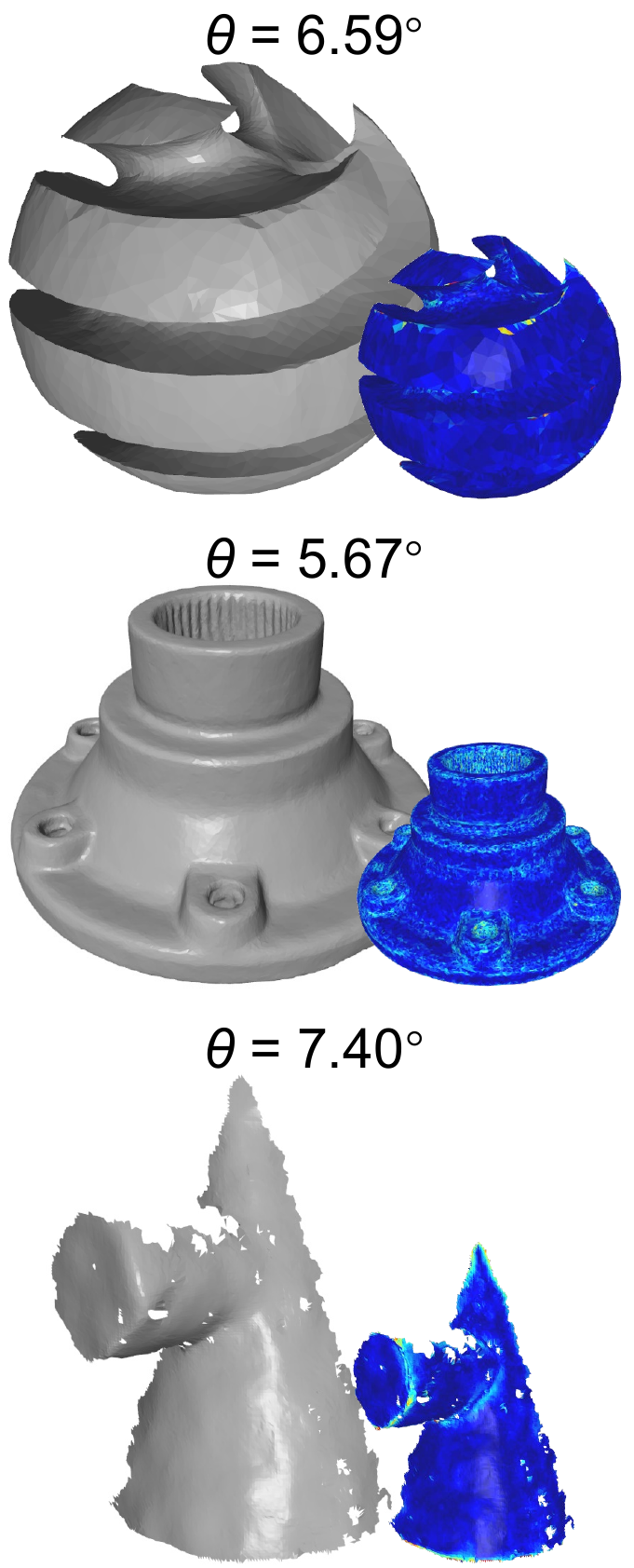}}
    \subfloat[Ours]{\label{fig:vsDNFNet-c}\includegraphics[width=0.16\textwidth]{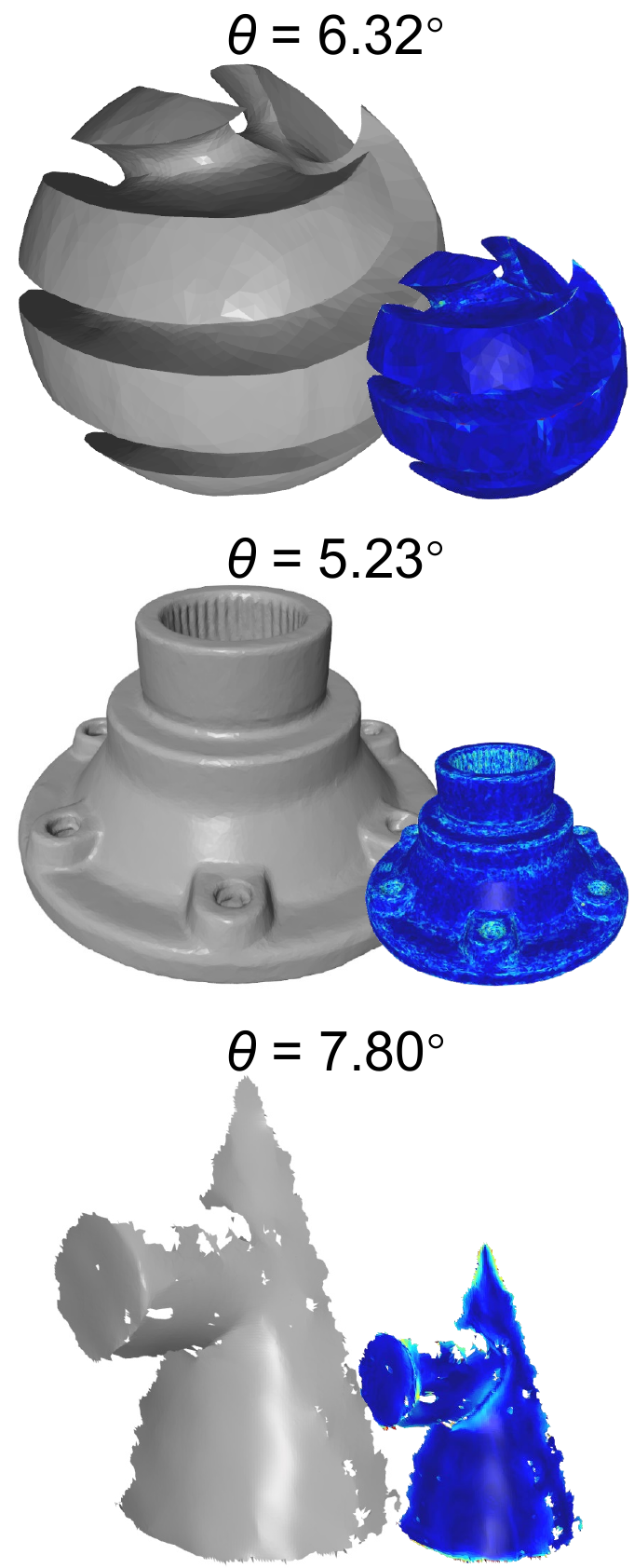}}
    \caption{Comparison between DNF-Net \cite{Li2020DNF} and our method TGV. From left to right: input noisy meshes, denoising results produced by DNF-Net and ours.
    The corresponding error maps, using the mean angular difference between the face normals of denoised meshes and ground truth meshes, are also demonstrated.
    } \label{fig:vsDNFNet}
\end{figure}

\subsection{TGV vs. DNF-Net}
Li et al. \cite{Li2020DNF}  recently proposed an end-to-end deep normal filtering network, named DNF-Net, which has received wide attention.
In Fig. \ref{fig:vsDNFNet}, we compare TGV with DNF-Net on three meshes (Sharpsphere, Carter, and Cone04).
To further visualize the mean angular difference ($\theta$) distribution for the tested meshes, we show the histogram of $\theta$ in Fig. \ref{fig:bar}.
For the mesh containing sharp features and smooth regions (Sharpsphere), our method clearly outperforms DNF-Net in terms of visual quality and the error metric $\theta$; see the top row in Fig. \ref{fig:vsDNFNet}.
For CAD mesh (Carter), both methods produce excellent feature-preserving results.
Nevertheless, the $\theta$ value of our result is lower than that of DNF-Net; see the middle row in Fig. \ref{fig:vsDNFNet}.
For scanned mesh (Cone04), the DNF-Net result has $\theta$ value lower than ours, however it contains bumps in smooth regions. In contrast, our result does not show such artifact, hence we believe visually, our result is better; see the last row in Fig. \ref{fig:vsDNFNet}.
Moreover, as Fig. \ref{fig:bar} shows, TGV consistently produces high quality results that contain more $\theta$ in the range of $[0^{\circ}, 2^{\circ}]$.
Therefore, we argue that TGV performs favorably against DNF-Net.

\subsection{Discussions}
In the following, we discuss the performance of our method in various aspects, including efficacy for irregular sampling, robustness against different levels of noise, and robustness on sampling density (mesh resolution).

\begin{figure}[htb]
    \centering
    \includegraphics[width=0.48\textwidth]{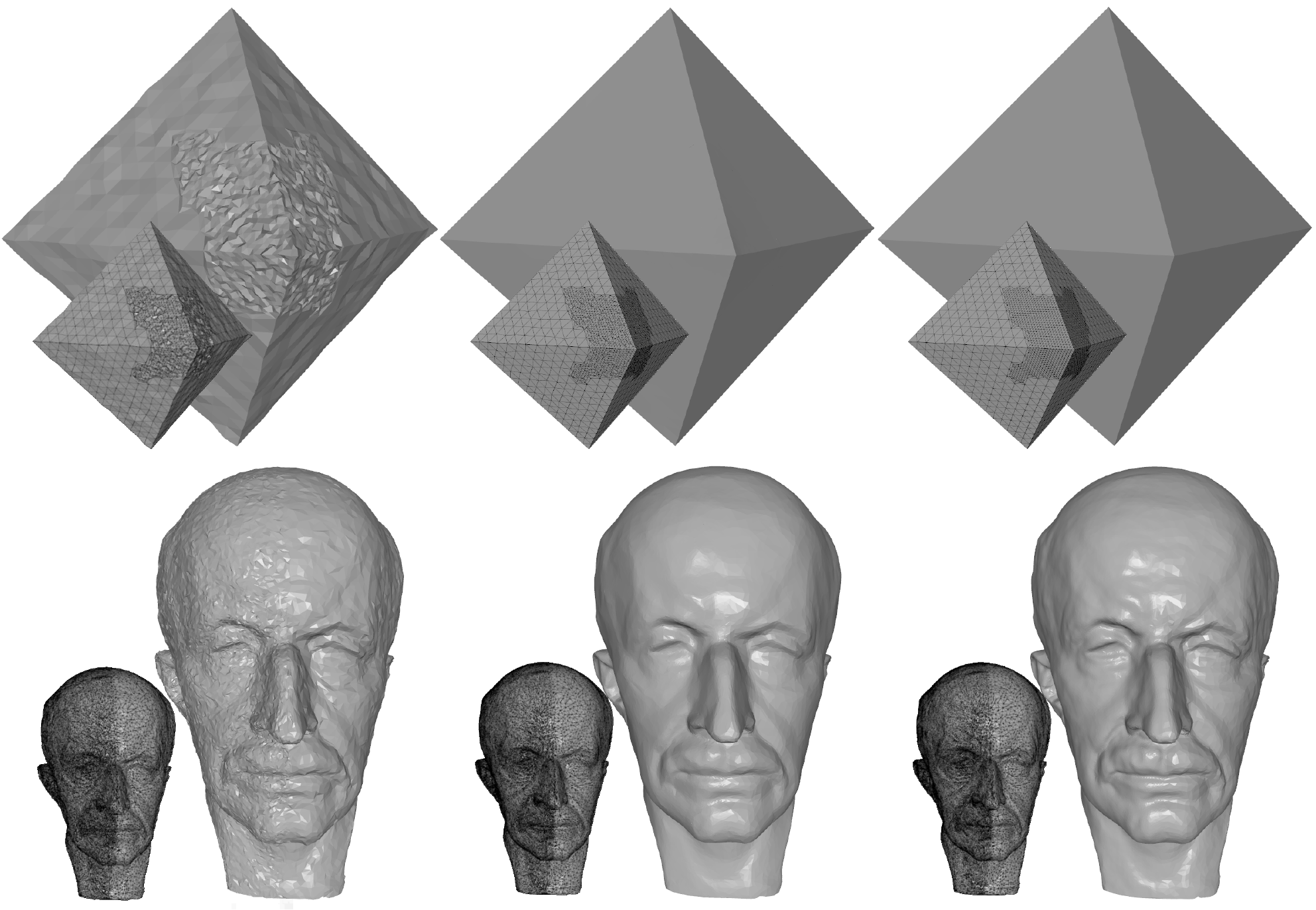}
    \caption{Denoising results for noisy input with non-uniform sampling. From left to right: input noisy meshes, denoising results, and the corresponding clean meshes.} \label{fig:sampling}
\end{figure}

\textbf{Irregular sampling}.
As we have rigorously defined the discrete operators used in our TGV normal filter \eqref{eq:TGV-NormalFilteringModel}, our method is robust against non-uniform sampling.
We demonstrate the robustness of our method against irregular sampling in Fig. \ref{fig:sampling}.
As we can see, although the noisy meshes are of varying density distributions, the obtained results still show compelling quality.

\begin{figure}[htb]
    \centering
    \subfloat[$\sigma=0.2\bar{l}_e$]{\label{fig:stressTesting-a}\includegraphics[width=0.12\textwidth]{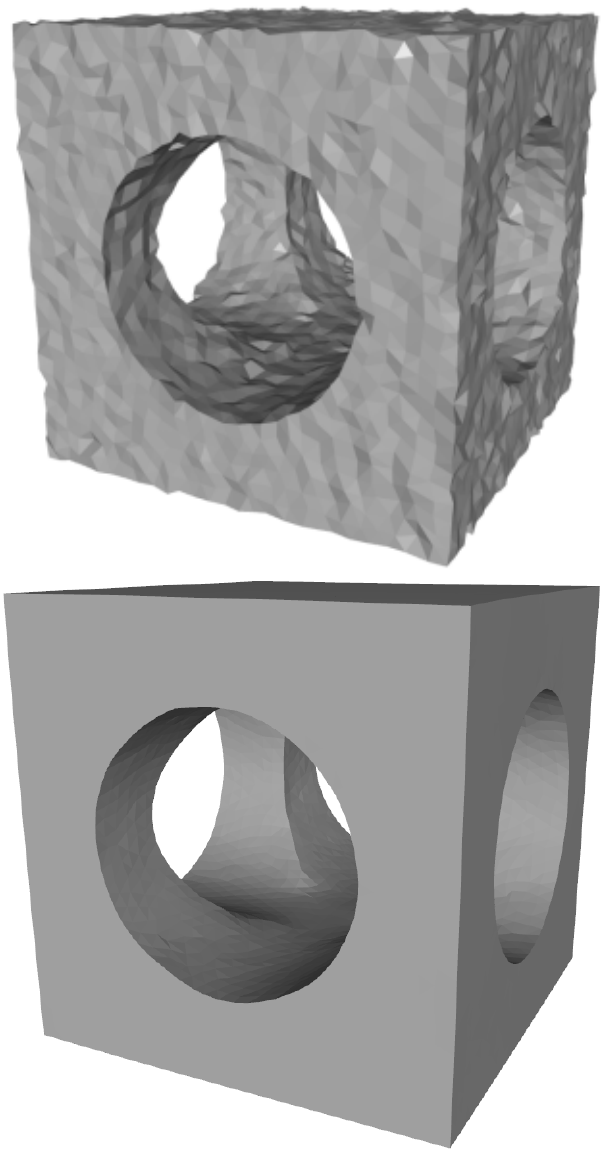}}
    \subfloat[$0.4\bar{l}_e$]{\label{fig:stressTesting-b}\includegraphics[width=0.12\textwidth]{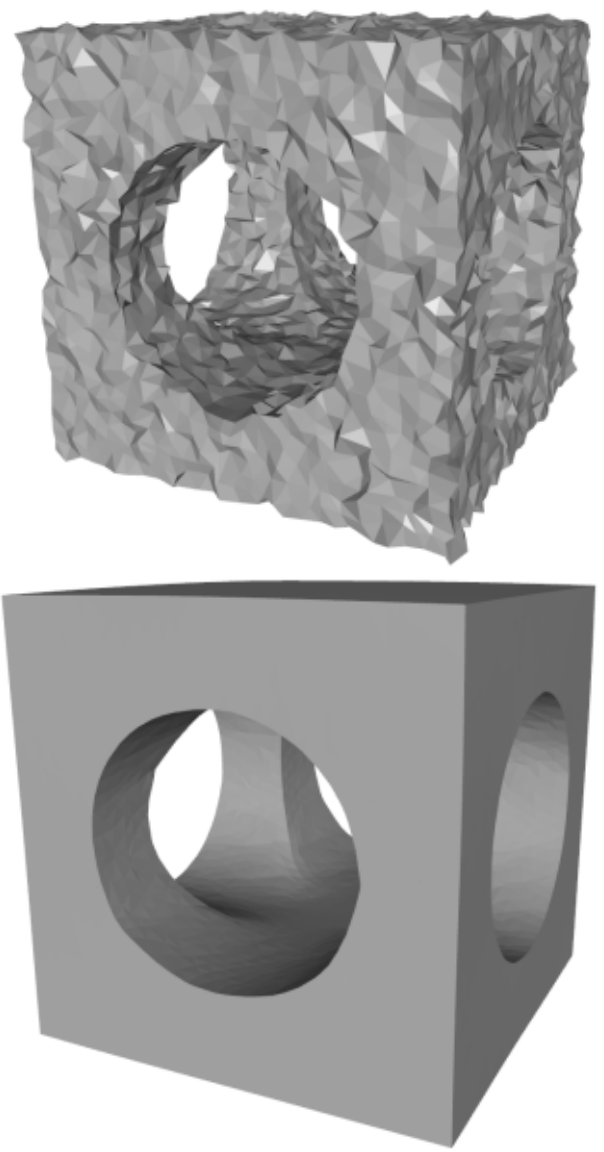}}
    \subfloat[$0.6\bar{l}_e$]{\label{fig:stressTesting-c}\includegraphics[width=0.12\textwidth]{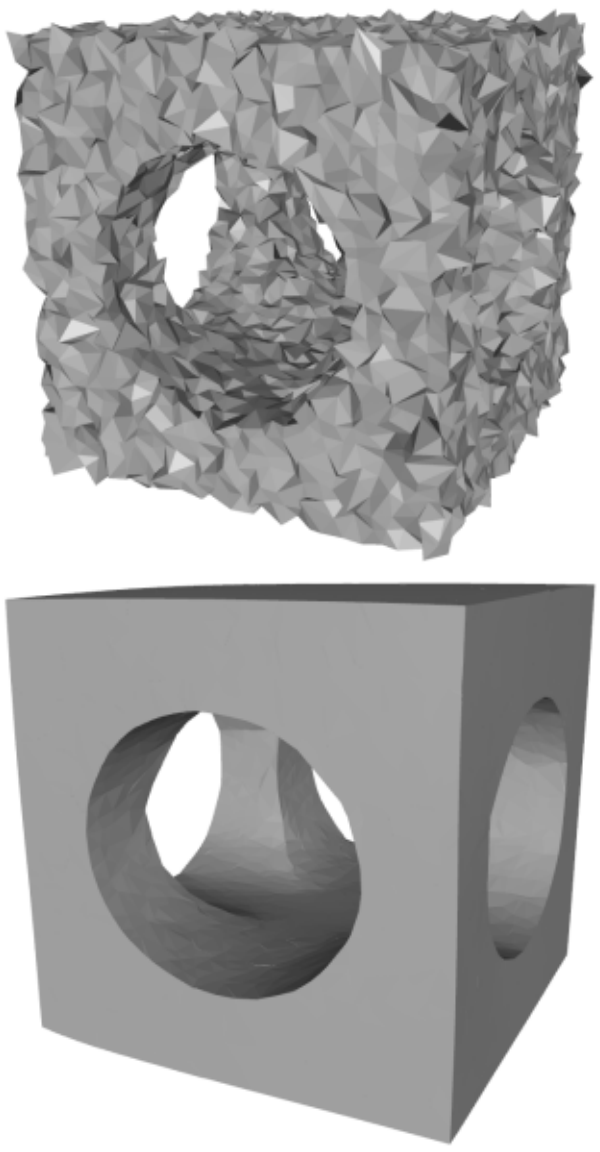}}
    \subfloat[$0.75\bar{l}_e$]{\label{fig:stressTesting-d}\includegraphics[width=0.12\textwidth]{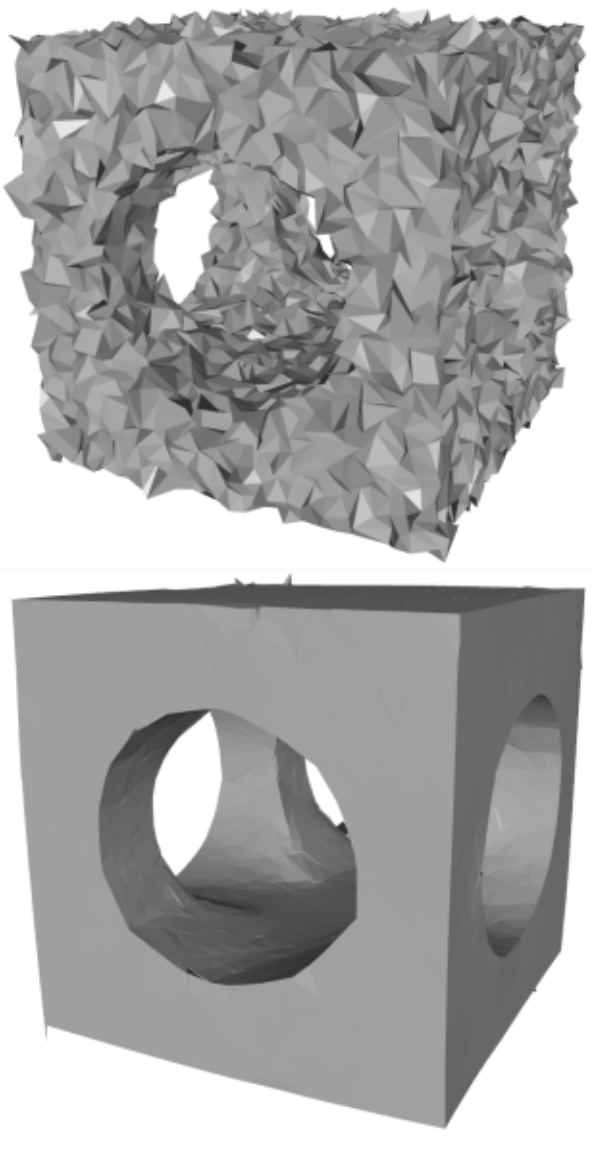}}
    \caption{Denoising results of Part, which is corrupted by different levels of noise. The top row shows the mesh after being corrupted with increasing level of noise, while the bottom row shows the corresponding denoising results.}
    \label{fig:stressTesting}
\end{figure}

\textbf{Stress test}. A stress test of our method with increasing level of noise is presented in Fig. \ref{fig:stressTesting}.
As can be seen, when the noise level is moderate, our method can remove noise effectively, while preserving sharp features and simultaneously recover smooth regions.
Moreover, our method can preserve sharp features even for the mesh under the highest level of noise; see Fig. \ref{fig:stressTesting-c}.
However, when the noise level increases to be larger than the feature size, our method fails to produce satisfactory results; see Fig. \ref{fig:stressTesting-d}.

\begin{figure}[]
    \centering
    \includegraphics[width=0.48\textwidth]{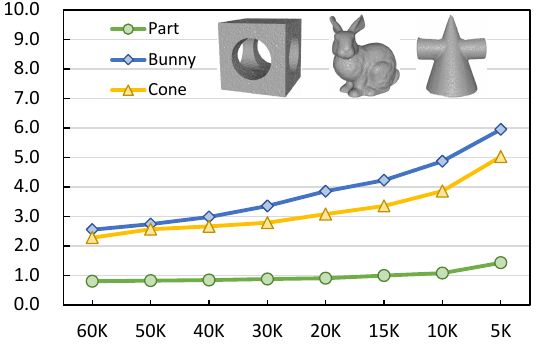}
    \caption{Error plot of mean angular difference ($\theta$) for three meshes (Part, Bunny, Cone) under different resolutions.  All the meshes are corrupted by Gaussian noise with $\sigma=0.1 \bar{l}_e$. The horizontal axis denotes the number of faces in the meshes, while the vertical axis denotes the $\theta$ value.
    }
    \label{fig:resolution}
\end{figure}

\textbf{Sampling density}. A robustness test of our method for varying mesh resolution is shown in Fig. \ref{fig:resolution}.
When the mesh resolution decreases, the $\theta$ value of our results does not change significantly for CAD mesh (Part) or scanned data (Cone).
For non-CAD mesh (Bunny), there is a slight jump of $\theta$ in the case of low-resolution, yet, overall it is still reasonable.
Thus, our method is robust against mesh resolution.

\section{Conclusion} \label{sec:7}
In this work, we present a numerical framework to discretize TGV for triangular meshes.
A normal filter based on vectorial TGV is proposed to smooth normal fields on meshes.
The optimization problem for the proposed filter is efficiently solved by variable-splitting and the augmented Lagrangian method.
Then, vertex positions are updated to match the filtered normal field.
We carefully evaluate our method in various aspects and compare it to the state-of-the-art methods.
Extensive experimental results show that our method has significant advantages in preserving sharp features, recovering smooth transition regions, as well as preventing various artifacts (e.g., staircase artifacts, over-smoothing or over-sharpening effects, and extra noise).
In summary, our method is highly effective for denoising CAD and man-made surfaces that contain sharp features and smooth transition regions.

There are many interesting directions for future research.
The proposed discretized TGV operator can be applied to other geometry processing problems, such as mesh segmentation, reconstruction, simplification, feature detection, etc.
Furthermore, we plan to investigate the possibility to extend our method to point clouds.

\section*{Acknowledgments}
This work was supported by NSF of China (Nos. 62072422, 12001144, 62025207, 62076227, and 61702467), National Key R$\&$D Program of China (No. 2020YFC1523102), NSF of Anhui Province, China (No. 2008085MF195), Youth Science and Technology Foundation of Gansu (No. 20JR5RA050), NSF of Zhejiang Province, China (No. LQ20A010007), and Zhejiang Lab (No. 2019NB0AB03).

\bibliographystyle{IEEEtran}
\bibliography{references}

\begin{IEEEbiography}[{\includegraphics[width=1in,height=1.25in,clip,keepaspectratio]{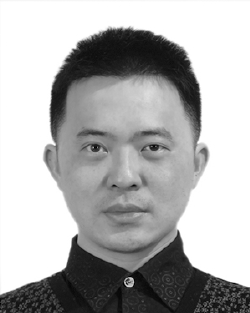}}]{Zheng Liu} is currently an associate professor in China University of Geosciences (Wuhan).
He received PhD from Central China Normal University in 2012.
From 2013 to 2014, he held a post-doctoral position with School of Mathematical Sciences, University of Science and Technology of China.
His research interests include geometry processing, computer graphics, 3D computer vision and deep learning.
\end{IEEEbiography}

\begin{IEEEbiography}[{\includegraphics[width=1in,height=1.25in,clip,keepaspectratio]{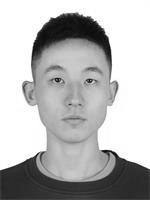}}]{Yanlei Li} is currently a M.S. candidate in China University of Geosciences (Wuhan).
He received B.S. degree from China University of Geosciences (Wuhan), in 2019.
His research interests include geometry processing and computer graphics.
\end{IEEEbiography}

\begin{IEEEbiography}[{\includegraphics[width=1in,height=1.25in,clip,keepaspectratio]{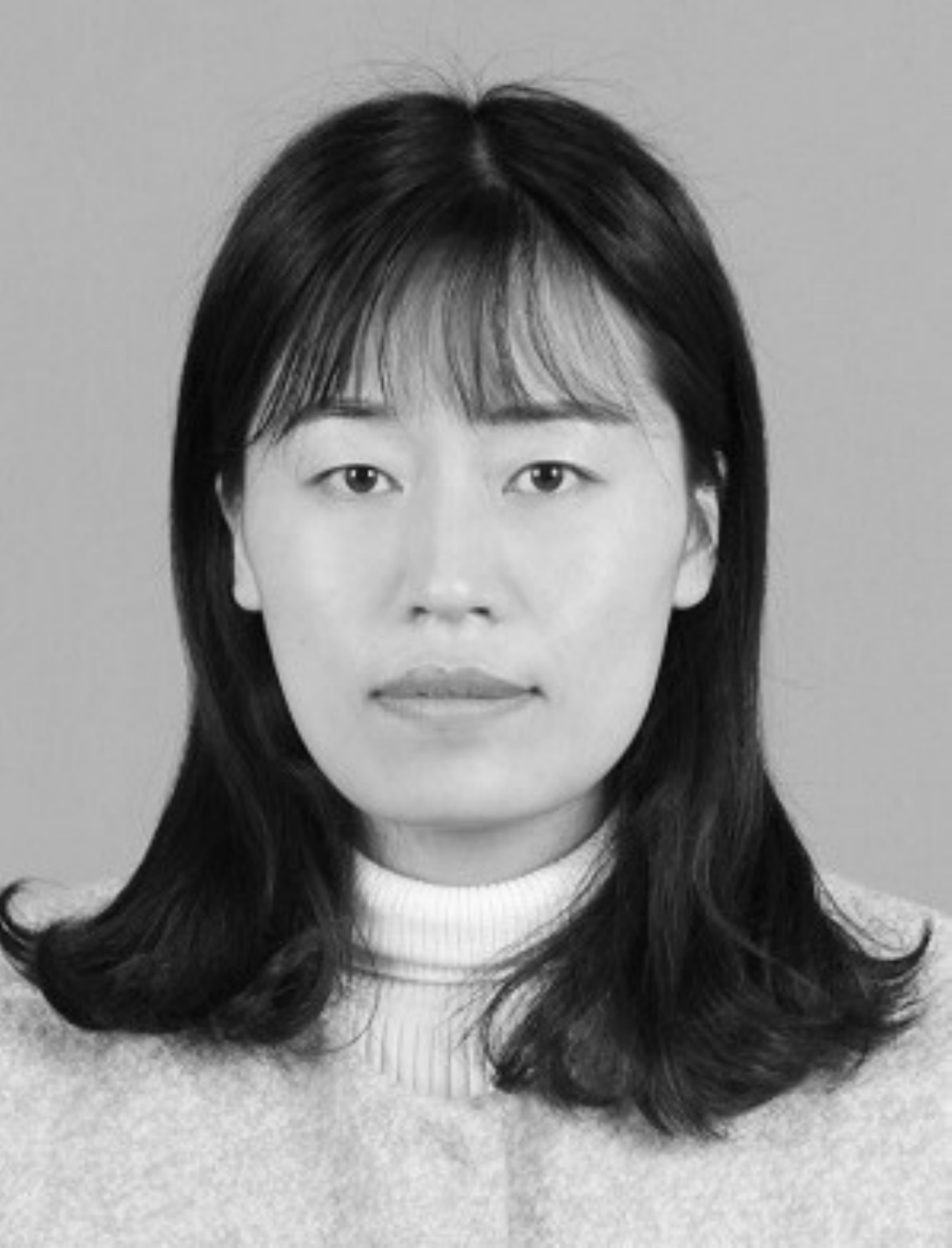}}]{Weina Wang} is currently a lecturer in the department of Mathematics, Hangzhou Dianzi University, China.
She received the Ph.D. degree from the University of Science and Technology of China, in 2018.
Her research interests include numerical optimization and image processing.
\end{IEEEbiography}

\begin{IEEEbiography}[{\includegraphics[width=1in,height=1.25in,clip,keepaspectratio]{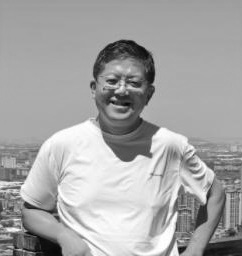}}]{Ligang Liu} is a professor at the University of Science and Technology of China. He received his PhD from Zhejiang University in 2001. He once worked at Microsoft Research Asia, Zhejiang University, and visited Harvard University. His research interests include computer graphics and geometry processing. He serves as the associated editors for journals including IEEE TVCG, IEEE CG$\&$A, CAGD, C$\&$G, The Visual Computer, etc. He served as the conference co-chair of GMP 2017 and the program co-chairs of conferences including Chinagraph 2020, SIAM GD 2019, GMP 2018, CAD/Graphics 2017, CVM 2016, SGP 2015, and SPM 2014. He serves as the steering committee member of GMP and the secretary of Asiagraphics Association.
\end{IEEEbiography}

\begin{IEEEbiography}[{\includegraphics[width=1in,height=1.25in,clip,keepaspectratio]{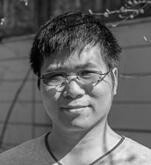}}]{Renjie Chen} is a professor at the University of Science and Technology of China (USTC). He holds a PhD degree from Zhejiang University, China. Before joining USTC, he was a postdoctoral fellow at the Technion--Israel Institute of Technology, a postdoctoral research associate at the University of North Carolina at Chapel Hill, a key researcher in the BeingThere Center in Nanyang Technological University, Singapore, and a senior researcher heading a research group working on 3D geometry and images at the Max Planck Institute for Informatics (MPII) in Saarbrucken, Germany. His research interests includes computer graphics, geometry modeling, computational geometry and glasses-free 3D display.
\end{IEEEbiography}


\end{document}